\documentclass[11pt,reqno]{amsart}
\usepackage{amsmath}
\usepackage{amsxtra}
\usepackage{mathrsfs}
\usepackage{amscd}
\usepackage{amsthm}
\usepackage{amsfonts}
\usepackage{amssymb}
\usepackage{latexsym}
\usepackage{epsf}
\usepackage{longtable}
\usepackage{epsfig}
\usepackage{epstopdf}
\usepackage{tabularx}
\usepackage{pdflscape}
\usepackage{diagbox}
\usepackage{tikz}
\usetikzlibrary{matrix,arrows,decorations.pathmorphing}

\def\Tr{\,{\rm Tr}\, }

\def\be{\begin{equation}}
\def\ee{\end{equation}}
\def\ba{\begin{eqnarray}}
\def\ea{\end{eqnarray}}

\newcommand{\C}{{\mathcal C}}

\renewcommand{\H}{{\mathcal H}}

\newcommand{\N}{{\mathcal N}}

\renewcommand{\P}{{\mathcal P}}

\newcommand{\RR}{\mathbb{R}}

\newcommand{\ZZ}{\mathbb{Z}}
\newcommand{\QQ}{\mathbb{Q}}
\newcommand{\CC}{\mathbb{C}}
\newcommand{\FF}{\mathbb{F}}
\newcommand{\PP}{\mathbb{P}}
\newcommand{\HH}{\mathbb{H}}
\newcommand{\MM}{\mathbb{M}}

\newcommand{\nn}{\nonumber}

\newcommand{\etabox}[2]{\underset{\ ~#2}{\mbox{\scriptsize $#1$}\ \framebox[15pt]{\phantom{a}}}}

\DeclareMathOperator{\ch}{ch}
\DeclareMathOperator{\Syl}{Syl}

\begin{document}

\title[Generalised Mathieu Moonshine]{Generalised Mathieu Moonshine}

\author[]{Matthias R.~Gaberdiel}
\address{ Institut f\"ur Theoretische Physik, ETH Z\"urich, CH-8093 Z\"urich, Switzerland}
\email{gaberdiel@itp.phys.ethz.ch}
%\urladdr{http://www.itp.phys.ethz.ch/people/gaberdim} 

\author[]{Daniel Persson}
\address{Fundamental Physics, Chalmers University of Technology,
  SE-412 96, Gothenburg, Sweden}
  \email{daniel.persson@chalmers.se}
  \urladdr{http://www.danper.se}
  
 \author[]{Henrik Ronellenfitsch}
  \address{Max-Planck-Institut f\"ur Dynamik und Selbstorganisation (MPIDS), 37077 G\"ottingen, Germany}
\email{henrik.ronellenfitsch@ds.mpg.de}

\author[]{Roberto Volpato}
\address{Max-Planck-Institut f\"ur Gravitationsphysik,
Am M\"uhlenberg 1, 14476 Golm, Germany}
\email{roberto.volpato@aei.mpg.de}
  
\begin{abstract}
The Mathieu twisted twining genera, i.e.\ the analogues of Norton's generalised
Moonshine functions, are constructed for the elliptic genus of K3. It is shown
that they satisfy the expected consistency conditions, and that their behaviour 
under modular transformations is controlled by a 3-cocycle in $H^3(M_{24},U(1))$,
just as for the case of holomorphic orbifolds.  This suggests
that a holomorphic VOA may be underlying Mathieu Moonshine.
\end{abstract}

\vspace{3cm}

\maketitle
%\newpage
\tableofcontents

\section{Introduction}
\subsection{Monstrous Moonshine}
In mathematics and physics the word `Moonshine' has come to represent various surprising 
and deep connections between a priori unrelated fields, such as number theory, representation 
theory of finite groups, algebra and quantum field theory. The first and most well-known  example of 
such a connection is, of course, Conway and Norton's \emph{Monstrous Moonshine conjecture} 
\cite{ConwayNorton}.
Their starting point was the observation of McKay that the first few Fourier coefficients of the modular invariant
$J$-function $J(\tau)$ are dimensions of representations of the Monster group $\mathbb{M}$, 
the largest finite simple sporadic group. Conway and Norton then conjectured that to each element 
$g\in\mathbb{M}$ of the Monster, one can associate a function $T_{g}(\tau)$ (the so-called McKay-Thompson series)
on the upper half-plane $\mathbb{H}_+$ that is invariant under some subgroup 
$\Gamma_{g}\subset SL(2,\mathbb{R})$. They also conjectured that the invariance group 
$\Gamma_g$  must be genus zero, meaning that, as a Riemann surface,
the quotient $\Gamma_g\backslash \mathbb{H}_+$ is topologically a sphere. Furthermore, they conjectured that 
$T_g(\tau)$ is in fact the Hauptmodul for $\Gamma_g$. 

Subsequently, Frenkel, Lepowsky and Meurman (FLM) \cite{FLM} constructed a graded 
$\mathbb{M}$-module $V^{\natural}=\bigoplus_{n=-1}^{\infty}V^{\natural}_n$, such that the dimension
of the graded subspaces $a(n)=\text{dim}\, V_n^{\natural}$ reproduce precisely the 
Fourier coefficients in the $q$-expansion of the $J$-function, $J(\tau)=\sum_{n=-1}^{\infty} a(n)q^{n}$, 
$q=e^{2\pi i \tau}$. The coefficients $a_g(n)$ in the Fourier
expansion of the McKay-Thompson series are then identified with the characters 
$\text{Tr}_{V_n^{\natural}}(g)$. Physically, the FLM-module is the $\mathbb{Z}_2$-orbifold of the open
bosonic string compactified on the Leech torus $(\mathbb{R}^{24}/\Lambda_{\text{Leech}})$, 
where $\Lambda_{\text{Leech}}$ is the Leech lattice. 
The FLM-construction provided important clues into the moonshine conjectures, 
but fell short of proving them. The final proof was found by 
Borcherds \cite{Borcherds}, making heavy use of the FLM-module $V^{\natural}$, but also introducing yet 
another set of ingredients to the story, namely generalised Kac-Moody algebras and automorphic denominator 
identities. Thus, by the time the original conjectures were proven, Monstrous Moonshine encompassed 
not only the realms of finite groups and modular forms, but also string theory and infinite-dimensional algebras,
for a nice review see \cite{Terrybook}.

\subsection{Mathieu Moonshine}
Recently, a very interesting new moonshine phenomenon was conjectured by Eguchi, Ooguri and Tachikawa (EOT) 
\cite{Eguchi:2010ej}: they observed that the first few Fourier coefficients of the elliptic genus of K3 
are dimensions of representations of the largest Mathieu group $M_{24}$. This suggests the existence of a 
new moonshine-type connection between Mathieu groups, Jacobi forms and K3 surfaces. The analogues of the
McKay-Thompson series, the \emph{twining elliptic genera} $\phi_g(\tau, z)$, $g\in M_{24}$, were constructed 
in a series of papers \cite{Cheng:2010pq,Gaberdiel:2010ch,Gaberdiel:2010ca,Eguchi:2010fg}, and it was 
shown that they are all weak Jacobi forms of weight $0$ and index $1$ (with multiplier system) for subgroups 
$\Gamma_0(N_g)\subset SL(2,\mathbb{Z})$, where $N_g$ is the order of the element $g$. 
(The elliptic genus $\phi_{\rm K3}$ itself corresponds to taking $g$ to be the identity element.) 
The compatibility with 
$M_{24}$-representations was checked up to the first $600$ coefficients 
\cite{Gaberdiel:2010ca,Eguchi:2010fg}; according to Gannon \cite{GannonMathieu} this is sufficient to prove
that all Fourier coefficients of the elliptic genus of K3 are dimensions of $M_{24}$ representations. 
It is also shown in \cite{GannonMathieu} that the multiplicities of the real $M_{24}$ representations
are always even, see also \cite{CHM}.

Although in a certain sense the above results establish the Mathieu Moonshine conjecture, there are many 
aspects of it that  are much less understood compared to Monstrous Moonshine. For instance, the 
genus zero property of Monstrous Moonshine does not seem to hold for the $M_{24}$-twining genera since 
some of the invariance groups $\Gamma_0(N_g)$ are not genus zero.\footnote{It was proposed in  
\cite{Cheng:2011ay} that the correct generalisation of the genus zero property is a certain 
Rademacher-summability condition, which is indeed satisfied in all cases.} 
More importantly, the analogue of the FLM module $V^{\natural}$ is not yet known for
Mathieu Moonshine, i.e.\ we do not know of any CFT with automorphism group $M_{24}$ whose elliptic genus
reproduces the elliptic genus of K3. In particular, none of the K3-sigma models have this property since
$M_{24}$ is not contained in the automorphism group of any of them \cite{Gaberdiel:2011fg}.

\subsection{Generalised  Monstrous Moonshine}
A few years after the original Monstrous Moonshine conjectures, Norton proposed a generalisation that
he dubbed \emph{Generalised Moonshine} \cite{Norton}. He suggested that for 
each commuting pair of 
Monster group elements $g,h\in\mathbb{M}$ there exists a function $f(g,h;\tau)$ that is also invariant under 
a genus zero subgroup of $SL(2,\mathbb{R})$. These functions generalise the McKay-Thompson series
to which they reduce for $g=e$. Furthermore, they transform into one another under a  
simultaneous action of $SL(2,\mathbb{Z})$ on 
$(\tau;g,h)\in \mathbb{H}_+\times \P(\MM)$, where $\P(\MM)\subset \mathbb{M}\times \mathbb{M}$ is the set of pairs of commuting elements of $\MM$.
Finally, the Fourier coefficients in the $q$-expansion of $f(g,h;\tau)$ are conjectured to be 
dimensions of  (projective) representations of the centraliser of $g$ in $\mathbb{M}$, 
$C_{\mathbb{M}}(g)=\{k\in \mathbb{M} : gk=kg\}$. Although 
this conjecture has been proven in special cases \cite{Dong:1997ea,Hohn,Tuite:1994ni}, the general case is still 
open.\footnote{Carnahan has announced a series of papers \cite{CarnahanI,CarnahanII,CarnahanIII,CarnahanIV} 
which he claims will lead to a complete proof.} 

The Generalised Moonshine conjecture was given a physical interpretation by Dixon, Gins\-parg 
and Harvey \cite{Dixon:1988qd} that was later elaborated upon by Ivanov and Tuite 
\cite{Tuite:1994ni,Ivanov:2002bn,Ivanov:2002an}. 
They showed that the Norton series $f(g,h;\tau)$ arises naturally as  the character in the twisted sector
$V_g^{\natural}$ of an orbifold of the Monster CFT $V^{\natural}$ by the element $g\in \mathbb{M}$, `twined'
by the group element $h$; in standard CFT language, they can therefore be interpreted as 
\be
f(g,h;\tau) = \etabox{g} h =\text{Tr}_{V_g^{\natural}}\left(h\, q^{L_0-1}\right)\ .
\ee
Many of the properties conjectured by Norton can be proved from holomorphic or\-bifold
considerations \cite{Dong:1997ea}.

\subsection{Generalised Mathieu Moonshine}

In this paper, we show that Norton's generalisation also applies to Mathieu Moonshine. 
More specifically, for every pair of commuting group elements $g,h\in M_{24}$, 
we construct `twisted twining genera' $\phi_{g,h} : \mathbb{H}_+ \times \mathbb{C} \to \mathbb{C}$
that either vanish or are weak Jacobi forms of weight $0$ and index $1$ for some 
$\Gamma_{g,h}\subset SL(2,\mathbb{Z})$. For $g=e$, they reduce to the twining genera of 
\cite{Cheng:2010pq,Gaberdiel:2010ch,Gaberdiel:2010ca,Eguchi:2010fg}, and they transform
under the modular group $SL(2,\mathbb{Z})$ into one another. The multiplier phases that appear
in these transformations behave as though these twisted twining genera were twisted twining characters
of a holomorphic orbifold; in particular, they are controlled by a  3-cocycle 
$\alpha\in H^3(M_{24},U(1))$ via a formula that was  first written down by Dijkgraaf and Witten 
in \cite{Dijkgraaf:1989pz}. Furthermore, the Fourier coefficients of the twisted elliptic genera
$f(g,e;\tau)$ equal dimensions of projective representations of the centraliser 
$C_{M_{24}}(g)$, and the central extension that characterises the projective representation is again
determined by the cohomology class $\alpha\in H^3(M_{24}, U(1))$. 

\smallskip

The idea for using the cohomology group $H^3(M_{24}, U(1))$ and the formalism of Dijkgraaf and Witten 
in order to understand the multiplier phases of the twisted twining genera was first suggested to us by 
Terry Gannon in 2011 \cite{Gannontalk}, see also \cite{GannonMathieu}.
 Mason has also speculated \cite{MasonCohomology} that 
$H^3(\mathbb{M}, U(1))$ should play a similar role in the context of Generalised Monstrous Moonshine, although 
this has, to our knowledge, not yet been worked out (partially, because $H^3(\mathbb{M}, U(1))$  is unknown).
\smallskip 

We should also mention that Mason has previously proposed a version of Generalised Moonshine for
$M_{24}$ \cite{MasonGen},  where the role 
of the Norton series $f(g,h;\tau)$  is played by  products of eta functions $\eta(\tau)$. The precise relation between our 
twisted twining genera and Mason's  eta products will be explained elsewhere \cite{GenMathieuII}.

\subsection{Outline}
The paper is organised as follows. In section \ref{sec_ttgenera} we introduce the twisted twining genera $\phi_{g,h}$, and
discuss their general properties, in particular the expected behaviour under modular transformations. Based on 
these properties we then explain
how many independent twisted twining genera there are, and list all of them explicitly, see table~\ref{t:res}.
The detailed derivation of these genera is illustrated in section \ref{s:cohomology}.
First, we review the structure of holomorphic orbifolds and explain in particular, the role
of the 3-cocycle in characterising the various transformation properties of the twisted twining characters. We then 
postulate that the transformation properties of the twisted twining genera are similarly constrained. In support of this assumption, we prove that there exists a unique 3-cocycle in $H^3(M_{24},U(1))$ compatible with the known properties of the (untwisted) twining genera of \cite{Cheng:2010pq,Gaberdiel:2010ch,Gaberdiel:2010ca,Eguchi:2010fg}.  This then allows us to find all  twisted twining genera explicitly. We also
check that,  up to the first 500 levels in each twisted sector, the resulting functions are compatible with the requirement that they arise from the appropriate
projective representation of the centraliser. In section~4 we subject our results to two independent consistency
checks. First, for group elements whose orbifold leads again to a K3 sigma-model, we calculate the twining genera
of the orbifold from the twisted twining genera of the original theory, and show that we reproduce answers from
\cite{Gaberdiel:2010ca,Eguchi:2010fg}. However, as it turns out, the relevant group element is sometimes different
from what one would have expected, and we explain this `relabelling' phenomenon in quite some detail in 
section~\ref{s:relabelling},  see also section~\ref{s:relex} for an explicit example. Secondly, we explain the 
vanishing of some of the twisted twining genera from a geometrical point of view, see section~\ref{sec:ttgex}.
Finally we end with some conclusions and open problems in section \ref{sec_conclusions}. We have 
relegated some of the more technical material to various appendices to which we 
refer throughout the main body of the paper. Moreover, the appendices contain the  
character tables of the centralisers and the decompositions of the twisted sectors, up to the first 20 levels. The decompositions up to 500 levels, as well as the details of the computations requiring computer support, are collected in the ancillary files in the arXiv repository of this paper.

\section{Twisted Twining Genera}
\label{sec_ttgenera}

Let us begin by introducing the twisted twining genera that are the main object of this paper. 
Suppose we are given a K3 sigma-model $\H$, i.e.\ a CFT describing string propagation on K3,
whose automorphism group contains two commuting elements 
$g,h\in M_{24}$. Then we can consider the orbifold of the sigma-model by $g$, and in particular, define the
$g$-twisted sector $\H_g$. Since $g$ and $h$ commute, $h$ gives rise to an action on $\H_g$, and we can
define the \emph{twisted twining genus} $\phi_{g,h}(\tau,z)$ by 
\be
\phi_{g,h}(\tau,z)=\Tr_{\H_g}\Big(h (-1)^{J_0+\bar{J}_0}q^{L_0-\frac{c}{24}}{\bar q}^{\bar L_0-\frac{\bar c}{24}}y^{J_0}\Big)\ ,
\label{eqn:twining-trace}
\ee 
where $q=e^{2\pi i \tau}$ and $y=e^{2\pi iz}$. 
We expect that $\phi_{g, h}\, :\, \mathbb{H}_+\times \mathbb{C}\rightarrow \mathbb{C}$ is holomorphic in both $\tau$ and $z$. 
Furthermore, since the elliptic genus is independent of the choice of the underlying K3 sigma-model, we expect that the same is true 
for  these twisted twining genera, i.e.\ we expect that (\ref{eqn:twining-trace}) does not depend on the choice of $\H$ (as long
as $g$ and $h$ are automorphisms of $\H$). By construction, for $g=e$ the identity element in $M_{24}$, the 
twisted twining genus $\phi_{e,h}$ agrees with the twining genus  $\phi_h$ considered in 
\cite{Cheng:2010pq,Gaberdiel:2010ch,Gaberdiel:2010ca,Eguchi:2010fg}; in particular, for $g=h=e$ 
$\phi_{e,e}$ is just the elliptic genus of K3. 

Unfortunately, while for some commuting pairs $(g,h)$ an actual K3 sigma-model
for which both $g$ and $h$ are automorphisms can be found, this is not true in general \cite{Gaberdiel:2011fg};
in fact, this problem already arises for the usual twining genera, i.e.\ for the pairs $(e,h)$. 
In the spirit of the EOT conjecture, we shall nevertheless
assume that twisted twining genera $\phi_{g,h}$ can be defined --- albeit not directly by a formula of the form 
(\ref{eqn:twining-trace}) ---  for \emph{all} commuting pairs
$g,h\in M_{24}$. The fact that our construction will be successful is an a posteriori justification for this assumption.

\subsection{Properties of the twisted twining genera}
\label{sec_ttprop}

The definition of the twisted twining genera $\phi_{g,h}$ in terms of (\ref{eqn:twining-trace}) suggests that they
should satisfy the following properties:
\begin{enumerate}
\item Elliptic and modular properties:
\begin{align} 
&\phi_{g,h}(\tau,z+ \ell \tau + \ell') = e^{-2 \pi i(\ell^2 \tau+ 2 \ell z)}\, \phi_{g,h}(\tau,z) &
\ell,\ell'\in \ZZ\\
&\phi_{g,h}\Bigl(\frac{a \tau + b}{c \tau + d} , \frac{z}{c \tau + d}\Bigr) =
\chi_{g,h}(\begin{smallmatrix} a & b \\ c & d  \end{smallmatrix})\,
e^{ 2 \pi i  \frac{c z^2}{c \tau + d} } \, \phi_{h^cg^a,h^dg^b}(\tau,z) \ ,
& \hspace*{-0.9cm} (\begin{smallmatrix} a & b \\ c & d  \end{smallmatrix} )\in SL(2,\ZZ) 
\label{modcond}
\end{align} 
for a certain multiplier $\chi_{g,h}:SL(2,\ZZ)\to  U(1)$.
In particular, each $\phi_{g,h}$ is a weak Jacobi form of weight 0 and index 1 with multiplier $\chi_{g,h}$ 
under a subgroup $\Gamma_{g,h}$ of $SL(2,\ZZ)$ (see \cite{EichlerZagier} for the definitions).
\item Invariance  under conjugation of the pair $g,h$ in $M_{24}$, 
\be \label{conj1} 
\phi_{g,h}(\tau,z)=\xi_{g,h}(k)\, \phi_{k^{-1}gk,k^{-1}hk}(\tau,z)\ ,\qquad
k\in M_{24}\ ,
\ee
where $\xi_{g,h}(k)$ is a phase.
\item If $g\in M_{24}$ has order $N$, the twisted twining genera $\phi_{g,h}$ have an expansion of the form 
\be
\phi_{g,h}(\tau, z) = \sum_{\substack{r\in\lambda_g+\ZZ/N\\ r\ge 0}} \Tr_{\H_{g,r}}\bigl(\rho_{g,r}(h)\bigr)
\ch_{h=\frac{1}{4}+r, \ell}(\tau, z)\ , \label{eqn:decomp}
\ee 
where $\lambda_g\in\mathbb{Q}$, and $\ch_{h, \ell}(\tau,z)$ are elliptic genera of
Ramond representations of the $\N = 4$ superconformal algebra at central charge $c=6$. (Here $\ell=\tfrac{1}{2}$,
except possibly for $h=\tfrac{1}{4}$, where $\ell=0$ is also possible --- if both $\ell=0,\tfrac{1}{2}$ 
appear for $r=0$, it is understood that there are two such terms in the above sum. We use the same conventions
for the elliptic genera as e.g.\ in \cite{Eguchi:1987wf}.) Furthermore, each vector 
space $\H_{g,r}$ is finite dimensional, and it carries a projective representation $\rho_{g,r}$ of the centraliser 
$C_{M_{24}}(g)$ of $g$ in $M_{24}$, such that 
\begin{equation}
\rho_{g,r}(g) = e^{2\pi i r} \ , \qquad 
\rho_{g,r}(h_1) \, \rho_{g,r}(h_2) = c_g(h_1,h_2) \, \rho_{g,r}(h_1 h_2) \ ,
\end{equation}
for all $h_1,h_2\in C_{M_{24}}(g)$. Here $c_{g}: C_{M_{24}}(g) \times C_{M_{24}}(g) \to U(1)$ 
is independent of $r$, and satisfies the cocycle condition
\be 
c_g(h_1,h_2)\, c_g(h_1h_2,h_3)=c_g(h_1,h_2h_3)\, c_g(h_2,h_3)
\ee
for all $h_1,h_2,h_3\in C_{M_{24}}(g)$.
\item For $g=e$, where $e$ is the identity element of $M_{24}$, the functions $\phi_{e,h}$ correspond to the 
twining genera considered in \cite{Cheng:2010pq}--\cite{Eguchi:2010fg}. In particular, $\phi_{e,e}$ is the K3 elliptic genus. 

\end{enumerate} 

Since the elliptic genus behaves essentially like the character of a holomorphic CFT --- in particular, it is modular invariant
and holomorphic by itself --- it is natural to believe that the same will be true for the twisted twining genera, i.e.\ that they
will be analogous to twisted twining characters of a holomorphic CFT. As we will review in more detail in 
section~\ref{s:cohomology}, the modular properties of the twisted twining characters of a holomorphic CFT are controlled by 
a $3$-cocycle $\alpha:G\times G\times G\to U(1)$ representing a cohomology class in the third cohomology group 
$H^3(G,U(1))$. (Some background material about group cohomology can be found in appendix~\ref{app:groupcoho}.)
We will therefore postulate that 
\begin{enumerate}\setcounter{enumi}{4}
\item The multipliers $\chi_{g,h}$, the phases $\xi_{g,h}$, and the 2-cocycles 
$c_g$ associated with the projective representations $\rho_{g,r}$ are completely determined 
(by the same formulas as for holomorphic orbifolds) in terms of a 
$3$-cocycle $\alpha$ representing a class in $H^3(M_{24},U(1))$.
\end{enumerate}

The third cohomology group of $M_{24}$  was only recently computed with the result  \cite{DutourEllis}\footnote{Note that 
for a finite group $G$ one has the isomorphisms 
$$H_{n-1}(G,\mathbb{Z})\cong H^n(G,\mathbb{Z}), \qquad \qquad H^{n}(G,\mathbb{Z})\cong H^{n-1}(G,U(1))\ ,$$
which in particular imply that $H_3(M_{24}, \mathbb{Z})\cong H^3(M_{24}, U(1)).$}
\be\label{cohomo}
H^3(M_{24}, U(1))\cong \ZZ_{12}\ .
\ee
The fact that this group is known explicitly plays a crucial role in our analysis. The specific cohomology class $[\alpha] \in H^3(M_{24},U(1))$ that is relevant in our context is uniquely determined by the condition that it reproduces the multiplier system for the 
twining genera $\phi_{e,h}$ as described in \cite{Gaberdiel:2010ca}, namely
\be \label{untwmult}
\chi_{e,h}(\begin{smallmatrix} a & b \\ c & d  \end{smallmatrix})=e^{\frac{2\pi i cd}{o(h)\ell(h)}}\ ,\qquad 
\left(\begin{smallmatrix} a & b \\ c & d  \end{smallmatrix}\right)\in \Gamma_0(o(h))\ .
\ee 
Here, $o(h)$ is the order of $h$ and $\ell(h)$ is the length of the smallest cycle, when $h\in M_{24}$ is regarded 
as a permutation of $24$ symbols \cite{Cheng:2011ay}.
Indeed, since $\ell({\rm 12B})=12$, it follows that 
$\alpha$ must correspond to the primitive generator of $H^3(M_{24}, U(1))$.
The main result of our paper can now be stated as follows:

\medskip

%\rule[0.1in]{15.3cm}{0.2mm} \\
\emph{There exists a unique set of functions $\phi_{g,h}$ (unique up to redefinitions by $(g,h)$-dependent but 
otherwise constant phases)
and a unique cohomology class $[\alpha]\in H^3(M_{24},U(1))$ such that all conditions (A)--(E) are satisfied
(condition (C) has been verified only for the first 500 representations $\rho_{g,r}$ for each $g\in M_{24}$).}
%\rule[0.1in]{15.3cm}{0.2mm} \\

\medskip
When $g=e$ is the trivial element of $M_{24}$, the existence of all representations $\rho_{e,r}$ fitting eq.~\eqref{eqn:decomp} has been recently proven in \cite{GannonMathieu}. When $g\neq e$,  condition (C) is, strictly speaking, still a conjecture. However, we provide compelling numerical evidence in favour of it, by verifying eq.~\eqref{eqn:decomp} for the first 500 representations $\rho_{g,r}$ in each twisted sector. A complete proof of \eqref{eqn:decomp} should be possible along the lines of \cite{GannonMathieu}.

In the following we want to illustrate how the set of functions $\phi_{g,h}$ can be determined, and which explicit form it takes. 
As it turns out, a surprisingly large number of the twisted twining genera vanish, as we shall
now explain.

\subsection{Cohomological obstructions}\label{sec_cohobstr}

For some pairs of commuting elements $g,h\in M_{24}$, the transformation properties above can only be 
satisfied if $\phi_{g,h}$ vanishes identically. In this case, we will say that the corresponding twisted twining genus is 
\emph{obstructed}. As will be shown in more detail in section~\ref{s:cohomology}, these obstructions are also 
controlled by the cohomology class of the cocycle $\alpha\in H^3(M_{24},U(1))$.\footnote{The fact that such obstructions might 
exist was first suggested to us by T. Gannon.}

In order to understand the origin of these obstructions, let us first derive some general consequences of our 
assumptions (A)--(E).  First we note that the $SL(2,\ZZ)$ action on the twisted twining characters can be extended to 
$GL(2,\ZZ)$ by setting
\be\label{complconj1}
\phi_{g,h}^*(\tau,z)= \chi_{g,h}(\begin{smallmatrix} 1 & 0 \\ 0 & -1  \end{smallmatrix})\, \, \phi_{g,h^{-1}}(\tau,z)\ ,
\ee
where $\phi_{g,h}^*(\tau,z)$ is obtained by taking the complex conjugate of all Fourier coefficients of 
$\phi_{g,h}(\tau,z)$
\be \label{stardef}
\phi_{g,h}^*(\tau,z)=\overline{\phi_{g,h}(-\bar\tau,-\bar z)}\ .
\ee  
The identity (\ref{complconj1}) follows from eq.~\eqref{eqn:decomp}, together with the observation that, for any 
projective  representation $R$ of a finite group $G$,  $\Tr_R(h^{-1})$ equals $\overline{\Tr_R(h)}$ up to a phase, which 
we have denoted by 
$\chi_{g,h}(\begin{smallmatrix} 1 & 0 \\ 0 & -1  \end{smallmatrix})$. As we will see, this phase also depends  on 
the $3$-cocycle $\alpha$.
We also note that since the $\N=4$ characters are invariant under $z\to -z$, we have the identity 
\be\label{selfconj} 
\phi_{g,h}(\tau,-z)=\phi_{g,h}(\tau,z)\ .
\ee 
With these preparations we can now describe two possible kinds of obstructions.
\smallskip

\noindent {\bf Obstruction 1:} Let us consider pairwise commuting $g,h,k$. Then, by \eqref{conj1}, we have
\be \phi_{g,h}(\tau,z)=\xi_{g,h}(k)\, \phi_{g,h}(\tau,z)\ , \label{eqn:obstruction1}
\ee and if
\be \xi_{g,h}(k)\neq 1\ ,
\ee 
it follows that $\phi_{g,h}(\tau,z)=0$.
\smallskip

\noindent {\bf Obstruction 2:}  Suppose there are $g,h,k\in M_{24}$, with $g$ and $h$ commuting and 
\be k^{-1}g^{-1}k=g\ ,\qquad k^{-1}h^{-1}k=h\ ,
\ee 
i.e.\ the commuting pair $(g,h)$ is conjugate within $M_{24}$ to the pair $(g^{-1},h^{-1})$. Then, 
by eqs.~\eqref{charge} and \eqref{conj} (see section~\ref{s:cohomology} for details), we obtain the relations
\begin{align} 
\phi_{g,h}(\tau,z)&=\chi_{g,h}(\begin{smallmatrix} -1 & 0 \\ 0 & -1  \end{smallmatrix})\, 
\phi_{g^{-1},h^{-1}}(\tau,-z) \label{eq1} \\
&=\chi_{g,h}(\begin{smallmatrix} -1 & 0 \\ 0 & -1  \end{smallmatrix})\, \xi_{g^{-1},h^{-1}}(k)\, \phi_{g,h}(\tau,-z)\ .
\end{align} 
Therefore, if
\be 
\chi_{g,h}(\begin{smallmatrix} -1 & 0 \\ 0 & -1  \end{smallmatrix})\, \xi_{g^{-1},h^{-1}}(k)\neq 1\ ,
\ee 
eq.~\eqref{selfconj} implies $\phi_{g,h}(\tau,z)=0$.

As we shall see, these two obstructions are responsible for the fact that most twisted twining genera  vanish.

\subsection{Classification of independent twisted twining genera}\label{s:classif}

Our next aim is to enumerate all possible independent twisted twining genera. Let us denote the set of 
commuting group elements by 
\be 
\P=\{(g,h)\in M_{24}\times M_{24}\mid gh=hg\} \ . 
\ee
This set carries an action of $GL(2,\ZZ)\times M_{24}$, given by
\be \label{groupa}
(g,h) \mapsto 
(k^{-1}(g^ah^c)k,k^{-1}(g^bh^d)k)\ ,\quad \left(\begin{smallmatrix}
a & b\\ c & d
\end{smallmatrix} \right) \in GL(2,\ZZ)\ ,\ \ k\in M_{24}\ ,
\ee 
and the twisted twining genera associated to different $(g,h)$'s in the same orbit are related to one another by 
modular transformations, see eqs.~(\ref{modcond}), (\ref{conj1}), and/or by complex conjugation, see
eq.~(\ref{complconj1}). We therefore want to describe the set of orbits of $\P$ under the  action (\ref{groupa}). 
\smallskip

First we note that the $GL(2,\ZZ)$ orbit of a pair $(g,h)$ consists of all possible pairs of generators
of the abelian group $\langle g,h\rangle\subset M_{24}$. Thus the orbits of 
$GL(2,\ZZ)\times M_{24}$ are in one-to-one correspondence with the conjugacy classes of abelian 
subgroups of $M_{24}$ generated by two elements, i.e.\
\be  
\bar \P =\P/(GL(2,\ZZ)\times M_{24})= 
\{ M_{24}-\text{conjugacy classes of groups }\langle g,h\rangle\subset M_{24}, gh=hg\}\ .
\ee

\begin{table}
$$ \begin{array}{cclccccccccc}
\# &  \text{Structure} & \text{Elements} & |C(g,h)| & \frac{|N(g,h)|}{|C(g,h)|} & \text{Orbits on }{\bf 24}  & \text{Max subgr.}\\\hline
1.&  \ZZ_2\times\ZZ_2 & ({\rm 2A})^3& 1536 &6&  1^ 8 \cdot    4^ 4   &\\
2.&  \ZZ_2\times\ZZ_2 & ({\rm 2A})^3&1536 &6&  2^ {12}   &\\
3.&  \ZZ_2\times\ZZ_2 & ({\rm 2A})^3&128 &6&  1^ 4 \cdot    2^ 6 \cdot    4^ 2   & \\
4.&  \ZZ_2\times\ZZ_2 & ({\rm 2B})^3& 3840 &6&  4^ 6 &\\
5.&  \ZZ_2\times\ZZ_2 & ({\rm 2B})^3& 96 &6& 4^ 6 &\\
6.&  \ZZ_2\times\ZZ_2 &({\rm 2B})^3&  64 &6&  4^ 6 & \\
7.&  \ZZ_2\times\ZZ_2 & ({\rm 2A})^2({\rm 2B})& 256 &2& 2^ 8 \cdot    4^ 2 & \\
8.&  \ZZ_2\times\ZZ_2 & ({\rm 2A})({\rm 2B})^2& 512 &2&2^ 4 \cdot    4^ 4  & \\
9.&  \ZZ_2\times\ZZ_2 &({\rm 2A})({\rm 2B})^2& 128 &2& 2^ 4 \cdot    4^ 4  &\\[3pt]
10.&  \ZZ_2\times\ZZ_4 & ({\rm 2A})^3({\rm 4A})^4& 64 & 8& 2^ 4 \cdot    8^ 2 &   1  \\
11.&  \ZZ_2\times\ZZ_4 & ({\rm 2A})^3({\rm 4A})^4& 64 &8&  4^ 6 & 2  \\
12.&  \ZZ_2\times\ZZ_4 & ({\rm 2A})^2({\rm 2B})({\rm 4A})^2({\rm 4B})^2&  32 &2& 2^ 2 \cdot    4^ 3 \cdot    8^ 1 &  7 \\
13.&  \ZZ_2\times\ZZ_4 & ({\rm 2A})({\rm 2B})^2({\rm 4A})^4& 64&8& 4^ 2 \cdot    8^ 2   &  8 \\
14.&  \ZZ_2\times\ZZ_4 & ({\rm 2A})({\rm 2B})^2({\rm 4A})^4&32 &8& 4^ 2 \cdot    8^ 2 &  8 \\
15.&  \ZZ_2\times\ZZ_4 & ({\rm 2A})({\rm 2B})^2({\rm 4C})^4& 32 &4& 4^ 2 \cdot    8^ 2 &  8 \\
16.&  \ZZ_2\times\ZZ_4 &({\rm 2A})({\rm 2B})^2({\rm 4C})^4& 16&4& 4^ 2 \cdot    8^ 2  &   9 \\
17.&  \ZZ_2\times\ZZ_4 & ({\rm 2A})^3({\rm 4B})^4& 64 &8& 1^ 4 \cdot    2^ 2 \cdot    8^ 2  &  1 \\
18.&  \ZZ_2\times\ZZ_4 &({\rm 2A})^3({\rm 4B})^4& 64 &  8& 2^ 4 \cdot    4^ 4 &   2 \\
19.&  \ZZ_2\times\ZZ_4 &({\rm 2A})^3({\rm 4B})^4& 16 & 8& 1^ 2 \cdot    2^ 3 \cdot    4^ 2 \cdot    8^ 1&  3 \\
20.&  \ZZ_2\times\ZZ_4 & ({\rm 2A})({\rm 2B})^2({\rm 4B})^4&64 &8& 2^ 4 \cdot    8^ 2 &  8\\
21.&  \ZZ_2\times\ZZ_4 & ({\rm 2A})({\rm 2B})^2({\rm 4B})^4& 16 & 8&  2^ 4 \cdot    8^ 2 &  9\\[3pt]
22.&  \ZZ_4\times\ZZ_4 & ({\rm 2A})({\rm 2B})^2({\rm 4A})^4({\rm 4C})^8& 16 &16&  8^ 1 \cdot   16^ 1 &  13, 15, 15 \\
23.&  \ZZ_4\times\ZZ_4 & ({\rm 2A})^3({\rm 4A})^8({\rm 4B})^4&16& 32& 2^ 2 \cdot    4^ 1 \cdot   16^ 1 &    10, 10, 17 \\
24.&  \ZZ_4\times\ZZ_4 & ({\rm 2A})^3({\rm 4A})^8({\rm 4B})^4& 16 &32&  4^ 2 \cdot    8^ 2 &   11, 11, 18  \\
25.&  \ZZ_4\times\ZZ_4 & ({\rm 2A})^3({\rm 4B})^{12}&16& 96&  1^ 4 \cdot    4^ 1 \cdot   16^ 1  &   17, 17, 17\\
26.&  \ZZ_4\times\ZZ_4 & ({\rm 2A})^3({\rm 4B})^{12}&16& 96&  4^ 6 &  18, 18, 18   \\[3pt]
27.&  \ZZ_2\times\ZZ_8 & ({\rm 2A})({\rm 2B})^2({\rm 4B})^4({\rm 8A})^8 &16 &8& 2^ 2 \cdot    4^ 1 \cdot   16^ 1 & 20\\
28.&  \ZZ_2\times\ZZ_6 & ({\rm 2A})^3({\rm 3A})^2({\rm 6A})^6& 12& 12& 1^ 2 \cdot    3^ 2 \cdot    4^ 1 \cdot   12^ 1 &  1 \\
29.&  \ZZ_2\times\ZZ_6 & ({\rm 2A})^3({\rm 3A})^2({\rm 6A})^6& 12 &12&  2^ 3 \cdot    6^ 3  &   2  \\
30.&  \ZZ_2\times\ZZ_6 & ({\rm 2B})^3({\rm 3B})^2({\rm 6B})^6&12 &12&  12^ 2 &  4  \\
31.&  \ZZ_2\times\ZZ_6 & ({\rm 2B})^3({\rm 3B})^2({\rm 6B})^6& 12 &12& 12^ 2 &  5  \\
32.&  \ZZ_2\times\ZZ_{10} & ({\rm 2B})^3({\rm 5A})^4({\rm 10A})^{12} & 20 &12&4^ 1 \cdot   20^ 1 & 4\\
33.&  \ZZ_3\times\ZZ_3 & ({\rm 3A})^8 & 9 &48&  1^ 3 \cdot    3^ 4 \cdot    9^ 1&\\
34.&  \ZZ_3\times\ZZ_3 & ({\rm 3A})^2({\rm 3B})^6& 9 &12&  3^ 2 \cdot    9^ 2 &
\end{array}
$$ \caption{The 34 conjugacy classes of abelian subgroups of rank $2$ in $M_{24}$.}\label{t:groups}
\end{table}

This description allows a complete classification of the orbits $[g,h]\in \bar \P$: there are $55$ such orbits,
$21$ of which correspond to cyclic subgroups, i.e.\ to subgroups of the form $[e,h]$. The associated
twisted twining genera are therefore just the twining genera $\phi_{e,h}$, for which explicit expressions were 
already derived in \cite{Gaberdiel:2010ca,Eguchi:2010fg}. Thus we only need to construct the remaining
$34$ genuinely twisted twining genera. We have tabulated the corresponding conjugacy classes of 
groups $\langle g,h\rangle$ in table~\ref{t:groups}. There we have described their structure as 
an abelian group, i.e.\ as $\ZZ_m\times\ZZ_n$, the $M_{24}$ classes of all its elements (excluding the identity), 
the order of the centraliser $C(g,h)$, the index $|N(g,h)|/|C(g,h)|$ of the centraliser in the normaliser of 
$\langle g,h\rangle$ in $M_{24}$, and the lengths of the orbits of 
$\langle g,h\rangle\subset M_{24}$ when acting as a group of permutations of $24$ objects. Finally, the 
last column gives  the conjugacy classes of the non-cyclic maximal subgroups. For example, group $24$ 
has three distinct non-cyclic maximal subgroups, all of the 
form $\ZZ_2\times \ZZ_4$; two of them are conjugated to group $11$ and one is conjugated to group $18$.

For $M_{24}$ it turns out that the orbits of $\P$ under $SL(2,\ZZ)\times M_{24}$ are exactly the same 
as those under $GL(2,\ZZ)\times M_{24}$, i.e.\footnote{For a generic group $G$ different from $M_{24}$, 
some $SL(2,\ZZ)\times G$ orbits might be strictly smaller than the $GL(2,\ZZ)\times G$ orbits.}
\be \label{riden}
\bar \P=\P/(GL(2,\ZZ)\times M_{24})=\P/(SL(2,\ZZ)\times M_{24})\ .
\ee
Thus the twisted twining genera in each orbit in $\bar\P$ are just related by
phases, see eqs.~(\ref{modcond}) and (\ref{conj1}), and we do not need to invoke eq.~(\ref{complconj1}).

\bigskip

\subsection{Modular properties of the twisted twining genera}\label{s:modprop}

For each commuting pair $(g,h)\in\P$, let us denote by 
$\tilde\Gamma_{g,h}\subset SL(2,\ZZ)\times M_{24}$ the group of elements 
$(\gamma,k)$ that leave the pair $(g,h)$ fixed or map it to its inverse $(g^{-1},h^{-1})$ 
\be \label{d1}
\tilde\Gamma_{g,h}=\Bigl\{\Bigl(\bigl(\begin{smallmatrix}
a & b\\ c & d
\end{smallmatrix}\bigr),k\Bigr)\in SL(2,\ZZ)\times M_{24}\mid 
\bigl(k^{-1}(g^ah^c)k,\,k^{-1}(g^bh^d)k\bigr)=(g,h)\text{ or }(g^{-1},h^{-1})\Bigr\}\ .
\ee
It follows from \eqref{modcond} and \eqref{conj1}, together  with \eqref{selfconj} as well
as (\ref{charge}) below, 
that the corresponding twisted twining genus $\phi_{g,h}$ will be invariant (up to a phase) under 
$\Gamma_{g,h} \equiv \pi(\tilde\Gamma_{g,h})$, where $\pi$ denotes the projection of 
$SL(2,\ZZ)\times M_{24}$ onto its first factor, i.e.\ onto $SL(2,\ZZ)$.

Each twisted twining genus $\phi_{g,h}$ belongs to a (finite-dimensional) vector representation of 
$SL(2,\ZZ)$, that is spanned by the functions $\{\phi_{(g,h)\gamma}(\tau,z)\}$, where 
$\gamma\in SL(2,\ZZ)$. Since every $\gamma\in \Gamma_{g,h}$ acts trivially (i.e.\ up to a phase) on $\phi_{g,h}$, 
the $SL(2,\ZZ)$ representation is  spanned by 
$\{\phi_{(g,h)\gamma}(\tau,z)\}_{\gamma\in \Gamma_{g,h}\backslash SL(2,\ZZ)}$, 
where $\gamma$ runs over a set of representatives for the cosets in $\Gamma_{g,h}\backslash SL(2,\ZZ)$. 
Each of these functions $\phi_{(g,h)\gamma}$ is a weak Jacobi form of weight $0$ and index $1$ for 
the congruence subgroup $\gamma^{-1} \Gamma_{g,h} \gamma \subseteq SL(2,\ZZ)$. 
We have tabulated for each orbit $[g,h]\in\bar\P$ the 
functions $\phi_{(g,h)\gamma}$ (starting from $\phi_{g,h}$)
and the invariance group $\Gamma_{g,h}$ in table~\ref{t:modulars}. (Note, however, that some of these functions vanish identically; in particular,
this will be the case if there is an obstruction.) Each $\phi_{g,h}$ is denoted by the $M_{24}$ class of $g$ and the $C_{M_{24}}(g)$-class of $h$, named as in appendix~\ref{app:char}.

\begin{table}
$$ \begin{array}{ccccccccccc}
\# &  \text{Structure} & \text{Functions}& \Gamma_{g,h}  \\\hline
1.&  \ZZ_2\times\ZZ_2 & \phi_{\rm 2A,2A_2} & \Gamma(1)  \\[-1pt]
2.&  \ZZ_2\times\ZZ_2 & \phi_{\rm 2A,2A_3} & \Gamma(1)  \\[-1pt]
3.& \ZZ_2\times\ZZ_2 & \phi_{\rm 2A,2A_5} & \Gamma(1)  \\[-1pt]
4.&  \ZZ_2\times\ZZ_2 & \phi_{\rm 2B,2B_2} & \Gamma(1) \\[-1pt]
5.&  \ZZ_2\times\ZZ_2 & \phi_{\rm 2B,2B_4} & \Gamma(1)   \\[-1pt]
6.&  \ZZ_2\times\ZZ_2 & \phi_{\rm 2B,2B_6} & \Gamma(1)\\ [-1pt]
7.&  \ZZ_2\times\ZZ_2 & \phi_{\rm 2B,2A_1},\ \phi_{\rm 2A,2B_3},\ \phi_{\rm 2A,2A_4} & \Gamma_0(2) \\[-1pt]
8.&  \ZZ_2\times\ZZ_2 &  \phi_{\rm 2A,2B_1},\ \phi_{\rm 2B,2A_2},\ \phi_{\rm 2B,2B_1} & \Gamma_0(2)  \\[-1pt]
9.&  \ZZ_2\times\ZZ_2 & \phi_{\rm 2A,2B_2},\ \phi_{\rm 2B,2A_3},\ \phi_{\rm 2B,2B_5} & \Gamma_0(2)   \\[-1pt]
10.&  \ZZ_2\times\ZZ_4 & \phi_{\rm 2A,4A_2},\ \phi_{\rm 4A,2A_2},\ \phi_{\rm 4A,4A_3} & \Gamma_0(2)  \\[-1pt]
11.&  \ZZ_2\times\ZZ_4 & \phi_{\rm 2A,4A_3},\ \phi_{\rm 4A,2A_3},\ \phi_{\rm 4A,4A_7} & \Gamma_0(2)   \\[-1pt]
12.&  \ZZ_2\times\ZZ_4 & \begin{matrix}
\phi_{\rm 2A,4A_4},\phi_{\rm 2A,4B_4},\phi_{\rm 2B,4A_1},\phi_{\rm 2B,4B_1},\phi_{\rm 4A,2A_1},\phi_{\rm 4A,2B_1},\\[-5pt]
\phi_{\rm 4B,2A_4},\phi_{\rm 4B,2B_2},\phi_{\rm 4B,4A_1},\phi_{\rm 4B,4A_2},\phi_{\rm 4A,4B_3},\phi_{\rm 4A,4B_4}
\end{matrix} & \Gamma_{\rm 2A,4A} \\[-1pt]
13.&  \ZZ_2\times\ZZ_4 & \phi_{\rm 2B,4A_2},\ \phi_{\rm 4A,2B_3},\ \phi_{\rm 4A,4A_5} & \Gamma_0(2) \\[-1pt]
14.&  \ZZ_2\times\ZZ_4 & \phi_{\rm 2B,4A_3},\ \phi_{\rm 4A,2B_2},\ \phi_{\rm 4A,4A_2} & \Gamma_0(2)  \\[-1pt]
15.&  \ZZ_2\times\ZZ_4 & \phi_{\rm 2A,4C_1},\phi_{\rm 4C,2A_2},\phi_{\rm 4C,2B_3},\phi_{\rm 2B,4C_2},
\phi_{\rm 4C,4C_4},\phi_{\rm 4C,4C_6}& \Gamma_0(4)  \\[-1pt]
16.&  \ZZ_2\times\ZZ_4 & \phi_{\rm 2A,4C_2},\phi_{\rm 4C,2A_1},\phi_{\rm 4C,2B_2},\phi_{\rm 2B,4C_3},
\phi_{\rm 4C,4C_3},\phi_{\rm 4C,4C_5}& \Gamma_0(4) \\[-1pt]
17.&  \ZZ_2\times\ZZ_4 & \phi_{\rm 2A,4B_2},\ \phi_{\rm 4B,2A_2},\ \phi_{\rm 4B,4B_8} & \Gamma_0(2)  \\[-1pt]
18.&  \ZZ_2\times\ZZ_4 & \phi_{\rm 2A,4B_3},\ \phi_{\rm 4B,2A_5},\ \phi_{\rm 4B,4B_9} & \Gamma_0(2) \\[-1pt]
19.&  \ZZ_2\times\ZZ_4 & \phi_{\rm 2A,4B_5},\ \phi_{\rm 4B,2A_3},\ \phi_{\rm 4B,4B_3} & \Gamma_0(2)   \\[-1pt]
20.&  \ZZ_2\times\ZZ_4 &  \phi_{\rm 2B,4B_2},\ \phi_{\rm 4B,2B_3},\ \phi_{\rm 4B,4B_1} & \Gamma_0(2) \\[-1pt]
21.&  \ZZ_2\times\ZZ_4 & \phi_{\rm 2B,4B_3},\ \phi_{\rm 4B,2B_1},\ \phi_{\rm 4B,4B_5} & \Gamma_0(2)  \\[-1pt]
22.&  \ZZ_4\times\ZZ_4 & \phi_{\rm 4A,4C_1},\phi_{\rm 4A,4C_2},\phi_{\rm 4C,4A_1},\phi_{\rm 4C,4A_2},\phi_{\rm 4C,4C_7},
\phi_{\rm 4C,4C_8}& \Gamma_{\rm 4A,4C}   \\[-1pt]
23.&  \ZZ_4\times\ZZ_4 & \phi_{\rm 4B,4A_3},\ \phi_{\rm 4A,4B_1},\ \phi_{\rm 4A,4A_1} & \Gamma_0(2)   \\[-1pt]
24.&  \ZZ_4\times\ZZ_4 & \phi_{\rm 4B,4A_4},\ \phi_{\rm 4A,4B_2},\ \phi_{\rm 4A,4A_4} & \Gamma_0(2)   \\[-1pt]
25.&  \ZZ_4\times\ZZ_4 & \phi_{\rm 4B,4B_4} & \Gamma(1) \\[-1pt]
26.&  \ZZ_4\times\ZZ_4 & \phi_{\rm 4B,4B_7} & \Gamma(1) \\[-1pt]
27.&  \ZZ_2\times\ZZ_8 & \phi_{\rm 2B,8A_{1,2}},\phi_{\rm 8A,2B_{1,2}},\phi_{\rm 4B,8A_{2,3}},\phi_{\rm 8A,4B_{1,3}},
\phi_{\rm 8A,8A_{2,8}},\phi_{\rm 8A,8A_{6,7}}& \Gamma_0(4)\\[-1pt]
28.& \ZZ_2\times\ZZ_6 & \phi_{\rm 2A,6A_2},\phi_{\rm 6A,2A_1},\phi_{\rm 6A,6A_1},\phi_{\rm 6A,6A_2} & \Gamma_0(3) \\[-1pt]
29.&  \ZZ_2\times\ZZ_6 & \phi_{\rm 2A,6A_3},\phi_{\rm 6A,2A_2},\phi_{\rm 6A,6A_3},\phi_{\rm 6A,6A_4} & \Gamma_0(3) \\[-1pt]
30.&  \ZZ_2\times\ZZ_6 & \phi_{\rm 2B,6B_2},\phi_{\rm 6B,2B_1},\phi_{\rm 6B,6B_2},\phi_{\rm 6B,6B_5} & \Gamma_0(3) \\[-1pt]
31.&  \ZZ_2\times\ZZ_6 & \phi_{\rm 2B,6B_3},\phi_{\rm 6B,2B_2},\phi_{\rm 6B,6B_1},\phi_{\rm 6B,6B_3}  & \Gamma_0(3) \\[-1pt]
32.&  \ZZ_2\times\ZZ_{10} & \begin{matrix} \phi_{\rm 2B,10A_1},\phi_{\rm 2B,10A_3},\phi_{\rm 10A,10A_1},\phi_{\rm 10A,10A_2},\phi_{\rm 10A,10A_3},\phi_{\rm 10A,10A_5},
\\[-5pt] \phi_{\rm 10A,10A_7},\phi_{\rm 10A,10A_9},\phi_{\rm 10A,10A_{10}},\phi_{\rm 10A,10A_{11}},
\phi_{\rm 10A,2B_2},\phi_{\rm 10A,2B_3}
\end{matrix}  & \Gamma_{\rm 2B,10A}  \\[-1pt]
33.&  \ZZ_3\times\ZZ_3 & \phi_{\rm 3A,3A_3} & \Gamma(1)  \\[-1pt]
34.&  \ZZ_3\times\ZZ_3 & \phi_{\rm 3A,3B_1},\phi_{\rm 3B,3A_1},\phi_{\rm 3B,3B_3},\phi_{\rm 3B,3B_4} &\Gamma_0(3)
\end{array}
$$ \caption{The independent functions and the group $\Gamma_{g,h}$ for each $[g,h]\in\bar P$.}\label{t:modulars}
\end{table}

Group $27$ is the only case where the pairs $(g,h)$ and $(g^{-1},h^{-1})$ are not conjugated within $M_{24}$. Thus, charge conjugation gives the identities $\phi_{{\rm 2B},{\rm 8A_1}}=\phi_{{\rm 2B},{\rm 8A_2}}$, $\phi_{{\rm 8A},{\rm 2B_1}}=\phi_{{\rm 8A},{\rm 2B_2}}$, and so on, and the respective functions are denoted in table \ref{t:modulars} by $\phi_{{\rm 2B},{\rm 8A_{1,2}}}$, $\phi_{{\rm 8A},{\rm 2B_{1,2}}}$, etcetera.

Because of (\ref{complconj1}), most of the functions satisfy a `reality' condition of the form
\be  
\phi_{g,h}^*=\zeta\, \phi_{g,h}\ ,
\ee 
for some constant phase $\zeta$, which implies that the Fourier coefficients of $\zeta^{1/2} \phi_{g,h}$ are all real. 
Note that even if $\phi_{g,h}^*$ and $\phi_{g,h}$ are not proportional to one another, they are necessarily related by 
a modular transformation because of (\ref{riden}); if this is the case, they are listed as 
distinct functions in the same orbit in $\bar \P$. 
\smallskip

Most of the modular groups $\Gamma_{g,h}$ are of the form $\Gamma(1)=SL(2,\ZZ)$ or
\be 
\Gamma_0(N):=\{ \left(\begin{smallmatrix}
a & b\\ c & d
\end{smallmatrix}\right) \in SL(2,\ZZ)\mid c\equiv 0 \mod N\}\ ,
\ee or conjugates of $\Gamma_0(N)$ in $SL(2,\ZZ)$. The exceptions are  the group in case $32$, where 
\be 
\Gamma_{\rm  2B,10A}=\bigcup_{i\in \ZZ/3\ZZ, j\in \ZZ/4\ZZ}\left( \begin{smallmatrix}
1 & 1\\ -5 & -4
\end{smallmatrix} \right)^i\left( \begin{smallmatrix}
-3 & -1\\ 10 & 3
\end{smallmatrix} \right)^j\Gamma_{2,10}\ ,\ee
is a subgroup of index $12$ in $SL(2,\ZZ)$ and 
\be
\Gamma_{2,10}:=\{ \left(\begin{smallmatrix}
a & b\\ c & d
\end{smallmatrix}\right) \in SL(2,\ZZ)\mid a\equiv 1,\ b\equiv 0\mod 2,\  c\equiv 0,\ d\equiv 1 \mod 10\}\ ,
\ee is the group of elements $\gamma\in SL(2,\ZZ)$ such that $(g,h)\cdot\gamma=(g,h)$;
 the group in case $12$, with
\be 
\Gamma_{\rm  \rm 2A,4A}=\{ \left(\begin{smallmatrix}
a & b\\ c & d
\end{smallmatrix}\right) \in SL(2,\ZZ)\mid b\equiv 0\mod 2, c\equiv 0 \mod 4\}\ ,
\ee 
which is a conjugate of $\Gamma_0(8)$ in $SL(2,\RR)$; and the group in case $22$, with
\be 
\Gamma_{\rm 4A,4C}=\langle \left(\begin{smallmatrix}
-1 & 1\\ -2 & 1
\end{smallmatrix}\right)\;, \  \left(\begin{smallmatrix}
1 & 2\\ 0 & 1
\end{smallmatrix} \right) \;  , \
\left(\begin{smallmatrix} 
1 & 0\\ 4 & 1
\end{smallmatrix} \right) \; , \
\left( \begin{smallmatrix}
3 & -2\\ -4 & 3
\end{smallmatrix} \right) \rangle\ .
\ee

\subsection{Explicit twining genera}\label{s:explicit}

We have now reduced our problem to finding $34$ functions $\phi_{g,h}$ that are weak Jacobi forms with respect to
$\Gamma_{g,h}$ up to some multiplier phases. As will be explained in the following section 
\ref{s:cohomology}, our assumption (E)  allows us to determine the precise form of these multipliers explicitly. Then
most of the $34$ functions vanish because of the obstructions, see table~\ref{t:res}. For those that do not, 
the modular properties are 
strong enough to determine them explicitly, see also  table~\ref{t:res}. These explicit results are one of the main
results of this paper, and their derivation is sketched in section~\ref{s:cohomology}, see
in particular section~\ref{sec_8A2Bex}, as well as appendix~\ref{app_detailsttgenera}. 
As will be explained below in section~\ref{s:proj}, these functions are also compatible
with the expected projective representation of the centraliser $C_{M_{24}}(g)$ on the $g$-twisted sector, see 
property (C) in section \ref{sec_ttgenera}.

\begin{table}
$$\begin{array}{c||c} \begin{array}{cccc}
\# &   \phi_{g,h} & \text{Obstr.}  \\\hline\\[-12pt]
1.&   0 & 1  \\
2.&   0 & 1  \\
3.&   0 & 1  \\
4.&  0 & 2 \\
5.&  0 & 2 \\
6.&  0 & 1 \\
7.&  0 & 1  \\
8.&   0 & 1  \\
9.&   0 & 1  \\
10.&  0 & 1  \\
11.&  0 & 1  \\
12.&  0 & 1 \\
13.&  \phi_{\rm 2B,4A_2}=4\frac{\eta(2\tau)^2}{\eta(\tau)^4}\vartheta_1(\tau,z)^2 & \text{no}\\
14.&  0 & 1 \\
15.&  0 & 1 \\
16.& 0 & 1 \\
17.&   0 & 1  
\end{array} &\begin{array}{cccc}
\# &  \phi_{g,h} & \text{Obstr.}  \\\hline\\[-12pt]
18.&  0 & 1  \\
19.&  0 & 1 \\
20.&  0 & 2 \\
21.&  0 & 1 \\
22.&  0 & 2 \\
23.&  \phi_{\rm 4B,4A_3}=2\sqrt{2}\frac{\eta(2\tau)^2}{\eta(\tau)^4}\vartheta_1(\tau,z)^2 &  \text{no} \\
24.&  \phi_{\rm 4B,4A_4}=2\sqrt{2}\frac{\eta(2\tau)^2}{\eta(\tau)^4}\vartheta_1(\tau,z)^2 & \text{no}\\
25.&  0 & 2 \\
26.&  0 & 2 \\
27.&   \phi_{\rm 2B,8A_{1,2}}=2\frac{\eta(2\tau)^2}{\eta(\tau)^4}\vartheta_1(\tau,z)^2 & \text{no} \\
28.&  0 & 2\\
29.&  0 & 2 \\
30.&   0 & 2  \\
31.&   0 & 2  \\
32.&  0 & 2 \\
33.& \phi_{\rm 3A,3A_3}= 0 & \text{no} \\
34.&  \phi_{\rm 3A,3B_1}=0 & \text{no} 
\end{array}
\end{array} 
$$
\caption{For each of the $34$ cases we give here the explicit result for the twisted twining genus, as well
as where applicable, the obstruction that it responsible for its vanishing.}\label{t:res}
\end{table}

\section{Holomorphic Orbifolds and Group Cohomology}\label{s:cohomology}

In this section we review the modular properties of holomorphic orbifolds since, 
according to our assumption (E), these are also relevant for the twisted twining genera of K3. 
We then explain how, in our context, the known multiplier phases of the twining genera determine
the underlying $3$-cocycle $\alpha\in H^3(M_{24}, U(1))$ uniquely, and how to obtain from this our explicit results given in table~\ref{t:res}. 
Finally, we shall check that these formulae indeed give  rise to the appropriate 
projective representations of the centraliser $C_{M_{24}}(g)$.

\subsection{Review of holomorphic orbifolds}
Suppose $\C$ is a self-dual VOA, and $G$ is a group of $\C$-automorphisms. We are interested in the 
`orbifold' of $\C$ by $G$.  To this end we consider the $G$-invariant sub-VOA of $\C$, 
\be 
\C^{G}=\{ {\psi \in {\C} }   :   g \psi = \psi, \, \forall g\in G\} \  .
\ee  
Each representation of $\C$ gives rise to a representation of $\C^{G}$. In addition, there are 
new representations of $\C^{G}$ that appear in the different twisted sectors $\C_A$; since $\C$ is holomorphic 
there is a unique twisted sector $\C_A$  for each conjugacy class $A$ in $G$.  

\subsubsection{Twisted twining characters} 
Each twisted sector $\C_A$ defines a representation of the $G$-invariant sub-VOA $\C^{G}$, but
it is typically not irreducible. In particular, for each $g_A$ in the conjugacy class $g_A\in[A]$,
we can project $\C_{g_A}$ onto a $\C^G$-invariant subspace using any 
(projective) character $\chi$ of $C_G(g_A)$. Thus we can label 
the irreducible representations of $\C^G$ by pairs $(A,\chi)$. In physics, by the orbifold of $\C$ by $G$
one usually means the theory where one chooses $\chi=1$ in each twisted sector, i.e.\ the full
partition function takes the form
\be
Z_{\rm orb}(\tau) = \sum_{A}\frac{1}{|C_G(A)|}\sum_{h\in C_G(g_A)}Z_{g_A,h}(\tau)\ ,
\ee  
where
\be \label{tracegh}
Z_{g, h}(\tau)\equiv \etabox{g} h =\text{Tr}_{\C _g}\left(\rho_g(h)\, q^{L_0-\frac{c}{24}}\right)\ ,
\ee
and $\rho_g$ denotes the action of $C_{G}(g)$ on $\C_g$. We should mention that this action is usually \emph{not}
canonically defined; for example, 
what is usually referred to as discrete torsion \cite{Vafa:1986wx} can be thought of as being an ambiguity in the 
definition of $\rho_g$.
Incidentally, the fact that the action is not canonically defined will be important to us later, see
in particular section~\ref{s:relabelling}. For the moment though, we want to assume that such an action has been
chosen.

\subsubsection{Projective representations}\label{sec:projrep}
As alluded to above, the twisted sectors $\C_g$ typically do not form genuine representations of $C_{G}(g)$ but only 
\emph{projective representations}. A projective representation $\rho$ of a finite group $G$ is characterised by the relation
(see also appendix~\ref{app_projreps} for some introductory comments)
\be
\rho(g_1)\, \rho(g_2)=c(g_1, g_2)\, \rho(g_1g_2)\ ,
\label{projectiverep}
\ee
where $c \, :\, G\times G \to U(1)$ is a (normalised) 2-cocycle representing a cohomology class in 
$H^2(G,U(1))$. Genuine representations of finite groups are classified, up to equivalence, by their characters 
$\text{Tr}_\rho(h)$. This is, however, no longer true for projective representations since  taking the trace over a 
projective representation does not lead to a class function, i.e.\ a function that is invariant 
under conjugation by elements in $G$. Instead, for 
projective representations one has 
\be
\text{Tr}_\rho (hgh^{-1}) = \frac{c(h,h^{-1}gh)}{c(g,h)}\, \text{Tr}_\rho(g)\ , 
\ee
as follows from the defining relation (\ref{projectiverep}) together with the cyclicity of the trace 
(see e.g.\ \cite{Willerton} for a nice discussion). For the case of the holomorphic orbifold, this then
results in the conjugation relation
\be
Z_{g, h}(\tau) = \frac{c_g(h,k)}{c_g(k, k^{-1}h k)} \, Z_{k^{-1}gk, k^{-1}hk}(\tau)\ ,
\label{parttransf}
\ee  
where the 2-cocycle $c_g$ represents a class in $H^{2}(C_G(g), U(1))$ specifying the 
projective representation of the centralizer $C_G(g)$ on $\C_g$, and we have assumed that $k$ and $g$ commute
so that both expressions are evaluated in the same twisted sector. (In fact, (\ref{parttransf}) is even true 
if $k$ and $g$ do not commute, see \cite{Dijkgraaf:1989pz}.)  If in addition  $k$ commutes with $h$,  (\ref{parttransf}) 
reduces to 
\be\label{obstr1}
Z_{g,h}(\tau)= \frac{c_g(h,k)}{c_g(k, h )} \, Z_{g, h}(\tau)\ ,
\ee
which therefore implies that $Z_{g,h}$ vanishes unless the 2-cocycle is \emph{regular},
\be
c_g(h,k)=c_g(k,h)\ .
\ee
This is the holomorphic orbifold version of the first type of obstruction discussed in section \ref{sec_cohobstr} above.

\subsubsection{The underlying cohomology}
The possible choices for  the $2$-cocycles $c_g$ 
in the various twisted sectors 
are constrained; for example this follows from demanding consistency of the OPE involving fields
from different twist sectors.\footnote{This origin of the constraint
was suggested in \cite{Roche:1990hs}.} In particular, it was argued in \cite{Dijkgraaf:1989pz} that
the various consistent choices are in one-to-one correspondence with elements in the third 
cohomology group $[\alpha]\in H^3(G, U(1))$, see also \cite{Bantay:1990yr,Roche:1990hs,Coste:2000tq,Frohlich:2009gb} for 
subsequent work. The $3$-cocycle $\alpha$ determines distinguished elements $c_g\in H^2(C_G(g),U(1))$
via the formula  \cite{Dijkgraaf:1989pz,Bantay:1990yr}
\be 
c_g(h_1, h_2)= \frac{\alpha(g, h_1, h_2)\, \alpha(h_1, h_2,(h_1h_2)^{-1} g(h_1h_2))}{\alpha(h_1, h_1^{-1}gh_1, h_2)}\ .
\label{ch0}
\ee 

Since $\alpha$ fixes all $2$-cocylces $c_g$ via (\ref{ch0}), it is clear that the choice of $\alpha$ determines
the behaviour of the twisted twining characters under conjugation (\ref{parttransf}). Furthermore, 
$\alpha$ also determines all the multiplier
phases that appear under modular transformations, since it follows from 
\cite{Bantay:1990yr} that\footnote{Note that (\ref{modular}) is obtained from \cite{Bantay:1990yr} upon
working with the inverse $3$-cohomology element. We also thank Terry Gannon for explaining
these formulae to us. It is believed (and we have verified this in simple 
cases) that (\ref{modular}) follows from the transformation rules for the irreducible orbifold 
characters, see  \cite[eqs.\ (5.23) and (5.24)]{Coste:2000tq},
but we have not proven this in complete generality.}
\be
\begin{array}{rcl}
{Z}_{g,h}(\tau+1)&=& c_g(g,h)\, {Z}_{g, gh}(\tau)\ ,\\[4pt]
{Z}_{g,h}(-1/\tau)&=& \overline{c_h(g, g^{-1})}\, {Z}_{h, g^{-1}}(\tau)\ .
\end{array}
\label{modular}
\ee 

The choice of a particular representative $\alpha$ in the class $[\alpha]\in H^3(G, U(1))$ corresponds to a choice of 
normalisation for the action of $h$ on the $g$-twisted sector. To see this, let $\tilde \alpha$ represent a 3-cocycle in the 
same cohomology class $[\alpha]\in H^3(G, U(1))$, so that $\tilde \alpha=\alpha \cdot \partial \beta$ (see eq.~\eqref{alphagauge})
for some 2-cochain $\beta$. Then the associate $2$-cocycles $c_g$ and $\tilde{c}_g$ differ (for each $g\in G$) 
only by a $2$-coboundary i.e.\ $\tilde c_g=c_g \cdot \partial\gamma_g$, with $\gamma_g(h)=\frac{\beta(h,g)}{\beta(g,h)}$, 
see  appendix \ref{app:groupcoho}. Thus, the 3-cocycle $\tilde \alpha$ is associated
with the modular properties of the functions $\tilde Z_{g,h}=\gamma_g(h)Z_{g,h}$, which, 
by \eqref{tracegh}, simply correspond to a different normalisation $\tilde \rho_g(h)=\gamma_g(h)\rho_g(h)$ for the action  of 
$h$ on the $g$-twisted sector.

\subsection{Application to Mathieu Moonshine}
\label{sec_proposal}
Given what we said before, it is now natural to postulate that for the twisted twining genera of K3, the
various multiplier phases also come from a $3$-cocycle $\alpha\in H^3(M_{24}, U(1))$. This is to say,
we postulate that  the projective representation appearing in (C) is the one determined by $c_g$, given
by (\ref{ch0}). Furthermore, 
$\chi_{g,h}\left(\begin{smallmatrix} a & b \\ c& d \end{smallmatrix}\right)$ in (\ref{modcond})
is determined by (\ref{modular}), and $\xi_{g,h}(k)$ in (\ref{conj1}) agrees with (\ref{obstr1}), where
in both cases $c_g$ is again the function determined by (\ref{ch0}) for a fixed $\alpha\in H^3(M_{24}, U(1))$. 
More explicitly, we therefore propose that 
\begin{align} &\phi_{g,h}\left(-\tfrac{1}{\tau},\tfrac{z}{\tau}\right)=\overline{c_h(g,g^{-1})}\, 
e^{2\pi i\frac{z^2}{\tau}}\, \phi_{h,g^{-1}}(\tau,z)\ ,\nn \\
%&\label{invS}\phi_{g,h}(-\frac{1}{\tau},\frac{z}{\tau})=c_g(h^{-1},h)e^{2\pi i\frac{z^2}{\tau}}\phi_{h^{-1},g}(\tau,-z)\ ,\\
&\phi_{g,h}(\tau+1,z)=c_g(g,h)\, \phi_{g,gh}(\tau,z)\ , 
%&\phi_{g,h}(\tau-1,z)=\frac{1}{c_g(g,g^{-1}h)}\phi_{g,g^{-1}h}(\tau,z)
%\\
\label{STmodular}
 \end{align}
with $c_g(h_1,h_2)$ given by (\ref{ch0}), and 
\be 
\label{conj} \phi_{g,h}(\tau,z)=\frac{c_g(h,k)}{c_g(k,k^{-1}hk)}\, \phi_{k^{-1}gk,k^{-1}hk}(\tau,z)\ ,\qquad k\in M_{24}\ .
\ee
Moreover, we postulate that 
\be
\phi_{g,h^{-1}}(\tau,z)=c_{g}(h,h^{-1})\, \phi_{g,h}^*(\tau,z)\ ,
\ee
where $\phi_{g,h}^*(\tau,z)$ is defined in (\ref{stardef}). 

Combining modular transformations with conjugation one obtains a whole set of relations among the twisted
twining genera, where the multipliers are given by
\begin{align}\label{transf1}
& \chi_{g,h}(\gamma_1\gamma_2)=\chi_{g,h}(\gamma_1)\,\chi_{(g,h)\gamma_1}(\gamma_2)\ ,& \gamma_1,\gamma_2\in SL(2,\ZZ)\ ,\\\label{transf2}
&\chi_{g,h}\left(\begin{smallmatrix}0 & -1\\ 1 & 0\end{smallmatrix}\right)=\frac{1}{c_h(g,g^{-1})}\ ,\\\label{transf3}
&\chi_{g,h}\left(\begin{smallmatrix}1 & 1\\ 0 & 1\end{smallmatrix}\right)=c_g(g,h)\ ,\\
\label{transf4}
&\chi_{g,h}\left(\begin{smallmatrix}1 & 0\\ 0 & -1\end{smallmatrix}\right)=\frac{1}{c_{g}(h,h^{-1})}\ ,\\
&\xi_{g,h}(k)=\frac{c_g(h,k)}{c_g(k,k^{-1}hk)}\ .
\end{align}

Since $SL(2,\ZZ)$ is generated by $S=\left(\begin{smallmatrix}
0 & -1\\ 1 & 0
\end{smallmatrix}\right)$ and $T=\left(\begin{smallmatrix}
1 & 1\\ 0 & 1
\end{smallmatrix}\right)$ with the only relations being $S^4=1$ and $(ST)^3=S^2$, it follows that eqs.~\eqref{transf1} -- \eqref{transf4} determine the multiplier $\chi_{g,h}(\gamma)$, $\gamma\in SL(2,\ZZ)$, provided that the following consistency conditions are satisfied
\be\label{consist1} \chi_{g,h}(S^4)=1\ ,\qquad \qquad \chi_{g,h}((ST)^3)=\chi_{g,h}(S^2)\ .
\ee Furthermore, since the actions of $SL(2,\ZZ)$ and $M_{24}$ on the set $\P$ commute with one another, there are two more consistency conditions
\be\label{consist2}
\xi_{g,h}(k)\, \chi_{k^{-1}gk,k^{-1}hk}(\gamma)=\chi_{g,h}(\gamma)\, \xi_{(g,h)\gamma}(k)
\qquad
\xi_{g,h}(k)\, \xi_{{k^{-1}gk,k^{-1}hk}}(k')=\chi_{g,h}(kk')\ ,
\ee  
for all $\gamma\in SL(2,\ZZ)$ and $k,k'\in M_{24}$. In fact, conditions \eqref{consist1} and \eqref{consist2} 
hold for a generic group $G$, and can be easily checked using standard identities for the cocycle $c_g$, in particular, 
eqs.~(3.2.5)--(3.2.9) of \cite{Roche:1990hs} (see also \cite{Bantay:1990yr}).

Note that the charge conjugation operator $C=S^2$ acts on the genera $\phi_{g,h}(\tau,z)$ by 
flipping the sign of the second argument $z$. In fact, writing $\tau=-\tfrac{1}{\tilde\tau}$ and 
$z=\tfrac{\tilde z}{\tilde\tau}$, we get from twice applying (\ref{STmodular})
\begin{align}\label{charge} 
\phi_{g,h}(\tau,z)&=\phi_{g,h}(-\tfrac{1}{\tilde\tau},\tfrac{\tilde z}{\tilde\tau})
=\frac{e^{2\pi i\frac{\tilde z^2}{\tilde\tau}}}{c_h(g,g^{-1})}\, \phi_{h,g^{-1}}(\tilde\tau,\tilde z)\\
&=\frac{e^{-2\pi i\frac{z^2}{\tau}}}{c_h(g,g^{-1})}\, \phi_{h,g^{-1}}(-\tfrac{1}{\tau},-\tfrac{z}{\tau})
=\frac{1}{c_h(g,g^{-1})c_{g^{-1}}(h,h^{-1})}\, \phi_{g^{-1},h^{-1}}(\tau,-z)\ ,\notag
\end{align}  so that
\be \chi_{g,h}\left(\begin{smallmatrix}-1 & 0\\ 0 & -1\end{smallmatrix}\right)=\frac{1}{c_h(g,g^{-1})c_{g^{-1}}(h,h^{-1})}\ .\ee
Together with \eqref{selfconj}, this then leads to 
\be \phi_{g,h}(\tau,z)=\frac{1}{c_h(g,g^{-1})\, c_{g^{-1}}(h,h^{-1})}\, \phi_{g^{-1},h^{-1}}(\tau,z)\ ,
\ee
thus reproducing (\ref{eq1}). This identity played a key role in the derivation of obstruction 2 in 
section~\ref{sec_cohobstr}.

While our proposal about the multiplier phases 
is certainly natural, we do not have a direct proof for it. However, the fact that with this 
assumption we shall find consistent answers is in our opinion very convincing evidence in 
favour of this ansatz.

\subsection{Determining the cohomology class}\label{s:multex}

In order to have  explicit formulae for the multipliers, the next step consists of determining 
$[\alpha]\in H^3(M_{24}, U(1))$. 
In this subsection, we will prove (with some computer support) that there exists a unique cohomology class $[\alpha]$ that 
reproduces \eqref{untwmult} for all the 
\emph{untwisted} twining genera $\phi_{e,g}(\tau,z)$.  Since this is an overdetermined problem, the existence of such class already represents a non-trivial consistency check for our proposal.
In the following subsections, we will then explain how the remaining twisted twining characters can be computed, once the class $[\alpha]$ is known.

First we note that the normalisation condition for $\alpha$, see
eq.~(\ref{alphnorm}), implies that $c_e(g,h)=1$ for all $g,h\in M_{24}$.
Thus $ \phi_{e,g}(\tau+1,z)=\phi_{e,g}(\tau,z)$, $\phi_{e,g^{-1}}(\tau,z)=\phi_{e,g}(\tau,z)$ and $\phi_{e,g}(\tau,z)=\phi_{e,k^{-1}gk}(\tau,z)$ for all 
$k\in M_{24}$, as expected. Furthermore, the representation $\rho_e$ in the decomposition \eqref{eqn:decomp} 
is a genuine representation of $C_{M_{24}}(e)=M_{24}$.

Next we observe that eq.~\eqref{STmodular} implies 
\be\label{Tmultip}
\phi_{g^a,g^b}(\tau+o(g^a),z)=\kappa(g^a,g^b)\, \phi_{g^a,g^b}(\tau,z)\ ,
\ee where
\be \kappa(g^a,g^b):= \prod_{k=1}^{o(g^a)}c_{g^a}(g^a,g^{b+ak}) = \prod_{k=1}^{o(g^a)}\alpha(g^a,g^{b+ak},g^a)\ .
\ee More precisely, the phases $\kappa(g^a,g^b)$ are the multipliers of $\phi_{e,g}$ under certain parabolic elements of the form \be\label{parabol} \gamma \left(\begin{matrix}
1 & n\\ 0 & 1
\end{matrix}\right)\gamma^{-1}\in \Gamma_0(o(g))\ ,\ee where $\gamma\in SL(2,\ZZ)$ is such that $(e,g)\gamma=(g^a,g^b)$ and $n=o(g^a)$. By comparing such a multiplier with the one expected from \eqref{untwmult}, one obtains the following condition
\be\label{expected} 
\kappa(g^a,g^b)
=e^{-\frac{2\pi i}{\ell(g^a)}}\ ,
\ee  for all $a,b\in \ZZ$,
where $\ell(g^a)$ is the length of the smallest cycle of $g^a$ when considered as a permutation of $24$ objects. 
It is easy to verify that $\kappa(g^a,g^b)$ only depends on the cohomology class of the 3-cocycle $\alpha$. Thus
we have to prove that there is a unique class $[\alpha]\in H^3(M_{24},U(1))$ such that \eqref{expected} is satisfied
for all $g\in M_{24}$, $a,b\in\ZZ$.   
\smallskip

In order to show that such an $\alpha$ exists, we first need two basic facts about group cohomology. 
Given two finite groups $G$ and $H$ and a homomorphism $\iota:H\to G$, we can define the 
induced map $\iota^*$ from $U(1)$-cochains $\alpha$ for $G$ to $U(1)$-cochains $\iota^*(\alpha)$ for $H$ 
\be 
\iota^*(\alpha)(g_1,\ldots,g_n)=\alpha(\iota(g_1),\ldots,\iota(g_n))\ ,
\ee
which induces a well-defined map
on cohomologies (\cite{Brown}, Chapter III.8). A particular case is when $H=G$ and the map $H\to G$ is 
conjugation $\iota_k(g)=k^{-1}gk$ by an element $k\in G$. In this case, the induced map on the cohomology 
$\iota_k^*$ acts as the identity, i.e.\
$[\iota^*(\alpha)]=[\alpha]$ (see \cite{Brown}, Chapter III Proposition 8.3). A second interesting case is when 
$H$ is a subgroup of $G$ with $\iota$ the natural inclusion; in this case, ${\rm Res}\equiv \iota^*$ is called a 
restriction map. In particular, if $H$ is a Sylow $p$-subgroup $\Syl_pG$ of $G$,\footnote{Recall that a $p$-group 
($p$ prime) is a finite group with order a power of $p$; a Sylow $p$-subgroup of a group $G$ is a 
$p$-subgroup of $G$ that is maximal, i.e.\ that is not properly contained in any other $p$-subgroup. For each $p$, all 
Sylow $p$-subgroups of $G$ are isomorphic and conjugate to one another inside $G$.} then the image of the restriction map 
\be 
{\rm Res}_p:H^n(G,U(1))\to H^n({\Syl}_p(G),U(1))
\ee 
is isomorphic to the $p$-part of $H^n(G,U(1))$ (\cite{Brown}, Chapter III Theorem 10.3). For $G=M_{24}$, using that  $H^3(M_{24},U(1))\cong \ZZ_{12}$ \cite{DutourEllis}, this implies 
\be {\rm Res}_2(H^3(M_{24},U(1)))\cong \ZZ_4\ ,\qquad 
{\rm Res}_3(H^3(M_{24},U(1)))\cong \ZZ_3\ ,
\ee 
while ${\rm Res}_p(H^3(M_{24},U(1)))$ is trivial for all primes $p>3$. 

\medskip 

With these preparations we can now turn to the proof that a unique class $[\alpha]\in H^3(M_{24}, U(1))$ 
satisfying  \eqref{expected} exists.
From the invariance of the cohomology classes under the conjugation map $\iota_k^*$, it follows that the multiplier in \eqref{expected} only depends on the conjugacy class of 
$g$. Next we note that in order to compute $\kappa(g^a,g^b)$, it is sufficient to consider the restriction of the cocycle $\alpha$ 
to the group $\langle g\rangle$ generated by $g$. For a cyclic group $\ZZ_N$, the third cohomology 
$H^3(\ZZ_N,U(1))$ is isomorphic to $\ZZ_N$, and a set of cocycle representatives for the cohomology classes are given by \cite{deWildPropitius:1995cf,Coste:2000tq}
\be 
\omega_q(g^a,g^b,g^c)=e^{2\pi i\frac{qa}{o(g)^2}([b]+[c]-[b+c])}\ ,\qquad q=0,\ldots,o(g)-1\ ,
\ee 
where $[\cdot]:\ZZ\to \{0,\ldots,o(g)-1\}$ is the reduction modulo $o(g)$.
\smallskip

\setcounter{footnote}{0}

Let us assume, without loss of generality, that 
$1\le a<o(g)-1$ and $0\le b<a$, and that $a$ divides $o(g)$; in fact, if this was not true, we can rewrite both
$g^a$ and $g^b$ as powers of another $\tilde{g}$ in $\langle g\rangle$ for which this is then true. 
The multiplier corresponding to a 3-cocycle $\omega_q$ is
\begin{align}\label{kappa} \kappa(g^a,g^b)=&\prod_{k=0}^{o(g^a)-1}\omega_q(g^a,g^{b+ak},g^a)=
e^{2\pi i\frac{qa}{o(g)^2}\sum_{k=0}^{o(g^a)-1}([b+ak]+[a]-[b+a(k+1)])}\\
=&\
e^{2\pi i\frac{qa}{o(g)^2}([b]+o(g^a)[a]-[b+ao(g^a)])}=e^{2\pi i\frac{qa}{o(g)}}\ ,\notag
\end{align} 
where we have used that  $o(g^a)=o(g)/a$ when $a$ divides $o(g)$. Thus, 
$\kappa(g^a,g^b)\equiv \kappa(g^a)$ is independent of $b$, and 
$\kappa(g^a)=\kappa(g)^a$ whenever $a$ divides $o(g)$. The latter property is consistent with \eqref{expected}
 thanks to the identity
\be 
\frac{1}{\ell(g^a)}\equiv \frac{a}{\ell(g)}\mod 1\ ,
\ee 
that holds for all elements $g\in M_{24}$ and for all divisors $a$ of $o(g)$.\footnote{Note that this property is special for 
$M_{24}$, and is not true for a generic permutation.}

It therefore follows that it is sufficient to prove \eqref{expected} for $a=1$ and $b=0$, and for one representative $g$ for each 
conjugacy class. Furthermore, thanks to $\kappa(g^a)=\kappa(g)^a$, we can restrict to classes whose order is a prime power. 
In other words, we only need to check \eqref{expected}  for a Sylow $p$-subgroup ${\rm Syl}_p(M_{24})$ and for each prime $p$. 
\smallskip

Using the software {\texttt{GAP}} \cite{GAP4} with the package {\texttt{HAP}} \cite{HAP} implemented, we have computed a basis of cocycle representatives for 
$H^3({\rm Syl}_2(M_{24}),U(1))$ and $H^3({\rm Syl}_3(M_{24}),U(1))$. Note that any 3-cocycle in the image of 
${\rm Res}_p$ has the property that, for any $g\in {\rm Syl}_p(M_{24})$, the corresponding multiplier $\kappa(g)$ depends 
only on the conjugacy class of $g$ in $M_{24}$. 
We have verified that there is a unique 
$\ZZ_4$ subgroup of $H^3({\rm Syl}_2(M_{24}),U(1))$ and a unique $\ZZ_3$ subgroup of 
$H^3({\rm Syl}_3(M_{24}),U(1))$ satisfying this property. Therefore these subgroups must
correspond to the restrictions ${\rm Res}_p(H^3(M_{24},U(1)))$, for $p=2$ and $p=3$, respectively. 
Finally, for both $p=2$ and $p=3$, we have verified that there is a unique cohomology class 
$[\alpha_p]\in {\rm Res}_p(H^3(M_{24},U(1)))$ that satisfies \eqref{expected} for all $g$ in ${\rm Syl}_p(M_{24})$. Thus, 
the unique class $[\alpha]\in H^3(M_{24},U(1))$ that has the property that ${\rm Res}_p([\alpha])=[\alpha_p]$ for all primes $p$, 
satisfies    \eqref{expected} for all $g\in M_{24}$. 

\medskip

We have thus proven that  there exists a unique class $[\alpha]\in H^3(M_{24},U(1))$  satisfying \eqref{expected}. It remains to 
show that this class exactly reproduces \eqref{untwmult} for all $\phi_{e,g}$.
First, we notice that the multiplier \eqref{untwmult} of  $\phi_{e,g}$ is trivial whenever the class of $g$ restricts to a class 
in $M_{23}$. Since $H^3(M_{23},U(1))$ is trivial, this is consistent with any choice of the cocycle $\alpha$. Thus, we only
need to check  \eqref{untwmult} for the classes that do not restrict to $M_{23}$. The corresponding characters $\phi_{e,g}$ are Jacobi forms under $\Gamma_0(o(g))$, where the possible orders $o(g)$ are $2,3,4,6$ and $12$. 
For $N=2,3,4$, the group $\Gamma_0(N)$ is generated by 
\be T=\left(\begin{matrix}
1 & 1\\ 0 & 1
\end{matrix}
\right)\ ,\qquad\qquad C=\left(\begin{matrix}
-1 & 0\\ 0 & -1
\end{matrix}
\right)\ ,\qquad\qquad \begin{pmatrix}
1 & 0\\ -N & 1
\end{pmatrix}=\begin{pmatrix}
0 &1 \\-1&0
\end{pmatrix}T^N\begin{pmatrix}
0 &-1 \\1&0
\end{pmatrix} \ee 
while $\Gamma_0(6)$ is generated by $T$, $C$, together with\be\begin{pmatrix}
1 &0 \\-6&1
\end{pmatrix}\ ,\qquad \qquad 
\begin{pmatrix}
-5 &3 \\-12&7
\end{pmatrix}=
\begin{pmatrix}
1 &0 \\2&1
\end{pmatrix}T^3\begin{pmatrix}
1 &0 \\-2&1
\end{pmatrix}\ .\ee
The correct modular properties \eqref{untwmult} under $T$ and $C$ are automatic for any normalised cocycle $\alpha$. The 
remaining generators are of the form \eqref{parabol} with $(e,g)\gamma=(g^a,g^b)$ and $n=o(g^a)$, and
hence  \eqref{untwmult} follows again from \eqref{expected}.

The group $\Gamma_0(12)$ is generated  by $T,C$ and\footnote{It is straightforward to check that $T,C,\alpha_1,\ldots,\alpha_4$ actually generate $\Gamma_0(12)$ by simply comparing them  to a  minimal set of generators obtained by some standard method, 
such as the Farey symbol algorithm.}
\begin{align}
&\alpha_1=\begin{pmatrix}
0 &1 \\-1&0
\end{pmatrix}\begin{pmatrix}
1 &12 \\0&1
\end{pmatrix}\begin{pmatrix}
0 &-1 \\1&0
\end{pmatrix}\ , &
&\alpha_2=\begin{pmatrix}
1 &0 \\3&1
\end{pmatrix}\begin{pmatrix}
1 &4 \\0&1
\end{pmatrix}\begin{pmatrix}
1 &0 \\-3&1
\end{pmatrix}\ ,\\
&\alpha_3=\begin{pmatrix}
1 &0 \\6&1
\end{pmatrix}\begin{pmatrix}
1 &1 \\0&1
\end{pmatrix}\begin{pmatrix}
1 &0 \\-6&1
\end{pmatrix}\ , &
&\alpha_4=\begin{pmatrix}
1 &0 \\2&1
\end{pmatrix}\begin{pmatrix}
1 &3 \\0&1
\end{pmatrix}\begin{pmatrix}
1 &0 \\-2&1
\end{pmatrix} \ .
\end{align}
The generators $\alpha_1,\ldots,\alpha_4$ are of the form \eqref{parabol} with $(e,g)\gamma=(g^a,g^b)$, but only for 
$\alpha_1,\alpha_2$ is the condition $n=o(g^a)$ satisfied, and thus only in these cases is condition \eqref{expected} 
sufficient to establish \eqref{untwmult}. The generators $\alpha_3,\alpha_4$ satisfy $(e,g)\alpha_i=(e,g^7)$, $i=3,4$, so one needs 
to conjugate by an element $k\in M_{24}$ such that $k^{-1} g^7k=g$. In order to compute the multiplier under 
these generators, one needs to determine the restriction of $[\alpha]\in H^3(M_{24},U(1))$ to the subgroup of $M_{24}$
generated by $g$ and $k$. Using \texttt{GAP}, we computed a basis for the third cohomology of the group $\langle g,k\rangle$ and 
looked for those classes that satisfy condition \eqref{expected} for all elements of the group. It turns out that there is only one 
such class, and hence it must correspond to the restriction of $[\alpha]\in H^3(M_{24},U(1))$ to the cohomology of 
$\langle g,k\rangle$.
Then it is straightforward to check that
this class reproduces indeed the multipliers \eqref{untwmult} under $\alpha_3$ and $\alpha_4$; this is true for both 
the $M_{24}$ conjugacy classes of order $12$. This completes the proof.

The {\texttt{GAP}}-files containing the computations described above are available online  \cite{thispaper}.

\subsection{Computation of the twisted twining characters: an example}
\label{sec_8A2Bex}

In the previous subsection, we proved that there is a unique cohomology class $[\alpha]\in H^3(M_{24},U(1))$ that reproduces the multipliers \eqref{untwmult} for all the \emph{untwisted} twining genera $\phi_{e,g}$. In this section, we shall explain how the remaining twisted twining genera $\phi_{g,h}$ can be determined, leading to the results collected in section~\ref{s:explicit}.

\smallskip

The first step is to compute the obstructions and the multipliers for a given twisted twining genus
$\phi_{g,h}$. To this end we note that it is sufficient to consider the restriction of the cocycle $\alpha$ to the normaliser 
$N(g,h)$ of $\langle g,h\rangle$. If this normaliser is contained in some Sylow $p$-subgroup, one can simply use the 
cocycle $\alpha_p$ determined in the previous subsection. Otherwise, one can use {\texttt{GAP}} to compute a basis for the cohomology of $N(g,h)$ 
and check which cohomology classes reproduce the correct multipliers (and in particular \eqref{expected}) for all genera of the form $\phi_{e,k}$, $k\in N(g,h)$. 
In all cases we considered, there was only one cohomology class with this property, which must therefore be the restriction of the 
class $[\alpha] \in H^3(M_{24},U(1))$ to the cohomology of $N(g,h)$. Once the restriction of $[\alpha]$ to $N(g,h)$ is determined, one can check the existence of obstructions for 
all 34 groups of table \ref{t:groups}. The results are collected in table \ref{t:res} and the {\texttt{GAP}}-files containing these computations are available online  \cite{thispaper}.

\smallskip

For the six unobstructed twisted twining 
genera $\phi_{g,h}$, one can compute the precise multiplier system. It turns out that this information is sufficient to 
determine $\phi_{g,h}$ up to normalisation. The normalisation can be fixed (up to a phase) by requiring that a 
decomposition \eqref{eqn:decomp} exists. In the remainder of this subsection, we illustrate how one of these non-zero 
twisted twining genera, namely the one
associated to $g\in {\rm 2B}$ and $h\in {\rm 8A}$, can be determined explicitly; the other
five cases are completely analogous and are dealt with in appendix~\ref{app_detailsttgenera}. Let us work with the conventions
that the three permutations 
\begin{align*}
(1,2,3,4,5,6,7,8,9,10,11,12,13,14,15,16,17,18,19,20,21,22,23) (24)\ ,\\
 (3,17,10,7,9)(4,13,14,19,5)(8,18,11,12,23)(15,20,22,21,16) (1) (2) (6) (24)\ , \\
(1,24)(2,23)(3,12)(4,16)(5,18)(6,10)(7,20)(8,14)(9,21)(11,17)(13,22)(15,19)\end{align*}
 generate $M_{24}\subset S_{24}$. As representative for $g\in {\rm 2B}$ we can then take
\be 
g=(1,10)(2,14)(3,8)(4,5)(6,22)(7,20)(9,18)(11,23)(12,24)(13,19)(15,16)(17,21)\ ,
\ee 
while for $h\in {\rm 8A}$ we consider
\be 
h=(2,14)(3,9,8,18)(4,6,21,19,15,24,20,11)(5,22,17,13,16,12,7,23) (1)(10)\ .
\ee 
The elements $g,h$ generate the abelian group $\ZZ_2\times\ZZ_8$ corresponding to case 27 of table \ref{t:groups}. 
There are two interesting conjugation relations, namely
\be (k^{-1}gh^4k\,,\,k^{-1}h^{-1}k)=(g,h)\ ,\qquad (r^{-1}gr\,,\,r^{-1}ghr)=(g,h)\ ,
\ee where 
\begin{equation}\label{62}
\begin{array}{l}
k=(1,10)(2,14)(3,18)(4,11)(5,13)(6,20)(7,12)(8,9)(15,19)(16,23)(17,22)(21,24)\ ,\\
r=(1,2)(3,9)(6,22)(8,18)(10,14)(11,23)(12,24)(13,19) (4) (5) (7) (15)(16)(17)(20)(21) \ .
\end{array}
\end{equation}
Thus, we can use the general formulae (\ref{STmodular}) 
and their inverses to deduce that\footnote{For brevity of presentation we omit the dependence of $z$, as well as 
the factors $e^{2\pi i\frac{cz^2}{c\tau+d}}$, in the following formulae.}
\begin{align}
\phi_{g,h}(\tau+1)&=c_g(g,h)\phi_{g,gh}(\tau)=c_g(g,h)\frac{c_g(gh,r)}{c_g(r,h)}\phi_{g,h}(\tau)=i\phi_{g,h}(\tau)\ ,\label{8a2bmultip1}
\end{align}
where we have used the explicit form of the multiplier factors from section~\ref{s:multex}.
We also have (recall that $g=g^{-1}$)
\begin{align}
\phi_{g,h}(\tfrac{\tau}{-4\tau+1})=&\frac{\phi_{h,g}(4-\frac{1}{\tau})}{c_h(g,g)}
=\frac{\prod_{i=1}^3c_h(h,gh^i)}{c_h(g,g)}\phi_{h,gh^4}(-\tfrac{1}{\tau})
=\frac{\prod_{i=1}^3c_h(h,gh^i)}{c_h(g,g)c_{gh^4}(h,h^{-1})}\phi_{gh^4,h^{-1}}(\tau)\notag\\
=&\frac{\prod_{i=1}^3c_h(h,gh^i)}{c_h(g,g)c_{gh^4}(h,h^{-1})}\frac{c_{g^4h}(h^{-1},k)}{c_{g^4h}(k,h)}\phi_{g,h}(\tau)=-\phi_{g,h}(\tau)\ .\label{8a2bmultip2}
\end{align}
This implies that $\phi_{g,h}(\tau,z)$ is a Jacobi form under the group $\Gamma_0(4)$ with the above multipliers. 

Let us consider now the dependence of $\phi_{g,h}(\tau,z)$ on $z$, for a fixed $\tau\in \HH_+$. The characters $\ch_{h,\ell}(\tau,z)$ of the $\N=4$ algebra are even functions of $z$ and $\ch_{h,\ell}(\tau,z=0)$ is independent of $\tau$ (in particular, $\ch_{h,\ell}(\tau,0)=0$ for $h>1/4$) \cite{Eguchi:1987wf}. By \eqref{eqn:decomp}, all twisted-twining genera must satisfy the same properties. Since the multiplier system for $\phi_{g,h}$ is non-trivial, the constant $\phi_{g,h}(\tau,0)$ must vanish and by $\phi_{g,h}(\tau,z)=\phi_{g,h}(\tau,-z)$ it follows that $\phi_{g,h}$ has a double zero at $z=0$ (mod $\ZZ+\tau\ZZ$).  Every (weak) Jacobi form  of index 1 has exactly two zeros (counting multiplicity) in each fundamental domain of $\CC/(\ZZ+\tau\ZZ)$ \cite{EichlerZagier}, so that these are the only zeroes of $\phi_{g,h}(\tau,z)$.   On the other hand, the function
\be 
\psi(\tau,z)\equiv \frac{\vartheta_1(\tau,z)^2}{\vartheta_2(\tau,0)^2}
\ee
is also a Jacobi form weight $0$ and index $1$  under $\Gamma_0(4)$ (with trivial multiplier) and has only a double zero at $z=0$
(mod $\ZZ+\tau\ZZ$). It therefore follows that 
\be 
F(\tau)\equiv \frac{\phi_{g,h}(\tau,z)}{\psi(\tau,z)}
\ee 
is independent of $z$ and is a modular function of weight $0$ under $\Gamma_0(4)$ (with multiplier system \eqref{8a2bmultip1}--\eqref{8a2bmultip2}) that 
is holomorphic in the interior of $\HH_+$. 
Let us consider the asymptotic behaviour of $F(\tau)$ at the boundary $\Gamma_0(4)\backslash(\QQ\cup\{\infty\})$ 
of the compactification  $\Gamma_0(4)\backslash\bar\HH_+$. This boundary consists of three points (cusps) of the form 
$\gamma_c\cdot \infty$, where  $\gamma_{\infty}=\left(\begin{smallmatrix}
1 & 0\\ 0 & 1
\end{smallmatrix}\right)$, $\gamma_{0}=\left(\begin{smallmatrix}
0 & -1\\ 1 & 0
\end{smallmatrix}\right)$
and $\gamma_{1/2}=\left(\begin{smallmatrix}
1 & 0\\ 2 & 1
\end{smallmatrix}\right)$.
The twisted twining genus $\phi_{g,h}$ has the Fourier expansion at these cusps
\begin{align}
\phi_{g,h}(\gamma_\infty(\tau,z))&= \phi_{g,h}(\tau,z)=A_\infty\,  q^{1/4}(2-y-y^{-1})+O(q^{5/4})\\
\phi_{g,h}(\gamma_0(\tau,z))&\sim
\phi_{h,g}(\tau,z)= A_0\, q^{1/8}(2-y-y^{-1}) +O(q^{3/8})\\
\phi_{g,h}(\gamma_{1/2}(\tau,z))&\sim 
\phi_{gh^2,h}(\tau,z)=A_{1/2}\, q^{1/4}(2-y-y^{-1}) +O(q^{3/4}) \ , 
\end{align}
where $\sim$ denotes equality up to a phase, and we have included the lowest non-negative powers of $q^r$ that are compatible 
with the multiplier system. It follows from \eqref{eqn:decomp} that  the constants $A_\infty$, $A_0$, $A_{1/2}$ equal
\be 
A_\infty=\Tr_{\H_{g,r=\frac{1}{4}}}(\rho_{g,\frac{1}{4}}(h))\ ,\quad
A_0=\Tr_{\H_{h,r=\frac{1}{8}}}(\rho_{h,\frac{1}{8}}(g))\ ,\quad
A_{1/2}=\Tr_{\H_{gh^2,r=\frac{1}{4}}}(\rho_{gh^2,\frac{1}{4}}(h))\ ,
\ee 
and from the Fourier expansion of $\phi_{h,e}$ we know that $\dim \H_{h,r=\frac{1}{8}}=1$, so that $|A_0|=1$.
On the other hand, at the cusps $\psi(\tau,z)$ satisfies 
\begin{align}
\psi(\gamma_\infty(\tau,z))&= \psi(\gamma_{1/2}(\tau,z))=\frac{\vartheta_1(\tau,z)^2}{\vartheta_2(\tau,0)^2}= \frac{1}{4}(2-y-y^{-1})+O(q)\\ 
\psi(\gamma_0(\tau,z))&= \frac{\vartheta_1(\tau,z)^2}{\vartheta_4(\tau,0)^2}= q^{1/4}(2-y-y^{-1})+O(q^{3/4}) \ , 
%\\ \psi(\gamma_{1/2}(\tau,z))&= \frac{\vartheta_1(\tau,z)^2}{\vartheta_2(\tau,0)^2}= \frac{1}{4}(2-y-y^{-1})+O(q)
\end{align} 
and we therefore obtain
\begin{align}
 F(\gamma_\infty\cdot \tau)&= 4\, A_\infty \, q^{1/4}+O(q^{5/4})\\
 F(\gamma_0\cdot \tau)&\sim A_0\, q^{-1/8}+O(q^{1/8})\\        
 F(\gamma_{1/2}\cdot \tau)&\sim 4\, A_{1/2}\, q^{1/4}+O(q^{3/4})\ .
 \end{align} Up to a phase, there is a unique modular form with the correct modular properties and the 
 expected Fourier expansion at the cusps, namely 
 \be 
 F(\tau)=8\frac{\eta(2\tau)^6}{\eta(\tau)^6}\ .
 \ee 
 Thus we conclude that, up to an overall phase, we have 
\be  \phi_{{\rm 2B,8A_{1,2}}}(\tau,z)=8\frac{\eta(2\tau)^6}{\eta(\tau)^6}\frac{\vartheta_1(\tau,z)^2}{\vartheta_2(\tau,0)^2}=2\frac{\eta(2\tau)^2}{\eta(\tau)^4}\vartheta_1(\tau,z)^2\ ,
\ee 
where we used the identity $\vartheta_2(\tau,0)^2=4\frac{\eta(2\tau)^4}{\eta(\tau)^2}$.
The other non-trivial cases can be worked out similarly; see appendix~\ref{app_detailsttgenera} for the details.

\subsection{Projective representations}\label{s:proj}

As we have argued above, see property (C) in section \ref{sec_ttgenera}, the twisted sector $\H_g$ should carry a projective representation of the 
centraliser $C_{M_{24}}(g)$, whose $2$-cocycle is  given by $c_g$ as determined in (\ref{ch0}). Given our 
explicit knowledge of all twisted twining genera as well as $c_g$, this can now be tested as in 
\cite{Gaberdiel:2010ca} (where the corresponding analysis was performed for the case of the twining genera):
if each $\H_{g,r}$ is a projective representation of $C_{M_{24}}(g)$, we can decompose it as
\be
\H_{g,r} = \bigoplus_{j} h_{g,r}^{(j)} \, R_j \ , 
\ee
where $h_{g,r}^{(j)}$ is the multiplicity with which the irreducible projective representation $R_j$
(whose projectivity is characterised by $c_g$)  appears in $\H_{g,r}$.
On the other hand, using the orthogonality of group characters (see appendix~\ref{app_projreps}), we can 
calculate $h_{g,r}^{(j)}$ from the knowledge of the twisted twining genera explicitly. The consistency condition 
is then that all $h_{g,r}^{(j)}$ are indeed (non-negative) integers. We have done this analysis for all
twisted sectors and for the first $500$ levels (i.e.\ the first $500$ values for $r$), and the multiplicities are 
indeed always non-negative integers.
The explicit results for the first $20$ levels as well as the (projective) character tables of all 
non-isomorphic centralisers $C_{M_{24}}(g)$ are given in appendix~\ref{app:char}. 
The decompositions for all $g$ and for the first $500$ levels are available online \cite{thispaper}.

 \bigskip
 
\section{K3 Orbifolds}
\label{sec_K3orb}

There are at least two further consistency checks on our proposal that can be fairly easily analysed.
Suppose we consider a K3 sigma-model $\C$ whose automorphism group contains the cyclic subgroup generated 
by $g$. Then we can consider the orbifold of $\C$ by $G=\langle g \rangle$. A priori, it is not guaranteed that
this orbifold is consistent --- indeed, since these symmetries effectively act asymmetrically, they generically
suffer from the level-matching problem, and hence may be inconsistent \cite{Narain:1986qm}. 
However, as was already mentioned in \cite{Gaberdiel:2012um}, 
the level-matching condition is satisfied provided that the twining genus $\phi_g$ has a trivial multiplier system.
This is in particular the case if $g\in M_{23}$. In that case, the resulting orbifold theory $\widehat{\C} = \C/G$
is again a K3 sigma model, as was also shown in \cite{Gaberdiel:2012um}.

Suppose then that the original $\C$ has, in addition to $g\in M_{23}$, another commuting group element $h$ in its 
automorphism group. Then $h$ also gives rise to a symmetry $\hat{h}$ of the orbifold theory
$\widehat{\C}$, and we can calculate the twining genus of $\hat{h}$ by the usual orbifold formula
\be\label{orbtwin}
\phi_{e,\hat{h}} = \frac{1}{o(g)} \sum_{i,j=0}^{o(g)-1}
\phi_{g^i,h g^j} \ .
\ee
This leads to a non-trivial consistency condition: given our explicit knowledge of all twisted twining genera, we can
calculate the right-hand-side explicitly, and this must agree, for every $h$, with \emph{one} of the twining genera of 
\cite{Gaberdiel:2010ca,Eguchi:2010fg}. We have checked that this is indeed true; an example is illustrated in section~\ref{s:relex}.

Naively, one may have guessed that $\hat{h}$ should simply agree with $h$, i.e.\ that the left-hand-side of (\ref{orbtwin})
is the twining genus associated to the {\em same} group element $h\in M_{24}$. However, this is in general not the 
case, as will be explained in the following section~\ref{s:relabelling}, see also section~\ref{s:relex} for an 
explicit example. Whenever this happens we will say that the group element $h$ has been {\em relabelled} in the
orbifold theory.
\medskip

The other consistency check is even more obvious. As was mentioned at the beginning of section~\ref{sec_ttgenera},
there exist K3 sigma-models with a commuting pair of automorphisms $(g,h)$. For them (\ref{eqn:twining-trace}) can 
 be calculated directly, and thus compared to our answers. 
We shall perform this computation for some simple cases in section~\ref{sec:ttgex}. In particular, we shall concentrate
on examples where our obstruction analysis predicts that the associated twisted twining genus must
vanish. The fact that we can reproduce this result using elementary methods is a good consistency check
on our analysis.

\subsection{The relabelling phenomenon}\label{s:relabelling}
Suppose that a K3 sigma-model $\C$ contains in its automorphism group an element $g$ such that the orbifold of 
$\C$ by $g$, $\hat\C=\C / \langle g \rangle$ is consistent. 
In general, each $g^r$-twisted sector carries a 
projective representation of $H\equiv C_{M_{24}}(g)$. Equivalently, we can think of the untwisted and twisted sectors as carrying 
genuine representations of some central extension $\tilde{H}$ of $H$. In fact, the central extension can be chosen to be of the form 
(see also appendix~\ref{app:centrorbi} for more details)
\be \label{centext}
1\to \langle \tilde Q\rangle \cong \ZZ_N \to \tilde{H} \stackrel{\rho}{\to} H \to 1\ .
\ee 
Here $\rho$ is the representation in the untwisted sector, i.e.\ on the spectrum of $\C$, $N$ is the order $o(g)$ of $g$ and the generator $\tilde Q$ (that we call the
\emph{quantum symmetry}) of the central $\ZZ_N$ acts by $e^{2\pi i r/N}$ on the $g^r$-twisted sector. Note that there is always a lift $\tilde{g}\in\tilde{H}$ of $g$ that acts by $e^{2\pi i k/N}$ on the states 
of conformal weight $h-1/4= k/N$. 
 
\smallskip

In the orbifold theory $\hat\C$, all states are invariant under the action of $\tilde{g}$ --- this is just the precise
way of imposing the usual requirement that the orbifold theory consists of  `$g$-invariant' states only. Since 
$\tilde{g}$ is in the center of $\tilde{H}$, the spectrum of 
$\hat\C$ carries a well-defined representation $\hat\rho$ of $\tilde{H}$, such that $\hat{\rho}(\tilde g)$ is the identity. In fact, 
it is easy to see that $\ker\hat{\rho}=\langle \tilde g\rangle$, because 
$\tilde{Q}$ acts non-trivially on the twisted sectors, and each non-trivial element in $H/\langle g\rangle$ lifts to an element of
$\tilde{H}$ that acts non-trivially on the $g$-invariant untwisted sector, which in turn is 
part of the spectrum of $\hat\C$. Thus, we have an exact sequence 
\be 
1\to \langle \tilde g\rangle \cong \ZZ_N\to \tilde{H}\stackrel{\hat\rho}{\to}\hat{H}\to 1\ .
\ee
The group
$\hat{H}\equiv\tilde{H}/\langle \tilde g\rangle$ defines a full-fledged symmetry of the orbifold CFT $\hat{\C}$. For the case at hand, both the original theory and the orbifold are K3 sigma models, and thus all of 
these symmetry groups should be subgroups of $M_{24}$,\footnote{We are assuming here that the
relevant K3 sigma model is `non-exceptional' in the terminology of \cite{Gaberdiel:2012um}.}
\be 
\hat{H}\subseteq M_{24}\ .
\ee
This assumption allows us to determine $\hat H$ (at least as an abstract group). In fact, the order of $\hat{H}$ is given by
\be\label{ordhatH}
|\hat{H}|=\frac{ |\tilde H|}{N}=|H|=|C_{M_{24}}(g)|\ .
\ee 
The quantum symmetry  $Q\equiv \hat{\rho}(\tilde{Q})$  must be in the same $M_{24}$-conjugacy class as $g$, since it has the same eigenvalues on 
the $24$-dimensional representation of RR ground states. Since
all elements of $\hat{H}$ commute with $Q$,  we can conclude by \eqref{ordhatH} that
\be \label{bigiso}
\hat{H}= C_{M_{24}}(Q)\cong C_{M_{24}}(g)=H\ .
\ee

This construction allows us to give a precise interpretation of \eqref{orbtwin}. For each symmetry $h\in H$ in the model $\C$, 
one chooses a lift $\tilde{h}\in \tilde H$, with $\rho(\tilde{h})=h$ (this amounts to choosing the phases of the twisted twining genera $\phi_{g,h}$).
Then the  formula for the twining genus of the symmetry $\hat\rho(\tilde h)$ in the orbifold model $\hat\C$ takes the form
\be\label{orbtwin2}
\phi_{e,\hat\rho(\tilde{h})} = \frac{1}{o(g)} \sum_{i,j=0}^{o(g)-1}
\phi_{{g}^i,\tilde{h} \tilde{g}^j} \ .
\ee

After these preparations we can now explain the relabelling phenomenon. 
While it follows from (\ref{bigiso}) that $\hat{H}$ and $H$ are isomorphic, they are not canonically isomorphic; in fact,
we only have a canonical isomorphism between
\be\label{smalliso} 
H/\langle g\rangle \cong \tilde H/\langle \tilde{g},\tilde{Q}\rangle\cong  \hat H/\langle Q\rangle\ ,
\ee
but this does not necessarily lift to a canonical isomorphism $\Phi: \hat{H}\to H$, with $\Phi(Q)=g$. Put differently, 
in general it is not possible to find an 
isomorphism $\Phi: \hat{H}\to H$, with $\Phi(Q)=g$ such that the following diagram commutes:
\[\begin{tikzpicture}[>=angle 90]
\matrix(a)[matrix of math nodes,
row sep=3em, column sep=1.5em,
text height=1.5ex, text depth=0.25ex]
{
&\tilde{H}\\
\hat{H} & & H
\\
& \tilde{H}/\langle \tilde{g},\tilde{Q}\rangle
 \\};
\path[->](a-1-2) edge node[above left]{$\hat \rho$} (a-2-1);
\path[->](a-1-2) edge node[above right]{$\rho$} (a-2-3);
\path[->](a-2-1) edge (a-3-2);
\path[->](a-2-3) edge (a-3-2);
\path[dashed,->](a-2-1) edge node[above]{$\Phi$} (a-2-3);
\end{tikzpicture}\]
 For example, this will be the case if for any lift $\tilde{h}$ with $\rho(\tilde{h})=h$, the corresponding
symmetry $\hat\rho(\tilde{h})$ is not in the same $M_{24}$ conjugacy class as $h$. In that case, the twining 
genus  obtained by formula \eqref{orbtwin2} does  not agree with $\phi_{e,h}$. 
Whenever this is the case, we shall say that the conjugacy class of $h$ has been \emph{relabelled} in the orbifold theory.

\subsection{An example}\label{s:relex}

As an explicit example of the relabelling phenomenon  let us consider the case where $g$ is 
in the class 2A of $M_{24}$ (so that $N=o(g)=2$). Then the group $\tilde{H}$ is a central extension of $H$ by the $\ZZ_2$ group 
generated by $\tilde Q$.
We denote by $\tilde{g}$ the lift of $g$ in $\tilde{H}$, such that $\tilde{g}$ and $\tilde{Q}$ have the following eigenvalues
\be \begin{array}{cclc}
\tilde Q & \tilde g & \text{sector} & \text{model}\\\hline\\[-10pt] 
+1 & \ +1 \ \ & \text{untwisted} & \C,\hat{\C}\\
+1 & -1 & \text{untwisted} & \C\\
-1 & +1 & \text{twisted},\ q^\ZZ & \hat{\C}\\
-1 & -1 & \text{twisted},\ q^{\ZZ+\frac{1}{2}} & -
\end{array}
\ee  
As expected from the discussion in the previous section, the group $\tilde{H}/\tilde Q$ is
isomorphic to the centraliser in $M_{24}$ of an element of class 2A. 
\smallskip

Now consider an element $h\in H$ in class 8A, with $h^4=g$. The trace of $h$ over the $24$-dimensional 
representation of ground states in the original model $\C$ is $+2$. Furthermore, since $hg$ is in the same class, 
also the trace over $hg$ equals $+2$, from which we conclude that on the ground states of $\C$ we have 
\be 
\Tr_{\H^0_{\tilde Q=1,\tilde g=1}}(h)=+2\ ,\qquad \Tr_{\H^0_{\tilde Q=1,\tilde g=-1}}(h)=0\ ,
\ee  where $\H^0_{\tilde Q=\pm 1,\tilde g=\pm 1}$ are the eigenspaces of $\tilde{Q}$ and $\tilde{g}$ at conformal weight $h_L-\frac{1}{4}=0$.
The lifts $\tilde{h}$ and $\tilde Q\tilde h$ in $\tilde{H}$ are such that 
$\tilde{h}^4=(\tilde Q\tilde{h})^4=\tilde{g}$. On dimensional grounds, the twisted sector at 
$h_L-1/4=0$ must correspond to a certain $8$-dimensional irreducible representation of 
$\tilde{H}$, where the lifts $\tilde{h}$ and $\tilde Q\tilde{h}$ have trace $\pm 2$. 
We notice that, up to a sign,  this trace is independent of the choice of the central extension, 
and is in agreement with the expectation from the twisted twining genus $\phi_{g,h}\equiv\phi_{h^4,h}$.  
On the ground states in the orbifold theory $\hat\C$, the trace of $\tilde{h}$ is now
\be 
\Tr_{\H^0_{\tilde Q=1,\tilde g=1}}(\tilde{h}) + \Tr_{\H^0_{\tilde Q=-1,\tilde g=1}}(\tilde{h})=2\pm 2\ ,
\ee 
and thus there is no choice for $\tilde{h}$ such that the trace is equal to $+2$, as required for an element in class 8A. 
In fact, the images $\hat h$ and $Q\hat{h}$ in $\hat{H}\cong C_{M_{24}}(Q)$ have order $4$ and belong to the classes 4A 
and 4B.  In particular, while the pair of commuting elements $g,h$ in the original model generate the cyclic group 
\be \langle g,h\rangle=\langle h\rangle\cong \ZZ_8\ ,\ee
both the groups $\langle Q, \hat{h}\rangle$ and $\langle Q, Q\hat{h}\rangle$ correspond to group 12 in our general list. Thus, there cannot exist an isomorphism $\Phi:\hat H\to H$ such that $\Phi(\hat\rho(\tilde h))=h$ or $\Phi(\hat\rho(\tilde Q\tilde h))=h$.

\smallskip

Let us verify the consistency of eq.~\eqref{orbtwin2} for this case, see the comments at the beginning of section~\ref{sec_K3orb}. 
The twining genera 
$\phi_{e,\hat\rho(\tilde{h})}$ and $\phi_{e,\hat\rho(\tilde{Q}\tilde{h})}$ are given by
\begin{align}
\phi_{e,\hat\rho(\tilde{h})}(\tau,z)&=\frac{1}{2}(\phi_{e,h}(\tau,z)+\phi_{e,h^5}(\tau,z)+\phi_{h^4,h}(\tau,z)+\phi_{h^4,h^5}(\tau,z))\ ,\\
\phi_{e,\hat\rho(\tilde{Q}\tilde{h})}(\tau,z)&=\frac{1}{2}(\phi_{e,h}(\tau,z)+\phi_{e,h^5}(\tau,z)-\phi_{h^4,h}(\tau,z)-\phi_{h^4,h^5}(\tau,z))\ .
\end{align}
Using the identity
\be 
\phi_{e,{\rm 8A}}(\tau,z)=\frac{1}{2}(\phi_{e,{\rm 4B}}(\tau,z)+\phi_{e,{\rm 4A}}(\tau,z))\ ,
\ee  it is easy to verify that 
\be
\phi_{e,\hat\rho(\tilde{h})}=\phi_{e,{\rm 4B}},\qquad \phi_{e,\hat\rho(\tilde{Q}\tilde{h})}=\phi_{e,{\rm 4A}}\ ,
\ee 
as expected. This example also shows that the consistency of eq.~\eqref{orbtwin2} gives, in general, highly non-trivial 
conditions on the twisted twining genera.

\smallskip

We have verified explicitly that a similar phenomenon also occurs for the orbifold by an element $g$ in class 4B.

\subsection{Computation of a twisted twining genus}\label{sec:ttgex}
In this section we shall calculate some twisted twining genera directly. In particular, we shall
concentrate on some simple cases where our general analysis predicts that they must vanish (because
of some obstruction). As we shall see, we can reach the same conclusion from first principles.

Let us consider a K3 $\sigma$-model $\C$ whose automorphism group contains two commuting elements
$g$, $h$ of order $m$ and $n$, respectively, generating the non-cyclic group $G=\langle g,h\rangle$. If these
symmetries have a geometric origin, i.e.\ if they are induced by automorphisms of the K3 target space that fix the 
holomorphic $2$-form, then 
the group  $G$ must be a subgroup of $M_{23}$ with at least five orbits over the $24$-dimensional representation,
as follows from the Mukai theorem \cite{Mukai}.
These conditions are satisfied by the groups 1 and 3 ($\ZZ_2\times \ZZ_2$), 
groups 17 and 19 ($\ZZ_2\times\ZZ_4$),  group 25 ($\ZZ_4\times \ZZ_4$), group 28 ($\ZZ_2\times \ZZ_6$) 
and group 33 ($\ZZ_3\times\ZZ_3$) of our table~\ref{t:groups}. In all these cases, the twisted twining 
genus vanishes, $\phi_{g,h}=0$, see table~\ref{t:res}. In the following we want to verify this independently. 
\smallskip

In the general $\ZZ_n\times\ZZ_m$ case, the twisted twining genus $\phi_{g,h}$ is expected to be 
a weak Jacobi form of weight $0$ and index $1$ under the congruence subgroup
\be \Gamma_{m,n}\equiv \{\left(\begin{smallmatrix}
a & b\\ c & d
\end{smallmatrix} \right)\in SL(2,\ZZ)\mid a\equiv 1,b\equiv 0\mod m,\ c\equiv 0,d\equiv 1\mod n\}\ .
\ee 
Note that this group is in general smaller than the group $\Gamma_{g,h}$ defined below (\ref{d1}), 
because we are not using the identifications under conjugation \eqref{conj1}. Since both $g$ and $h$ are elements of $M_{23}$,
the multiplier system for $\phi_{g,h}$ must be trivial (since $H^3(M_{23},U(1))\cong 0$). 

Because of the triviality of the multiplier system, it follows that any modular image $\phi_{g',h'}$ of $\phi_{g,h}$, where $(g',h')=(g,h)\cdot \gamma$ for some $\gamma\in SL(2,\ZZ)$, has a Fourier expansion of the form
\be\label{fourexp} \phi_{g',h'}(\tau,z)=\sum_{n=0}^\infty\sum_\ell c_{g',h'}(n,\ell) \, q^{\frac{n}{o(g')}}\, y^\ell\ ,
\ee 
where $o(g')$ is the order of $g'$. Let us assume
\be\label{ciszero} 
c_{g',h'}(0,\ell)=0 \quad \forall \ell \in\ZZ\ ,
\ee 
for any modular image $(g',h')=(g,h)\cdot \gamma$; we will prove this below. Then it follows that 
\be 
\phi_{g,h}(\tau,0)=\sum_\ell c_{g,h}(0,\ell)=0\ .
\ee 
Because of (\ref{selfconj}),  $\phi_{g,h}(\tau,z)=\phi_{g,h}(\tau,-z)$, and we can conclude that $\phi_{g,h}$ has a double zero at 
$z=0$ (mod $\ZZ+\tau\ZZ$). Since Jacobi forms of index 1 have only two zeroes on each fundamental domain of $\CC/(\ZZ+\tau\ZZ)$ \cite{EichlerZagier}, $\phi_{g,h}$  is nowhere else vanishing. Next we recall that $\vartheta_1(\tau,z)^2/\eta(\tau)^2$ is a Jacobi form of weight $0$ and index $1$ under $SL(2,\ZZ)$ 
(with a non-trivial multiplier). It has a double zero at $z=0$ (mod $\ZZ+\tau\ZZ$), and is non-vanishing elsewhere, so that
\be 
F_{g,h}(\tau)=\frac{\eta(\tau)^2\phi_{g,h}(\tau,z)}{\vartheta_1(\tau,z)^2}
\ee 
is a modular function of weight $0$ under $\Gamma_{m,n}$ (with non-trivial multiplier), which is 
holomorphic in the interior of the upper-half plane ${\mathbb{H}}_+$. Let us consider the 
asymptotics of $F_{g,h}$ at the boundary $\Gamma_{m,n}\backslash (\QQ\cup \infty) $ of the 
compactification $\Gamma_{m,n}\backslash\bar {\mathbb{H}}_+$. The boundary is the union of points (cusps)  
of the form $\gamma(\infty)$, where $\gamma$ runs through the representatives of the double cosets 
$\Gamma_{m,n}\backslash SL(2,\ZZ)/\Gamma^{\infty}$. Here, $\Gamma^{\infty}$ is the parabolic subgroup of $SL(2,\ZZ)$ that fixes $\infty\in \bar\HH_+$ and is generated by $\left(\begin{smallmatrix}1 & 1\\0 & 1\end{smallmatrix}\right)$.  The Jacobi form  $\vartheta_1(\tau,z)^2/\eta(\tau)^2$ has the same 
asymptotic behaviour at each cusp, namely
\be 
\frac{\vartheta_1(\gamma\cdot(\tau,z))^2}{\eta(\gamma\cdot\tau)^2}\stackrel{\tau\to\infty}{\sim} q^{1/6}(2-y-y^{-1})+O(q^{7/6})\ ,
\ee 
up to a constant coefficient. By \eqref{fourexp} and \eqref{ciszero}, it then follows that
\be 
F_{g,h}(\gamma\cdot\tau)\sim \frac{\eta(\tau)^2\phi_{g',h'}(\tau,z)}{\vartheta_1(\tau,z)^2}
\stackrel{\tau\to\infty}{\sim} O(q^{\frac{1}{o(g')}-\frac{1}{6}})\ ,
\ee 
where $(g',h')=(g,h)\gamma$ and $o(g')$ is the order of $g'$. For all the groups we are interested in, we have 
$o(g')\le 6$, and the inequality is strict for at least one $g'$. It follows that $F_{g,h}$ has no poles and at least one 
zero at the boundary of $\Gamma_{m,n}\backslash\bar {\mathbb{H}}_+$. Since it is holomorphic in the interior, it must 
therefore vanish  identically, and we conclude that 
\be 
\phi_{g,h}(\tau,z)=0\ ,
\ee 
as predicted by our results.
\smallskip

It therefore only remains to prove \eqref{ciszero}. The coefficient $c_{g,h}(0,\ell)$ corresponds to the trace of $h$ (or, more generally, 
of the lift $\tilde h$ to a central extension of the group $\langle g,h\rangle$) on the RR $g$-twisted states with conformal 
weight $h_L=h_R=1/4$ and weight $\ell$ under the $\mathfrak{su}(2)_L$ subalgebra of the left-moving $\N=4$ 
algebra. 
The $g$-twisted ground states are $g$-invariant (the conformal weight satisfies $h_L-1/4\in\ZZ$), so they belong to the spectrum of the orbifold theory $\C/\langle g\rangle$.
  
Any sigma-model whose target space is a K3 manifold $X$ contains $24$ RR ground states corresponding to the harmonic forms on 
$X$. Thus, the real vector space spanned by these states can be identified with the real cohomology $H^*(X,\RR)$ on $X$. 
The orbifold model $\hat\C=\C/\langle g\rangle$ corresponds to a non-linear sigma-model on the geometric orbifold 
$\hat X=X/\langle g\rangle$, which is a singular K3 surface. In particular, for each point of $X$ that is fixed by some non-trivial subgroup 
$\langle g^r\rangle\subseteq \langle g\rangle$, the orbifold $\hat X$ has a singularity of type $A_{n-1}$, where $n$ is the order of $g^r$.
By blowing up all these singularities, one obtains again a smooth K3 manifold. 

The RR ground states in the untwisted sector of 
$\hat\C$ are obtained by projecting onto the $g$-invariant subspace of $H^*(X,\RR)$, while the ground states in the twisted 
sectors are the Poincar\'e duals of the cycles corresponding to the exceptional divisors in the blow-up of $\hat X$. 
More precisely, the resolution of each $A_{n-1}$ singularity of $\hat\C$ gives rise to $n-1$ rational curves $\PP^1$, 
representing elements in the homology of the blow-up. The cohomology classes dual to these cycles span an $(n-1)$-dimensional
subspace of RR ground states for $\hat\C$, one in each $g^{ir}$ twisted sector, for $i=1,\ldots,n-1$. 
Thus, we arrive at the usual statement that the ground states of the $g^k$-twisted sector are in one to one correspondence 
with the $g^k$-fixed points of the target space.  

Any other symplectic automorphism $h$ of $X$ that commutes with $g$ acts as a permutation on the $g^r$-fixed points. This action induces a 
permutation on the singularities of $\hat X$, and therefore on the twisted RR ground states dual to the exceptional 
cycles.\footnote{More precisely, one should choose a lift $\tilde h$ of $h$ to a central extension of the group of 
symmetries of $\C$; then, $\tilde h$ acts on the $g$-twisted sector by a permutation times an overall phase that 
depends on the particular lift. Since the phase is not important for our argument, we will neglect this subtlety.} Therefore, the trace of $h$ over the $g$-twisted sector only gets a non-zero contribution from the singularities 
of $\hat X$ that are fixed by $h$, i.e.\ only from the points of $X$ that are fixed by both $g$ and $h$. It is known (see, for example  
Proposition 1.5 of \cite{Mukai}) that a finite group of symplectic automorphisms of a K3 manifold $X$ that fixes at least one 
point must be a subgroup of $SL(2,\CC)$. On the other hand, all finite abelian subgroups of $SL(2,\CC)$ are cyclic. Thus we 
conclude that if $\langle g,h\rangle$ is abelian but not cyclic, the trace of $h$ over the $g$-twisted ground fields is $0$ and 
\eqref{ciszero} follows.
\smallskip

Our results also agree with the analysis of \cite{Sen:2010ts}, where it was shown that the twisted twining genus $\phi_{g,h}$ 
for a pair of commuting symplectic automorphisms $g,h$ of a K3 manifold $X$ is given by a sum of contributions corresponding 
to the  points fixed by both $g$ and $h$. If the abelian group $\langle g,h\rangle$ is not cyclic, then there are no fixed points 
and $\phi_{g,h}=0$.

\section{Conclusions}
\label{sec_conclusions}

\subsection{Summary} Inspired by the generalised Monstrous Moonshine idea of Norton \cite{Norton} we have
in this paper established  `generalised Mathieu Moonshine' for the elliptic genus of K3.  
More specifically, we have found, for each commuting pair 
$(g,h)\in M_{24}$, explicit expressions for the twisted twining genera
$\phi_{g,h} : \mathbb{H}_+\times \mathbb{C} \to \mathbb{C}$, and we have verified
that these functions satisfy conditions (A), (B), (D)  of section \ref{sec_ttprop}, and given 
convincing evidence that condition (C) is also met. In particular, 
the twisted twining genera $\phi_{g,h}(\tau, z)$ are weak Jacobi forms of weight 0 and index 1 for certain subgroups 
$\Gamma_{g,h}\subset SL(2,\mathbb{Z})$ with a multiplier system
$ \chi_{g,h}  :  \Gamma_{g,h} \to U(1)$. 

One of the key insights of our work is that these multiplier phases are all determined in terms of a class
$[\alpha]\in H^3(M_{24}, U(1))$ via a formula (see eq.~(\ref{ch0})) that was first derived by 
Dijkgraaf and Witten in the context of holomorphic orbifolds \cite{Dijkgraaf:1989pz}. We have also shown that the twisted 
twining genera $\phi_{g,h}$ are compatible with a decomposition of the $g$-twisted sector into (projective) representations 
of the centraliser $C_{M_{24}}(g)$, up to the first 500 levels. The particular central extension of $C_{M_{24}}(g)$ associated with this projective 
representation is also determined by the cohomology class $[\alpha]\in H^3(M_{24}, U(1))$. 

As it turned out, many of the twisted twining genera  vanish, and in most cases this follows from the structure of the class $[\alpha]\in H^3(M_{24}, U(1))$. While these sorts of cohomological obstructions
should also arise in the context of other holomorphic orbifolds (in particular in Monstrous Moonshine), our results provide, 
to our knowledege, the first example where they have been explicitly verified. We have also confirmed the vanishing
of some of these twisted twining genera independently, using geometrical arguments.

\subsection{Open problems and future work}

The elliptic genus of K3 is closely related to the physics of $1/4$ BPS-dyons in 
$\mathcal{N}=4$ string theory. The generating function of elliptic genera of symmetric products of K3's is the famous Igusa cusp 
form $\Phi_{10}$, whose inverse is the partition function of $\mathcal{N}=4$ dyons \cite{Dijkgraaf:1996it,Shih:2005uc}. This 
fact lies at the heart of the recent progress in understanding wall-crossing of multi-centered BPS-states in $\mathcal{N}=4$ string 
theory using mock modular forms \cite{Dabholkar:2012nd}. It is therefore natural to wonder about the corresponding physical 
interpretation of the twisted twining genera $\phi_{g,h}$. In the special case of a trivial twist $g=e$, it has been shown that for 
some elements $h$ the twining genera $\phi_{e,h}$  correspond to BPS-indices in CHL-orbifolds 
\cite{Cheng:2010pq,Govindarajan:2010fu,Govindarajan:2010cf,Govindarajan:2011em}. We  may therefore
expect that the twisted twining genera $\phi_{g,h}$ should be related to the counting of `twisted dyons' in CHL-models. 
Indeed for some pairs $(g,h)$ we have verified this by showing that the twisted CHL-indices of Sen \cite{Sen:2010ts} are 
compatible with our $\phi_{g,h}$. It would be  interesting to determine more generally whether $\phi_{g,h}$ have a 
physical interpretation in terms of CHL-models or some generalisation thereof.
\smallskip

In Borcherds' proof of the Monstrous Moonshine conjecture he introduced a new class of infinite-dimensional 
(super-)Lie algebras that he called generalised Kac-Moody algebras (GKM) \cite{Borcherds}. The key idea was that the 
multiplicative lift of the modular $J$-function could be interpreted as the denominator formula for a specific GKM, now 
commonly called the Monster Lie algebra, whose root system carries an action of the Monster group $\mathbb{M}$.
The generalised moonshine ideas of Norton \cite{Norton} suggest that similar GKMs should exist for each class 
$[g]\in \mathbb{M}$ \cite{CarnahanI,CarnahanII,Duncan:2009sq}. In the context of Mathieu Moonshine the role of 
generalised Kac-Moody algebras has so far not been understood. Although there is a GKM associated with the elliptic 
genus of K3 \cite{GritsenkoNikulin}, this algebra does not seem to carry any natural action of $M_{24}$ due to the fact that its denominator formula 
is $1/\sqrt{\Phi_{10}}$ rather than $1/\Phi_{10}$. For some conjugacy classes $[g]\in M_{24}$ the 
associated second-quantised twining genera $\Phi_g$ also give rise to denominator formulas of 
GKMs \cite{GritsenkoClery}, which should presumably be identified with the wall-crossing algebras found  in 
CHL-models \cite{Cheng:2008fc,Cheng:2008kt,Govindarajan:2008vi,Govindarajan:2009qt,Govindarajan:2010fu}. 
Although these observations are very suggestive, the precise role of  the GKMs for Mathieu Moonshine remains to be 
understood.

A crucial point in our analysis was that the multiplier phases are controlled by a class
$[\alpha]\in H^3(M_{24}, U(1))$, from which also the various obstructions could be read off. It follows from the work of 
\cite{Dijkgraaf:1989pz,Coste:2000tq} that similarly
a cohomology class in $H^3(\mathbb{M}, U(1))$ should underly generalised Monstrous Moonshine, and it 
would be very interesting to see whether this point of view may lead to new insights. In particular, some of Norton's 
generalised Moonshine functions $f(g,h;\tau)$ are known to vanish, and it would be natural to expect that this is
due to some cohomological obstructions similar to those we have found in this paper.\footnote{We thank
Terry Gannon for suggestions along these lines, see also \cite{GannonMathieu}.}
Some related ideas were already put forward some time ago by Mason \cite{MasonCohomology}, but very little has 
been done explicitly, mainly because the group $H^3(\mathbb{M}, U(1))$ seems to be poorly understood. 

 In a similar vein,
it would also be interesting to investigate the 
generalised versions of the recently discovered Umbral Moonshine observations \cite{Cheng:2012tq}, of which the 
$M_{24}/{\rm K3}$-moonshine is a special case. In particular, one might expect that there is a relation between the  
Rademacher summability condition of \cite{Cheng:2011ay} and the cohomological obstructions we have described here.
\smallskip

The fact that there is a class in $H^3(M_{24}, U(1))$  determining the properties of the twisted twining genera 
$\phi_{g,h}$ represents strong evidence for the idea that a holomorphic VOA $\mathcal{H}$ underlies Mathieu Moonshine. 
Indeed, the original motivation for suspecting the relevance of the cohomology group $H^3(M_{24}, U(1))$ was based on the 
formal analogy with holomorphic orbifolds  \cite{Dijkgraaf:1989pz,Coste:2000tq}. In a sense we are therefore in a 
similar situation now as for Monstrous Moonshine before the discovery of the  Monster VOA $V^{\natural}$ \cite{FLM}. 
However, the story appears to be more subtle since the natural expectation that $\mathcal{H}$ would 
arise as a superconformal $\sigma$-model on K3 does not hold \cite{Gaberdiel:2011fg}. On the other hand, this is perhaps 
not too surprising, given the fact that the modular $J$-function corresponds to the {\em full} partition function of a CFT, while the elliptic 
genus of K3 only receives contribution from a {\em subset} of the physical states of the $\mathcal{N}=(4,4)$ superconformal theory on 
K3. Hence it is tempting to speculate that $\mathcal{H}$ should correspond to something like the algebra of BPS-states of 
string theory on K3 (in the sense of Harvey and Moore \cite{Harvey:1995fq,Harvey:1996gc}). In particular, one might
therefore hope that $\H$ can be constructed using some kind of topological sigma model, e.g.\ 
the `half-twisted' topological $\sigma$-model on K3 \cite{Kapustin:2005pt}.
Mathematically, this sigma model (or rather its large radius limit) can be viewed as a 
bundle of VOAs known as the chiral de Rham complex, $\Omega_{X}^{ch}$ \cite{MSV}, and in 
\cite{BorisovLibgober} it was shown that the elliptic genus of K3 can be obtained by taking the graded trace over the corresponding
cohomology $H^{*}(\Omega_{X}^{ch})$. It is therefore tempting to speculate that one may be able to construct an action of 
$M_{24}$ on  the cohomology of $\Omega_{X}^{ch}$, at least for some choice of K3 surface $X$.
\smallskip

We hope to return to these and related issues in future work.

\section*{Acknowledgements} 

We are indebted to Terry Gannon for the suggestion that the 3rd cohomology of $M_{24}$ should control
the modular properties of the twisted twining genera, as well as for many useful explanations and directions to
the relevant literature. We are furthermore grateful to Mathieu Dutour and Graham Ellis for helpful correspondence on using the computer package \texttt{HAP} for calculating the cohomology of finite groups. We also thank  Nils Carqueville, Miranda Cheng, Jeff Harvey, Stefan Hohenegger, Axel Kleinschmidt, Sameer Murthy, Daniel Roggenkamp, Ashoke Sen and Don Zagier for useful discussions, as well as the anonymous referee for many useful suggestions that 
considerably improved the exposition of the paper. D.P. would like to thank the Albert Einstein Institute in Golm  for hospitality while part of this work was carried out. The work of M.R.G. is 
partially supported by the Swiss National Science Foundation, and the visit of Henrik Ronellenfitsch at ETH was supported
by the Pauli Center.

\appendix 

\section{Details on the unobstructed twisted twining genera}
\label{app_detailsttgenera}

In this appendix, following the example of section \ref{sec_8A2Bex}, we will determine the twisted twining genera $\phi_{g,h}$ 
for the groups 13, 23, 24, 33 and 34 in Table 1. The genus $\phi_{\rm 2B,8A_{1,2}}$ for group 27 was obtained in section~\ref{sec_8A2Bex}, 
while for all the other groups $\phi_{g,h}$ vanishes due to some obstruction.

\subsection{The characters $\phi_{\rm 2B,4A_2}$ (group 13) $\phi_{\rm 4B,4A_3}$ (group 23) and $\phi_{\rm 4B,4A_4}$ (group 24)} 

In this section we will consider the twisted twining genera $\phi_{g,h}$ for groups 13, 23 and 24; as we will see, it turns out that these three characters have exactly the same modular properties. For group 13 $g$ is in class 2B and $h$ in class 4A, and we can choose
\begin{align}
&g=(1,10)(2,14)(3,8)(4,16)(5,15)(6,12)(7,21)(9,18)(11,13)(17,20)(19,23)(22,24)\ ,\notag\\
&h= (1,2)(3,18)(4,7,15,17)(5,20,16,21)(6,19,24,11)(8,9)(10,14)(12,23,22,13)\ .
\end{align} For the groups 27 and 28, $g$ is in class 4B and $h$ in class 4A and we can choose
\be 
g=(3,8)(4,21,15,20)(5,17,16,7)(6,19,24,11)(9,18)(12,23,22,13)\ ,
\ee for both groups (from now on, we drop the trivial cycles). 
 For group 23 we can take 
\be h=  (1,10)(2,14)(3,9)(4,11,5,13)(6,17,12,21)(7,22,20,24)(8,18)(15,19,16,23)\ .
\ee
and for group 24
\be
h= (1,2,10,14)(3,9,8,18)(4,5,15,16)(6,13)(7,21,17,20)(11,22)(12,19)(23,24)\ ,
\ee 
In all the three cases we have the relations 
\be 
(k^{-1}gk,k^{-1}ghk)=(g,h)\ ,\qquad
(r^{-1}gh^{-2}r,r^{-1}hr)=(g,h)\ ,
\ee 
where, for group 13
\begin{align}
&k=(1,9,2,3)(6,13,22,19)(7,21)(8,10,18,14)(11,24,23,12)(17,20)\ ,\\
&r= (3,8)(4,11,5,13)(6,20,12,7)(9,18)(15,19,16,23)(17,24,21,22)\ ,
\end{align}
for group 23
\begin{align}
&k=(1,2,10,14)(3,9,8,18)(5,16)(6,11,24,19)(7,17)(12,23,22,13)\ ,\\
&r=(1,2)(3,8)(6,12)(7,20)(9,18)(10,14)(17,21)(22,24)\ ,
\end{align}
and for group 24
\begin{align}
&k=(3,9)(4,12,5,6)(7,11,20,13)(8,18)(15,22,16,24)(17,19,21,23)\ ,\\
&r= (1,3)(2,9)(4,20)(5,7)(8,10)(14,18)(15,21)(16,17)\ .
\end{align}
Using the explicit knowledge of the 3-cocycle $\alpha$ we can derive
the modular properties of the twisted twining genus $\phi_{g,h}$, including the multiplier system. It turns out that in all these three cases $\phi_{g,h}$ is invariant under $\Gamma_0(2)$ with the same multiplier
\be \phi_{g,h}(\tau+1)=c_g(g,h)\frac{c_g(gh,k)}{c_g(k,h)}\phi_{g,h}(\tau)=i\phi_{g,h}(\tau) \ ,
\ee
\begin{align} \phi_{g,h}(\frac{\tau}{2\tau+1})&=\frac{1}{c_{h}(h,h^{-1}g)c_{h}(h,h^{-2}g)}\frac{c_{h}(h^{-2}g,r)}{c_{h}(r,g)}\phi_{g,h}(\tau)\notag\\ &=-i\phi_{g,h}(\tau) \ .
\end{align}
By arguments similar to the ones in section \ref{sec:ttgex}, $\phi_{g,h}(\tau,z)$ has only a double zero at $z=0$ (mod $\ZZ+\tau\ZZ$). This implies that  $\phi_{g,h}$ can be expressed as
\be 
\phi_{g,h}(\tau,z)=\frac{\vartheta_1(\tau,z)^2}{\vartheta_2(\tau,0)^2}F(\tau)
\ee 
for some $\Gamma_0(2)$-invariant function $F(\tau)$ that is holomorphic in $\HH_+$. 
The group $\Gamma_0(2)$ has two cusps at $\infty$ and $0$.
To proceed we study the asymptotic behavior at infinity
\be \phi_{g,h}(\tau,z)= A_\infty \, q^{1/4}(2-y-y^{-1})+\cdots\ee
\be
\frac{\vartheta_1(\tau,z)^2}{\vartheta_2(\tau,0)^2}=\frac{1}{4}\, (2-y-y^{-1})+\cdots \ ,
\ee 
which implies that $F(\tau)\sim 4A_\infty\, q^{1/4}+\cdots $ has a zero at $\tau\to\infty$. The behavior at $\tau=0$ is 
\be 
\phi_{g,h}(\gamma_0(\tau,z))\sim \phi_{h,g^{-1}}(\tau,z)= A_0 \, q^{1/8} \, (2-y-y^{-1})+\cdots 
\ee
\be
\frac{\vartheta_1(\tau,z)^2}{\vartheta_4(\tau,0)^2}= q^{1/4}\, (2-y-y^{-1})+\cdots \ ,
\ee 
implying that $F(\gamma_0\cdot\tau)\sim A_0\, q^{-1/8}+\cdots$ has a pole at $0$. 
A function with the correct modular properties and behaviour at the cusps is
\be 
F(\tau)=4A_\infty\, \frac{\eta(2\tau)^6}{\eta(\tau)^6}\ ,
\ee 
so that
\be
 \phi_{g,h}(\tau,z)=4A_\infty\frac{\eta(2\tau)^6}{\eta(\tau)^6}\frac{\vartheta_1(\tau,z)^2}{\vartheta_2(\tau,z)^2}=A_\infty \frac{\eta(2\tau)^2}{\eta(\tau)^4}\vartheta_1(\tau,z)^2\ .
\ee 
The constants $A_\infty$ for the three cases are determined (up to the phase) by ensuring a decomposition into projective characters of 
$C_{M_{24}}(g)$. In particular, $A_\infty=4$ for group 13 and $A_\infty=2\sqrt{2}$ for groups 23 and 24.

\subsection{The cases $\phi_{\rm 3A,3A_3}$ (group 33) and $\phi_{\rm 3A,3B_1}$ (group 34)}
\label{app_modularobstruction}

 In this section we show 
 that the unobstructed twisted twining genera $\phi_{\rm 3A,3A_3}$ and $\phi_{\rm 3A,3B_1}$ must
 vanish by modular arguments. 

Let us consider first the group $\left<g,h\right>=\left<{\rm 3A,3B_1}\right>$ (type 34), generated by
\be 
g=(2,23,11)(3,20,10)(5,14,9)(6,16,13)(7,19,22)(12,24,18)
\ee 
in class 3A and
\be 
h=(1,15,17)(2,19,16)(3,18,5)(4,21,8)(6,11,7)(9,10,24)(12,14,20)(13,23,22)
\ee 
in class 3B. The following conjugation relation holds 
\be 
(k^{-1}gk,k^{-1}ghk)=(g,h)\ ,
\ee
 where
\be 
k=(2,7,16)(3,20,10)(4,21,8)(5,9,14)(6,11,22)(13,23,19)\ . 
\ee 
The modular properties of the associated twisted twining genus are
\be 
\phi_{g,h}(\tau+1)=c_g(g,h)\frac{c_g(gh,k)}{c_g(k,h)}\phi_{g,h}(\tau)=e^\frac{4\pi i}{3}\phi_{g,h}(\tau)\ ,
\ee 
and
\be 
\phi_{g,h}(\frac{\tau}{-3\tau+1})=\prod_{i=0}^2c_h(h,h^ig^{-1})\phi_{g,h}(\tau)=e^\frac{4\pi i}{3}\phi_{g,h}(\tau)\ ,
\ee 
where we have used the explicit form for the multiplier phases as determined in terms of the 
3-cocycle $\alpha\in H^3(M_{24}, U(1))$. We conclude that, 
up to the multiplier, $\phi_{g,h}$ is invariant under $\Gamma_0(3)$. Furthermore, as in the previous subsection and in section \ref{sec:ttgex}, for each fixed $\tau\in\HH_+$, $\phi_{g,h}(\tau,z)$ has only a double zero at $z=0$ (mod $\ZZ+\tau\ZZ$). Since $\frac{\vartheta_1(\tau,z)^2}{\eta(\tau)^2}$ is a weak Jacobi form of weight $0$ and index $1$  with non-trivial multiplier under $SL(2,\ZZ)$ and only a double zero at $z=0$, we have
\be 
\phi_{g,h}(\tau,z)=\frac{\vartheta_1(\tau,z)^2}{\eta(\tau)^2}F_{g,h}(\tau) 
\ee 
for some function $F_{g,h}(\tau)$ that is $\Gamma_0(3)$-invariant (up to a multiplier).  The asymptotics for $\tau\to \infty$ is 
\be 
\phi_{g,h}(\tau,z)\sim A_\infty \, q^\frac{2}{3}+\cdots\ ,
\ee 
for some (possibly vanishing) constant $A_\infty$, while the asymptotic behavior at $0$ is given by
\be 
\phi_{h,g}(\tau,z)\sim A_0 \, q^\frac{2}{9}+\cdots\ .
\ee 
Since both at $\infty$ and $0$ we have
\be 
\frac{\vartheta_1(\tau,z)^2}{\eta(\tau)^2}\sim q^{1/6}(2-y-y^{-1})+\cdots \ ,
\ee 
we deduce that $F_{g,h}$ has a zero at both $0$ and $\infty$ and no poles; hence it must vanish. We therefore conclude
\be 
\phi_{{\rm 3A,3B_1}}(\tau,z)=0\ ,
\ee
as claimed.

The $({\rm 3A,3A_3})$-case (group 33) is analogous. Here, the group is generated by
\be 
g=(2,23,11)(3,20,10)(5,14,9)(6,16,13)(7,19,22)(12,24,18)
\ee 
and
\be 
h=(2,22,13)(4,21,8)(5,14,9)(6,23,7)(11,19,16)(12,18,24) \ ,
\ee
and 
we have the relations
\be 
(k^{-1}gk,k^{-1}ghk)=(g,h)\ ,\qquad (r^{-1}hr,r^{-1}g^{-1}r)=(g,h)\ee
where
\be 
k=(1,17,15)(3,9,24)(5,18,20)(6,16,13)(7,22,19)(10,14,12)\ ,
\ee
\be 
r=(3,4,10,21)(5,18)(6,16,19,7)(8,20)(9,24,14,12)(11,22,23,13)\ .
\ee  
The modular properties of $\phi_{g,h}$ are
\be 
\phi_{g,h}(\tau+1)=c_g(g,h)\, \frac{c_g(gh,k)}{c_g(k,h)}\, \phi_{g,h}(\tau)=e^\frac{4\pi i}{3}\, \phi_{g,h}(\tau)\ ,
\ee
\be \phi_{g,h}(-1/\tau)=\frac{1}{c_h(g,g^{-1})}\frac{c_h(g^{-1},r)}{c_h(r,h)}\phi_{g,h}(\tau)=\phi_{g,h}(\tau)\ .
\ee 
Thus $\phi_{g,h}$ is invariant (up to a multiplier) under $SL(2,\ZZ)$, and $\phi_{g,h}\sim A\, q^{2/3}$ at $\infty$, so that 
\be 
F_{{\rm 3A,3A_3}}(\tau)\equiv \phi_{g,h}(\tau,z)\frac{\eta(\tau)^2}{\vartheta_1(\tau,z)^2}\sim A \, q^{1/2}+\cdots \ .
\ee 
Thus, $F_{{\rm 3A,3A_3}}$ has a zero and no poles and therefore must vanish, thus leading to 
\be 
\phi_{{\rm 3A,3A_3}}(\tau, z)=0\ .
\ee

\section{Some group cohomology}\label{app:groupcoho}

In this section, we will review the basic definitions and properties of group cohomology, see for example \cite{Brown} for an introduction to the subject.

In general, group cohomology is defined in terms of a finite group $G$ and a $G$-module $A$, i.e. an abelian group with an 
action of $G$ satisfying certain compatibility properties. For our purposes, it is sufficient to consider $A=U(1)$ with the trivial 
action of $G$. An $n$-cochain $\psi$ (with $n\ge 0$) for $G$ with values in $U(1)$ is simply a function from $G\times\ldots \times G$ 
($n$ times) to $U(1)$; a $0$-cochain is just an element of $U(1)$. On the space $C^n(G,U(1))$ of $n$-cochains one can define 
a \emph{coboundary} operator $\partial_n:C^n(G,U(1))\to C^{n+1}(G,U(1))$ by
\begin{multline} 
\partial_n \psi \,(g_1,\ldots, g_{n+1})=\psi(g_2,\ldots,g_{n+1})\\ \times \prod_{i=1}^n\psi(g_1,\ldots,g_{i-1},g_ig_{i+1},g_{i+2},\ldots, g_{n+1})^{(-1)^i}\psi(g_1,\ldots,g_{n})^{(-1)^{n+1}}\ .
\end{multline} 
The cochains in the kernel of $\partial_n$ are called $n$-cocycles (or co-closed $n$-cochains), while the cochains in the 
image of $\partial_{n-1}$ are called $n$-coboundaries. The coboundary operator satisfies $\partial_{n+1}\circ\partial_n=1$ 
(the trivial element of $U(1))$), so that we can define the cohomology, as usual, as the quotient of the space of cocycles modulo 
coboundaries
\be 
H^n(G,U(1))=\frac{\ker\partial_n}{{\rm Im}\partial_{n-1}}\ ,
\ee 
where $n>0$, and we use the convention that $H^0(G,U(1))=U(1)$, and that every 
$0$-cochain is coclosed, ie. $\partial_0=1$.

In particular, for a $1$-cochain $\gamma$, the condition $\partial_1\gamma=1$ is simply 
$\gamma(g)\gamma(h)=\gamma(gh)$ for all $g,h\in G$, so that $H^1(G,U(1))$ is just the group of homomorphism from 
$G$ to $U(1)$.  A 2-cochain $\beta \, :\, G\times G \, \rightarrow U(1)$ is co-closed (and hence
defines a cocycle) provided it satisfies 
\be 
\beta(g_1, g_2 g_3)\, \beta(g_2, g_3)= \beta(g_1g_2, g_3)\, \beta(g_1, g_2) 
\label{2cocycle}
\ee  
for $g_1, g_2, g_3\in G$. The second cohomology $H^2(G,U(1))$ then consists of the co-closed 2-cochains, 
modulo the ambiguity 
\be 
\beta(g_1, g_2) \rightarrow \beta(g_1, g_2) \frac{\gamma(g_1)\gamma(g_2)}{\gamma(g_1g_2)}\ , 
\label{betagauge}
\ee  
where $\gamma$ is an arbitrary 1-cochain, i.e.\ an arbitrary function 
$\gamma \, :\, G\, \to \, U(1)$. The second cohomology of a group $G$ classifies the inequivalent projective 
representations of $G$, see appendix \ref{app_projreps}. A 3-cochain $\alpha$,
\be 
\alpha\, :\, G\times G\times G\, \rightarrow\, U(1)
\ee   
is closed provided it satisfies
\be 
\alpha(g_1, g_2, g_3)\, \alpha(g_1, g_2 g_3, g_4)\, \alpha(g_2, g_3, g_4)
=\alpha(g_1g_2, g_3, g_4)\, \alpha(g_1, g_2, g_3g_4)\ .
\label{3cocycle}
\ee  
In the cohomology group $H^3(G,U(1))$ the 3-cocycles are then identified modulo 
\be 
\alpha(g_1, g_2,g_3)\rightarrow \alpha(g_1, g_2, g_3) \,
\frac{\beta(g_1, g_2g_3)\beta(g_2, g_3)}{\beta(g_1 g_2, g_3)\beta(g_1, g_2)}\ .
\label{alphagauge}
\ee  
Note that the multiplying factor is trivial if $\beta$ is closed, i.e.\ if it satisfies the 2-cocycle condition 
\eqref{2cocycle}. We shall always work with normalised cocycles, i.e.\ we shall assume that 
\be\label{alphnorm}
\alpha(e,g_1,g_2) = \alpha(g_1,e,g_2) = \alpha(g_1,g_2,e) = 1 
\ee
for all $g_1,g_2\in G$, where $e$ denotes the identity element of $G$.

Given a 3-cocycle $\alpha$, we can define, for any $h\in G$, a 2-cochain $c_h: G\times G \rightarrow U(1)$ via
\be 
c_h(g_1, g_2)= \frac{\alpha(h, g_1, g_2)\, \alpha(g_1, g_2,(g_1g_2)^{-1} h(g_1g_2))}{\alpha(g_1,g_1^{-1} h g_1, g_2)}\ .
\label{ch}
\ee 
It is shown in \cite{Dijkgraaf:1989pz} that $c_h$ defines a 2-cocycle of the centraliser 
$C_G(h)\subset G$ (i.e.\ the subgroup of all elements $g_1,g_2$ which commute with $h$). When $g_1,g_2\in C_G(h)$, we have the simplified expression
\be c_h(g_1, g_2)= \frac{\alpha(h, g_1, g_2)\, \alpha(g_1, g_2,h)}{\alpha(g_1, h, g_2)}\ ,\qquad g_1,g_2\in C_G(h)\ .
\ee

\noindent Under the `gauge transformation' \eqref{alphagauge}, $c_h$ transforms as 
\be 
c_h(g_1, g_2)\rightarrow c_h(g_1, g_2) \frac{\gamma_h(g_1)\gamma_h(g_2)}{\gamma_h(g_1g_2)}\ ,\qquad g_1,g_2\in C_G(h)\ ,
\label{chgauge}
\ee  
where we defined the 1-cochain $\gamma_h$ by
\be 
\gamma_h(g)\equiv \frac{\beta(g,h)}{\beta(h,g)}\ .
\ee  
This is indeed of the form \eqref{betagauge}, and hence, 
for all $h\in G$, $c_h$  defines a map
\be 
c_h\, :\, H^3(G, U(1))\, \rightarrow H^2(C_G(h), U(1))\ .
\ee

\section{Projective representations of finite groups}
\label{app_projreps}
This section follows \cite{Coste:2000tq, Karpilovsky1993}.
A projective representation is a map $\rho: G \rightarrow {\rm End}(V)$ from a group
$G$ to the endomorphism group of some complex vector space $V$ satisfying
\be
\rho(g) \, \rho(h) = \beta(g,h) \, \rho(gh)\ , \quad \beta(g,h) \in U(1)\ .
\ee
It follows from the associativity of the group multiplication that the phase
$\beta(g,h)$ satisfies the 2-cocycle condition
\be
\beta(g,h) \, \beta(gh, k) = \beta(h,k) \, \beta(g, hk)\ .
\ee
There is the freedom of rescaling the maps $\rho(g)$ as 
\be
\rho(g) \mapsto \gamma(g) \rho(g)\ , \quad \gamma(g) \in U(1)\ .
\ee
For the 2-cocylce $\beta$ this leads to the rescaling by a coboundary
\be
\beta(g,h) \rightarrow \beta(g,h)\,  \gamma(gh)\,  \gamma(g)^{-1} \, \gamma(h)^{-1}\ .
\ee In particular, one can always choose a rescaling such that $\rho(e)$ is the identity matrix in ${\rm End}(V)$; with this choice, the cocycle $\beta$ is normalised.
Two projective representations whose 2-cocycles differ only by a coboundary
are called projectively equivalent, since their representation
matrices are related by a rescaling.
Therefore, the inequivalent projective representations of $G$ are captured
by the second group cohomology $H^2(G, U(1))$.

As for genuine linear representations, one may define the character of a projective
representation as the trace
\be
\chi(g) = \Tr_V(\rho(g))\ .
\ee
Due to the modified composition law, projective characters are in general not
class functions, but differ by a phase on different representatives of one
conjugacy class. However, it is always possible to choose a cocycle
representative such that the characters are indeed class functions.

Assuming that $gh=hg$, we have, using the projective composition law,
\be
\chi(g) = \chi(h^{-1} g h) = \frac{\beta(gh, h^{-1})}{\beta(h^{-1}, gh)} \, \chi(g)\ .
\ee
Using the cocycle condition, it is easy to see that this condition on the characters
is equivalent to the obstuction of the first kind (\ref{eqn:obstruction1}).
Also, the obstruction does not depend on the specific cocycle representative $\beta(g,h)$
or the class representative $g$. Classes whose projective characters vanish
are called $\beta$-irregular.
It may be shown that to each cocycle $\beta(g,h)$, there exist exactly
as many linearly inequivalent irreducible projective representations as there are
$\beta$-regular conjugacy classes.

Two irreducible characters belonging to representations with the same 
cocycle $\beta(g,h)$ satisfy an orthogonality relation of the form
\be
\sum_{[g]} |K_g|^{-1} \, \chi(g)^*\, \tilde\chi(g) =
\begin{cases} 1 \quad \chi = \tilde\chi \\ 0\quad \text{otherwise,} \end{cases}
\ee
where the sum is taken over a set of ($\beta$-regular) class representatives of $G$, and
$|K_g|$ is the size of the conjugacy class of $g$.

There is a natural connection between projective representations and central extensions
of a group $G$. We say that $\tilde G$ is a central extention of $G$ if there is an
abelian subgroup $A\subset Z(\tilde G)$ of the center $Z(\tilde{G})$ of $\tilde{G}$ such that the short sequence
\be
 0 \rightarrow A \rightarrow \tilde G \rightarrow G \rightarrow 0
\ee
is exact, or equivalently that $G = \tilde G / A$. The simplest
central extension is of course \linebreak 
$\tilde G = A \times G$, but this is in general 
not the only possibility. It can be shown that
the non-isomorphic central extensions of $G$ by $A$ are also classified
by $H^2(G, A)$. Each set of projectively inequivalent projective representations
of $G$ corresponds to a set of genuine representations of some extension $\tilde G$.

In some sense there exists a maximal central extension $S$ for each $G$, 
the Schur cover or `Darstellungsgruppe'. It is characterised by the property that any 
projective representation of $G$ corresponds to a genuine representation of $S$.
The Schur cover is, in general, not unique, unless $G$ is a perfect group, i.e. it coincides with the commutator subgroup. Otherwise, the Schur covers are related by isoclinism. 

\subsection{Central extension and orbifolds}\label{app:centrorbi}

Let $\C$ be a K3 sigma-model and suppose that its automorphism group contains an element $g$ such that the orbifold of 
$\C$ by $g$, $\hat\C=\C / \langle g \rangle$ is consistent. In this subsection, we will show that all projective representations of $H\equiv C_{M_{24}}(g)$ on the $g^r$-twisted sectors are equivalent to genuine representations of a central extension $\tilde{H}$ of the form described in section \ref{s:relabelling}, eq.~\eqref{centext}.

By \eqref{Tmultip} and \eqref{kappa}, the level-matching 
condition is satisfied if and only if the restriction of the class $[\alpha]\in H^3(M_{24},U(1))$ to $H^3(\langle g\rangle,U(1))$ is trivial.
Each $g^r$-twisted sector carries a, in general projective, 
representation $\rho_{g^r}$ of $H\equiv C_{M_{24}}(g)$, with underlying 2-cocycle $c_{g^r}$ given by \eqref{ch0}. 
The various cocycles $c_{g^r}$ are related to one another by the identity 
\be c_{g^r}(h,k)=\frac{f_{g,r}(h)f_{g,r}(k)}{f_{g,r}(hk)} \, c_g(h,k)^r\ ,\qquad h,k\in C(g)\ ,
\ee where
\be f_{g,r}(h)=\prod_{i=1}^{r-1}c_h(g,g^i)\ .
\ee
Here we have used the definition \eqref{ch0} of $c_g$ and applied repeatedly the condition that the 3-cocycle $\alpha$ is closed.
Thanks to the level-matching condition, we can choose the representative cocycle $\alpha$ to be identically $1$ when restricted to 
$\langle g\rangle$, so that $f_{g,r}(g^i)=1$ for all $i$. Furthermore, by changing $\alpha$ by a coboundary $\partial\beta$ as 
in \eqref{alphagauge}, the functions $f_{g,r}(h)$ transform as
\be
 f_{g,r}(h)\to f_{g,r}(h)\, \frac{\beta(g^r,h)\beta(h,g)^r}{\beta(h,g^r)\beta(g,h)^r}\ .
\ee  
Thus, we can choose $\beta$ in such a way that $f_{g,r}(h)=1$ for all $h\in C(g)$, so that
\be 
c_{g^r}(h,k)= c_g(h,k)^r\ ,\qquad h,k\in C(g)\ .
\ee 
In particular, since $c_e(h,k)=1$, $c_g(h,k)$ is always an $N$-th root of unity, where $N=o(g)$.

\section{Centralisers $C_{M_{24}}(g)$ and (projective) character tables}\label{app:char}

This section describes the projective character tables of the centraliser $H\equiv C_{M_{24}}(g)$ for each conjugacy class of $M_{24}$,  
corresponding to the projective equivalence class determined by the 2-cocycle $c_g$. The procedure to compute these tables is
as follows. 

Let ${\rm Schur}(H)$ denote a Schur cover of $H$, i.e.\ a central extension of $H$ fitting into the exact sequence
\be 
1\to M(H)\to {\rm Schur}(H)\to H\to 1\ ,
\ee 
where $M(H)= H_2(H,\ZZ)=(H^2(H,U(1)))^*$ is the Schur multiplier. The Schur cover ${\rm Schur}(H)$ and its character table 
can be easily computed (we used {\texttt{GAP}} \cite{GAP4} for this). Any irreducible projective representation 
$\rho$ of $H$ is equivalent to a genuine irreducible representation $\tilde\rho$  of ${\rm Schur}(H)$ and vice versa. 
Thus, one can associate a class of $H^2(H,U(1))$ to each irreducible representation of ${\rm Schur}(H)$.  The projective 
character table we are interested in is obtained from the one of ${\rm Schur}(H)$ by keeping only the representations 
associated to the class $[c_g]$, and by choosing a lift in ${\rm Schur}(H)$ of each conjugacy class of 
$H$ (different choices of the lifts lead to projectively equivalent representations).

It only remains to determine which irreducible representations of ${\rm Schur}(H)$ correspond to the class $[c_g]$. 
Consider an irreducible projective representation $\rho$ of $H$ associated with the 2-cocycle $c_g$. For any commuting pair 
$h,k\in H$, $hk=kh$,  we have 
\be 
\rho(h)\rho(k)\rho(h)^{-1}\rho(k)^{-1}=\frac{c_g(h,k)}{c_g(k,h)}\ .
\ee 
(Note that $\frac{c_g(h,k)}{c_g(k,h)}$ only depends on the class $[c_g]$ and not on the representative 
cocycle $c_g$.) Thus, any representation $\sigma$ of ${\rm Schur}(H)$ in the class $[c_g]$ must satisfy
\be\label{commutator} 
\sigma(\tilde{h}\tilde{k}\tilde{h}^{-1}\tilde{k}^{-1})=\frac{c_g(h,k)}{c_g(k,h)}\ ,
\ee  
where $\tilde{h},\tilde{k}$ are some lifts of $h,k$ in ${\rm Schur}(H)$. It is easy to see that the 
commutator $\tilde{h}\tilde{k}\tilde{h}^{-1}\tilde{k}^{-1}$ is an element of $M(H)$, and that it does not depend of the choice 
of the lifts. In fact, the commutators of this form generate $M(H)$, and  if a representation $\sigma$ satisfies 
\eqref{commutator} for a set of generators of $M(H)=(H^2(H,U(1)))^*$, then it must be associated with the class $[c_g]$. 
This condition can be easily tested for each irreducible representation of ${\rm Schur}(H)$, given our explicit knowledge 
of the cocycle $c_g$.

\subsection{Character Tables}

In this section we collect some information about the centralisers $C_{M_{24}}(g)$ of representatives of each $M_{24}$ conjugacy class $[g]$, in particular a (rough) description of the group structure and the group order. Our notation is as follows:
\smallskip

\begin{tabular}{rl}
$G=N.Q$ & group $G$ containing a normal subgroup $N$ and such that $G/N\cong Q$ \\
$N\rtimes Q$ &  (left) semidirect product of $N$ and $Q$ \\
$\ZZ_n$ & cyclic group of order $n$ \\
$S_n$ & group of permutations (symmetric group) of $n$ objects \\
$A_n$ & group of even permutations (alternating group) of $n$ objects \\
$D_n$ & dihedral group of order $n$ \\
$PSL_n(q)$ & projective special linear group over the $n$-dim.\ vector space over \\ 
& the finite field $\FF_q$
\end{tabular}
\smallskip

Furthermore, for each class $[g]$, we specify whether the representation carried by the $g$-twisted sector $\H_g$ is 
(projectively equivalent to) a linear representation, or whether it is genuinely projective; in the latter case we describe 
a minimal central extension with respect to which the representation is linear. Finally, we provide the character table of 
$C_{M_{24}}$ for the relevant projective equivalence class. The names of the conjugacy classes are chosen to correspond to the ATLAS names of $M_{24}$ classes, with additional indices
in case of redundancy. We adopt the following notation for algebraic numbers
$$ \varepsilon_n=e^{\frac{2\pi i}{n}}\ , \qquad\qquad e_p^{\pm}=\frac{1}{2}(1\pm i\sqrt{p})\ .
$$

\bigskip

{\bf Classes 11A, 12A, 12B, 14AB, 15AB, 21AB, 23AB}

\smallskip

When $g$ is in one of these classes, the centralizer $C_{M_{24}}(g)$ is the cyclic group $\langle g\rangle$ and all 
representations in the $g$-twisted sector are (projectively equivalent to) linear representations. Thus, all irreducible 
representations are one dimensional, and a useful choice for the corresponding (projective) characters $\chi_k$, 
$k=1,\ldots,o(g)$ is
\be\label{cyclicchar} \chi_k(g^a)=e^{2\pi i\frac{a}{o(g)}(k-\frac{1}{\ell(g)})}\ ,
\ee $a=0,\ldots,o(g)-1$, where $\ell(g)$ is the length of the shortest cycle of $g$ when considered as a permutation of $24$ objects. Indeed,  the characters \eqref{cyclicchar} correspond to the standard definition  $\rho_g(g^a)=e^{2\pi ia(L_0-\tilde L_0)}$ for the action of $\langle g\rangle$ in the $g$-twisted sector.

\begin{landscape}
{\bf Class 2A}

\medskip

$C_{M_{24}}(2A)\cong \ZZ_2^4.(\ZZ_2^3\rtimes PSL_3(2))$
\hspace{40pt}
Order: $2^{10}\cdot 3\cdot 7=21504$

Genuinely projective rep.
\hspace{40pt}
Central extension: $\ZZ_2.C_{M_{24}}(2A)$

\begin{table}[h]
\scalebox{0.6}{
$
\begin{array}{lrrrrrrrrrrrrrrrrrrrrrrrrrrrrrr}
&21504 & 21504 & 1536 & 1536 & 384 & 512 & 256 & 256 & 64 & 128 & 128 & 64 & 128 & 64 & 32 & 64 & 32 & 32 & 16 & 16 & 16 & 24 & 24 & 12 & 12 & 12 & 14 & 14 & 14 & 14 \\
&\mathrm{1A} & \mathrm{2A_{1}} & \mathrm{2A_{2}} & \mathrm{2A_{3}} & \mathrm{4A_{1}} & \mathrm{2B_{1}} & \mathrm{2A_{4}} & \mathrm{2B_{3}} & \mathrm{4B_{2}} & \mathrm{2A_{5}} 
& \mathrm{2B_{2}} & \mathrm{4A_{2}} & \mathrm{4B_{1}} & \mathrm{4A_{3}} & \mathrm{4C_{1}} & \mathrm{4B_{3}} & \mathrm{4B_{4}} & \mathrm{4A_{4}} & \mathrm{4B_{5}} & \mathrm{4C_{2}}
 & \mathrm{8A_{1}} & \mathrm{3A_{1}} & \mathrm{6A_{1}} & \mathrm{6A_{2}} & \mathrm{6A_{3}} & \mathrm{12A_{1}} & \mathrm{7A_{1}} & \mathrm{14A_{1}} & \mathrm{7B_{1}} & \mathrm{14B_
{1}}\\ \hline 
\chi_{1} & 8 & -8 & 0 & 0 & 0 & 0 & 0 & 0 & 0 & 0 & 0 & 0 & 4 & 0 & 0 & 0 & 0 & 0 & 0 & 0 & 2 & -2 & 2 & 0 & 0 & 0 & -1 & 1 & -1 & 1 \\
\chi_{2} & 8 & 8 & 0 & 0 & 4 & 0 & 0 & 0 & 0 & 0 & 0 & 0 & 0 & 0 & 0 & 0 & 0 & 0 & 0 & 0 & 0 & -2 & -2 & 0 & 0 & -2 & -1 & -1 & -1 & -1 \\
\chi_{3} & 8 & 8 & 0 & 0 & 4 & 0 & 0 & 0 & 0 & 0 & 0 & 0 & 0 & 0 & 0 & 0 & 0 & 0 & 0 & 0 & 0 & 1 & 1 & \sqrt{3} & \sqrt{3} & 1 & -1 & -1 & -1 & -1 \\
\chi_{4} & 8 & 8 & 0 & 0 & 4 & 0 & 0 & 0 & 0 & 0 & 0 & 0 & 0 & 0 & 0 & 0 & 0 & 0 & 0 & 0 & 0 & 1 & 1 & -\sqrt{3} & -\sqrt{3} & 1 & -1 & -1 & -1 & -1 \\
\chi_{5} & 24 & -24 & 0 & 0 & 0 & 0 & 0 & 0 & 0 & 0 & 0 & 0 & -4 & 0 & 0 & 0 & 0 & 0 & 0 & 0 & 2 & 0 & 0 & 0 & 0 & 0 & e^-_7 & -e^-_7 & e^+_7 & -e^+_7 \\
\chi_{6} & 24 & -24 & 0 & 0 & 0 & 0 & 0 & 0 & 0 & 0 & 0 & 0 & -4 & 0 & 0 & 0 & 0 & 0 & 0 & 0 & 2 & 0 & 0 & 0 & 0 & 0 & e^+_7 & -e^+_7 & e^-_7 & -e^-_7 \\
\end{array}
$}\\[5pt]
\caption{Projective character table for the centralizer of class 2A.}
\end{table}

\newpage
{\bf Class 2B}

\medskip

$C_{M_{24}}(2B)\cong \ZZ_2^6\rtimes S_5$ 
\hspace{40pt}
Order: $2^{9}\cdot 3\cdot 5=7680$

Genuinely projective rep.
\hspace{40pt}
Central extension: $\ZZ_2.C_{M_{24}}(2B)\cong (\ZZ_2.\ZZ_2^6)\rtimes S_5$

\begin{table}[h]
\scalebox{0.6}{
$
\begin{array}{lrrrrrrrrrrrrrrrrrrrrrrrrrrrrrr}
 & 7680& 512& 256& 3840& 512& 7680& 96& 32& 96& 32& 24& 24& 12& 12& 12& 128& 64& 32& 32& 64& 64& 128& 20& 20& 20& 20& 16& 16& 16& 16 \\
% & \mathrm{1A} & \mathrm{2A_1} & \mathrm{2A_2} & \mathrm{2B_2} & \mathrm{2B_3} & \mathrm{2B_1} & \mathrm{2B_4} & \mathrm{4A_2} & \mathrm{4C_1} & \mathrm{4C_2} & \mathrm{3B_1} & \mathrm{6B_1} & \mathrm{6B_2} & \mathrm{6B_3} & \mathrm{12B_1} & \mathrm{2A_3} & \mathrm{4A_1} & \mathrm{4B_1} & \mathrm{4A_2} & \mathrm{2B_5} & \mathrm{4B_3} & \mathrm{2B_5} & \mathrm{5A_1} & \mathrm{10A_2} & \mathrm{10A_1} & \mathrm{10A_3} & \mathrm{4B_2} & \mathrm{8A_1} & \mathrm{4C_3} & \mathrm{8A_2} \\
&\mathrm{ 1A } & \mathrm{  2B_{1} } & \mathrm{  2A_{1} } & \mathrm{  2B_{2} } & \mathrm{  2A_{2} } & \mathrm{  2B_{3} } & \mathrm{  2B_{4} } & \mathrm{  4A_{1} } & \mathrm{  4C_{1} } & \mathrm{  4C_{2} } & \mathrm{  3B_{1} } & \mathrm{  6B_{1} } & \mathrm{  6B_{2} } & \mathrm{  6B_{3} } & \mathrm{  12B_{1} } & \mathrm{  2B_{5} } & \mathrm{  4A_{2} } & \mathrm{  4B_{1} } & \mathrm{  4A_{3} } & \mathrm{  4B_{2} } & \mathrm{  2B_{6} } & \mathrm{  2A_{3} } & \mathrm{  5A_{1} } & \mathrm{  10A_{1} } & \mathrm{  10A_{2} } & \mathrm{  10A_{3} } & \mathrm{  4C_{3} } & \mathrm{  8A_{1} } & \mathrm{  4B_{3} } & \mathrm{  8A_{2} } \\
\hline

\chi_{1} & 4 & 0 & 0 & 4 & 0 & -4 & 2 & 0 & -2 & 0 & -1 & 1 & -1 & -1 & 1 & 0 & 2 & 0 & 0 & -2 & 0 & 0 & -1 & 1 & 1 & -1 & 0 & 2e^-_1 & 0 & 2e^+_1 \\
\chi_{2} & 4 & 0 & 0 & -4 & 0 & -4 & 2 & 0 & 2 & 0 & -1 & 1 & 1 & -1 & -1 & 0 & 2 & 0 & 0 & 2 & 0 & 0 & -1 & -1 & 1 & 1 & 0 & -2e^+_1 & 0 & -2e^-_1 \\
\chi_{3} & 4 & 0 & 0 & 4 & 0 & -4 & -2 & 0 & 2 & 0 & -1 & 1 & -1 & 1 & -1 & 0 & 2 & 0 & 0 & -2 & 0 & 0 & -1 & 1 & 1 & -1 & 0 & -2e^-_1 & 0 & -2e^+_1 \\
\chi_{4} & 4 & 0 & 0 & -4 & 0 & -4 & -2 & 0 & -2 & 0 & -1 & 1 & 1 & 1 & 1 & 0 & 2 & 0 & 0 & 2 & 0 & 0 & -1 & -1 & 1 & 1 & 0 & 2e^+_1 & 0 & 2e^-_1 \\
\end{array}
$}\\[5pt]
\caption{Projective character table for the centralizer of class 2B.}
\end{table}

\newpage

{\bf Class 3A}

\medskip

$C_{M_{24}}(3A)\cong \ZZ_3.A_6$
\hspace{40pt}
Order: $2^{3}\cdot 3^3\cdot 5=1080$
\hspace{40pt}
Linear rep.

\begin{table}[h]
\scalebox{0.7}{
$\begin{array}{lrrrrrrrrrrrrrrrrr}
&1080 & 1080 & 1080 & 24 & 24 & 24 & 9 & 9 & 12 & 12 & 12 & 15 & 15 & 15 & 15 & 15 & 15 \\ 
&\mathrm{1A} & \mathrm{3A_{1}} & \mathrm{3A_{2}} & \mathrm{6A_{1}} & \mathrm{6A_{2}} & \mathrm{2A_{1}} & \mathrm{3A_{3}} & \mathrm{3B_{1}} & \mathrm{12A_{1}} & \mathrm{12A_{2}} & \mathrm{4A_{1}} & \mathrm{15A_{1}} & \mathrm{5A_{1}} & \mathrm{15B_{1}} & \mathrm{5A_{2}} & \mathrm{15A_{2}} & \mathrm{15B_{2}}\\ \hline 
\chi_{1} & 1 & 1 & 1 & 1 & 1 & 1 & 1 & 1 & 1 & 1 & 1 & 1 & 1 & 1 & 1 & 1 & 1\\
\chi_{2} & 3 & -3e_3^+ & -3e_3^- & e_3^+ & e_3^- & -1 & 0 & 0 & -e_3^+ & -e_3^- & 1 & \alpha & e_{-5}^+ & \bar{\alpha} & e_{-5}^- & \beta & \bar{\beta}\\
\chi_{3} & 3 & -3e_3^+ & -3e_3^- & e_3^+ & e_3^- & -1 & 0 & 0 & -e_3^+ & -e_3^- & 1 & \bar{\beta} & e_{-5}^- & \beta & e_{-5}^+ & \bar{\alpha} & \alpha\\
\chi_{4} & 3 & -3e_3^- & -3e_3^+ & e_3^- & e_3^+ & -1 & 0 & 0 & -e_3^- & -e_3^+ & 1 & \beta & e_{-5}^- & \bar{\beta} & e_{-5}^+ & \alpha & \bar{\alpha}\\
\chi_{5} & 3 & -3e_3^- & -3e_3^+ & e_3^- & e_3^+ & -1 & 0 & 0 & -e_3^- & -e_3^+ & 1 & \bar{\alpha} & e_{-5}^+ & \alpha & e_{-5}^- & \bar{\beta} & \beta\\
\chi_{6} & 5 & 5 & 5 & 1 & 1 & 1 & -1 & 2 & -1 & -1 & -1 & 0 & 0 & 0 & 0 & 0 & 0\\
\end{array}$
}\\[5pt]
\caption{Character table for the centralizer of class 3A. Here,
$\alpha = -e^{2\pi i \frac{2}{15}}-e^{2\pi i \frac{8}{15}}$ and
$\beta = -e^{2\pi i \frac{1}{15}}-e^{2\pi i \frac{4}{15}}$.}
\end{table}

\newpage
{\bf Class 3B}

\medskip

$C_{M_{24}}(3B)\cong \ZZ_3\times PSL_3(2)$
\hspace{40pt}
Order: $2^3\cdot 3^2\cdot 7=504$
\hspace{40pt}
Linear rep.

\begin{table}[h]
\scalebox{0.7}{
$\begin{array}{lrrrrrrrrrrrrrrrrrr}
&504 & 504 & 504 & 9 & 9 & 9 & 21 & 21 & 21 & 21 & 21 & 21 & 24 & 24 & 24 & 12 & 12 & 12 \\ 
&\mathrm{1A_{1}} & \mathrm{3B_{1}} & \mathrm{3B_{2}} & \mathrm{3A_{1}} & \mathrm{3B_{3}} & \mathrm{3B_{4}} & \mathrm{7A_{1}} & \mathrm{21A_{1}} & \mathrm{21A_{2}} & \mathrm{7B_{1}} & \mathrm{21B_{1}} & \mathrm{21B_{2}} & \mathrm{2B_{1}} & \mathrm{6B_{1}} & \mathrm{6B_{2}} & \mathrm{4C_{1}} & \mathrm{12B_{1}} & \mathrm{12B_{2}}\\ \hline 
\chi_{1} & 1 & 1 & 1 & 1 & 1 & 1 & 1 & 1 & 1 & 1 & 1 & 1 & 1 & 1 & 1 & 1 & 1 & 1\\
\chi_{2} & 1 & -e_3^+ & -e_3^- & 1 & -e_3^+ & -e_3^- & 1 & -e_3^+ & -e_3^- & 1 & -e_3^+ & -e_3^- & 1 & -e_3^+ & -e_3^- & 1 & -e_3^+ & -e_3^-\\
\chi_{3} & 1 & -e_3^- & -e_3^+ & 1 & -e_3^- & -e_3^+ & 1 & -e_3^- & -e_3^+ & 1 & -e_3^- & -e_3^+ & 1 & -e_3^- & -e_3^+ & 1 & -e_3^- & -e_3^+\\
\chi_{4} & 3 & 3 & 3 & 0 & 0 & 0 & -e_7^+ & -e_7^+ & -e_7^+ & -e_7^- & -e_7^- & -e_7^- & -1 & -1 & -1 & 1 & 1 & 1\\
\chi_{5} & 3 & 3 & 3 & 0 & 0 & 0 & -e_7^- & -e_7^- & -e_7^- & -e_7^+ & -e_7^+ & -e_7^+ & -1 & -1 & -1 & 1 & 1 & 1\\
\chi_{6} & 3 & -3e_3^+ & -3e_3^- & 0 & 0 & 0 & -e_7^+ & \alpha & \beta & -e_7^- & \bar{\beta} & \bar{\alpha} & -1 & e_3^+ & e_3^- & 1 & -e_3^+ & -e_3^-\\
\end{array}$
}\\[5pt]
\caption{Character table for the centralizer of class 3B. Here,
$\alpha = e^{2\pi i \frac{2}{21}}+e^{2\pi i \frac{8}{21}}+e^{2\pi i \frac{11}{21}}$ and
$\beta = e^{2\pi i \frac{1}{21}}+e^{2\pi i \frac{4}{21}}+e^{2\pi i \frac{16}{21}}$. }
\end{table}

\newpage

{\bf Class 4A}

\medskip

$C_{M_{24}}(4A)$
\hspace{40pt}
Order: $2^{7}\cdot 3=384$

Genuinely projective rep.
\hspace{40pt}
Central extension: $\ZZ_2.C_{M_{24}}(4A)$

\begin{table}[h]
\scalebox{0.7}{
$
\begin{array}{lrrrrrrrrrrrrrrrrrrrrrrrrrr}
&384& 32& 64& 16& 12& 64& 64& 32& 16& 16& 12& 32& 64& 16& 32& 12& 16& 12& 32& 16& 64& 32& 384& 64& 384& 384\\
&\mathrm{ 1A_{1} } & \mathrm{  2A_{1} } & \mathrm{  2A_{2} } & \mathrm{  4B_{1} } & \mathrm{  3A_{1} } & \mathrm{  2B_{3} } & \mathrm{  2A_{3} } & \mathrm{  2B_{2} } & \mathrm{  4C_{1} } & \mathrm{  4A_{1} } & \mathrm{  6A_{1} } & \mathrm{  4A_{2} } & \mathrm{  4A_{3} } & \mathrm{  4B_{2} } & \mathrm{ 2B_{1} } & \mathrm{  12A_{1} } & \mathrm{  4C_{2} } & \mathrm{  12A_{2} } & \mathrm{  4B_{3} } & \mathrm{  4A_{4} } & \mathrm{  4A_{5} } & \mathrm{  4B_{4} } & \mathrm{  4A_{6} } & \mathrm{  4A_{7} } & \mathrm{  4A_{8} } & \mathrm{  2A_{4} }
\\\hline
\chi_{1}& 2 & 0 & 0 & \sqrt{2} & -1 & -2 & 0 & 0 & 0 & i \sqrt{2} & 1 & 0 & 0 & \sqrt{2} & 0 & i & 0 & -i & 0 & i \sqrt{2} & -2 i & 0 & -2 i & 0 & 2 i & -2 \\
\chi_{2}& 2 & 0 & 0 & -\sqrt{2} & -1 & -2 & 0 & 0 & 0 & -i \sqrt{2} & 1 & 0 & 0 & -\sqrt{2} & 0 & i & 0 & -i & 0 & -i \sqrt{2} & -2 i & 0 & -2 i & 0 & 2 i & -2 \\
\chi_{3}& 2 & 0 & 0 & \sqrt{2} & -1 & -2 & 0 & 0 & 0 & i \sqrt{2} & 1 & 0 & 0 & -\sqrt{2} & 0 & -i & 0 & i & 0 & -i \sqrt{2} & 2 i & 0 & 2 i & 0 & -2 i & -2 \\
\chi_{4}& 2 & 0 & 0 & -\sqrt{2} & -1 & -2 & 0 & 0 & 0 & -i \sqrt{2} & 1 & 0 & 0 & \sqrt{2} & 0 & -i & 0 & i & 0 & i \sqrt{2} & 2 i & 0 & 2 i & 0 & -2 i & -2 \\
\chi_{5}& 4 & 0 & 0 & 0 & 1 & -4 & 0 & 0 & 0 & 0 & -1 & 0 & 0 & 0 & 0 & -i & 0 & i & 0 & 0 & -4 i & 0 & -4 i & 0 & 4 i & -4 \\
\chi_{6}& 4 & 0 & 0 & 0 & 1 & -4 & 0 & 0 & 0 & 0 & -1 & 0 & 0 & 0 & 0 & i & 0 & -i & 0 & 0 & 4 i & 0 & 4 i & 0 & -4 i & -4 \\
\end{array}
$}\\[5pt]
\caption{Projective character table for the centralizer of class 4A.}
\end{table}

\newpage

{\bf Class 4B}

\medskip

$C_{M_{24}}(4B)\cong ((\ZZ_4\times\ZZ_4)\rtimes \ZZ_4)\rtimes \ZZ_2$
\hspace{40pt}
Order: $128$

Genuinely projective rep.
\hspace{40pt}
Central extension: $\ZZ_2.C_{M_{24}}(4B)$

\begin{table}[h]
\scalebox{0.7}{
$\begin{array}{lrrrrrrrrrrrrrrrrrrrrrrrrrr}
&128& 16& 128& 128& 64& 64& 128& 16& 16& 16& 16& 32& 16& 16& 32& 32& 32& 16& 16& 16& 16& 64& 16& 64& 64& 64\\
&\mathrm{1A} & \mathrm{ 8A_{1}} & \mathrm{ 4B_{2}} & \mathrm{ 4B_{6}} & \mathrm{ 2B_{3}} & \mathrm{ 4B_{1}} & \mathrm{ 2A_{1}} & \mathrm{ 2B_{1}} & \mathrm{ 4B_{3}} & \mathrm{ 2A_{3}} & \mathrm{ 4B_{5}} & \mathrm{ 2B_{2}} & \mathrm{ 8A_{2}} & \mathrm{ 8A_{3}} & \mathrm{4A_{2}} & \mathrm{ 4A_{1}} & \mathrm{ 2A_{4}} & \mathrm{ 4B_{7}} & \mathrm{ 4B_{4}} & \mathrm{ 4A_{3}} & \mathrm{ 4A_{4}} & \mathrm{ 2A_{2}} & \mathrm{ 8A_{4}} & \mathrm{ 4B_{9}} & \mathrm{ 2A_{5}} & \mathrm{ 4B_{8} }\\\hline
\chi_{1} &2 & -2 & -2 & -2 & 2 & 2 & 2 & 0 & 0 & 0 & 0 & 0 & 0 & 0 & 0 & 0 & 0 & \sqrt{2} & \sqrt{2} & -\sqrt{2} & -\sqrt{2} & 0 & 0 & 0 & 0 & 0 \\
\chi_{2} &2 & -2 & -2 & -2 & 2 & 2 & 2 & 0 & 0 & 0 & 0 & 0 & 0 & 0 & 0 & 0 & 0 & -\sqrt{2} & -\sqrt{2} & \sqrt{2} & \sqrt{2} & 0 & 0 & 0 & 0 & 0 \\
\chi_{3} &2 & 0 & -2 & -2 & -2 & -2 & 2 & 0 & 0 & 0 & 0 & 0 & 0 & 0 & 0 & 0 & 0 & \sqrt{2} & -\sqrt{2} & -\sqrt{2} & \sqrt{2} & 0 & 2 & 0 & 0 & 0 \\
\chi_{4} &2 & 0 & -2 & -2 & -2 & -2 & 2 & 0 & 0 & 0 & 0 & 0 & 0 & 0 & 0 & 0 & 0 & -\sqrt{2} & \sqrt{2} & \sqrt{2} & -\sqrt{2} & 0 & 2 & 0 & 0 & 0 \\
\chi_{5} &2 & 0 & -2 & -2 & -2 & -2 & 2 & 0 & 0 & 0 & 0 & 0 & 0 & 0 & 0 & 0 & 0 & \sqrt{2} & \sqrt{2} & \sqrt{2} & \sqrt{2} & 0 & -2 & 0 & 0 & 0 \\
\chi_{6} &2 & 0 & -2 & -2 & -2 & -2 & 2 & 0 & 0 & 0 & 0 & 0 & 0 & 0 & 0 & 0 & 0 & -\sqrt{2} & -\sqrt{2} & -\sqrt{2} & -\sqrt{2} & 0 & -2 & 0 & 0 & 0 \\
\end{array}$}\\[5pt]
\caption{Projective character table for the centralizer of class 4B}
\end{table}

\newpage

{\bf Class 4C}

\medskip

$C_{M_{24}}(4C)\cong \ZZ_4 \times S_4$
\hspace{40pt}
Order: $2^{5}\cdot 3=96$

Genuinely projective rep.
\hspace{40pt}
Central extension: $\ZZ_2.C_{M_{24}}(4C)\cong \ZZ_4\times\ZZ_2.S_4$

\begin{table}[h]\scalebox{0.8}{
$
\begin{array}{lrrrrrrrrrrrrrrrrrrrr}
&96& 16& 12& 16& 32& 96& 16& 12& 16& 32& 96& 16& 12& 16& 32& 96& 16& 12& 16& 32\\
&\mathrm{1A} & \mathrm{ 2A_{1}} & \mathrm{ 3B_{1}} & \mathrm{ 4A_{1}} & \mathrm{ 2A_{2}} & \mathrm{ 4C_{1}} & \mathrm{ 4C_{3}} & \mathrm{ 12B_{1}} & \mathrm{ 4C_{7}} & \mathrm{ 4C_{4}} & \mathrm{ 4C_{2}} & \mathrm{ 4C_{5}} & \mathrm{ 12B_{2}} & \mathrm{4C_{8}} & \mathrm{ 4C_{6}} & \mathrm{ 2B_{1}} & \mathrm{ 2B_{2}} & \mathrm{ 6B_{1}} & \mathrm{ 4A_{2}} & \mathrm{ 2B_{3}}\\\hline
\chi_{1} &2 & 0 & -1 & \sqrt{2} & 0 & 2 & 0 & -1 & i \sqrt{2} & 0 & 2 & 0 & -1 & i \sqrt{2} & 0 & 2 & 0 & -1 & \sqrt{2} & 0 \\
\chi_{2} &2 & 0 & -1 & -\sqrt{2} & 0 & 2 & 0 & -1 & -i \sqrt{2} & 0 & 2 & 0 & -1 & -i \sqrt{2} & 0 & 2 & 0 & -1 & -\sqrt{2} & 0 \\
\chi_{3} &2 & 0 & -1 & \sqrt{2} & 0 & -2 & 0 & 1 & -i \sqrt{2} & 0 & -2 & 0 & 1 & -i \sqrt{2} & 0 & 2 & 0 & -1 & \sqrt{2} & 0 \\
\chi_{4} &2 & 0 & -1 & -\sqrt{2} & 0 & -2 & 0 & 1 & i \sqrt{2} & 0 & -2 & 0 & 1 & i \sqrt{2} & 0 & 2 & 0 & -1 & -\sqrt{2} & 0 \\
\chi_{5} &2 & 0 & -1 & \sqrt{2} & 0 & -2 i & 0 & i & \sqrt{2} & 0 & 2 i & 0 & -i & -\sqrt{2} & 0 & -2 & 0 & 1 & -\sqrt{2} & 0 \\
\chi_{6} &2 & 0 & -1 & -\sqrt{2} & 0 & -2 i & 0 & i & -\sqrt{2} & 0 & 2 i & 0 & -i & \sqrt{2} & 0 & -2 & 0 & 1 & \sqrt{2} & 0 \\
\chi_{7} &2 & 0 & -1 & \sqrt{2} & 0 & 2 i & 0 & -i & -\sqrt{2} & 0 & -2 i & 0 & i & \sqrt{2} & 0 & -2 & 0 & 1 & -\sqrt{2} & 0 \\
\chi_{8} &2 & 0 & -1 & -\sqrt{2} & 0 & 2 i & 0 & -i & \sqrt{2} & 0 & -2 i & 0 & i & -\sqrt{2} & 0 & -2 & 0 & 1 & \sqrt{2} & 0 \\
\chi_{9} &4 & 0 & 1 & 0 & 0 & 4 & 0 & 1 & 0 & 0 & 4 & 0 & 1 & 0 & 0 & 4 & 0 & 1 & 0 & 0 \\
\chi_{10} &4 & 0 & 1 & 0 & 0 & -4 & 0 & -1 & 0 & 0 & -4 & 0 & -1 & 0 & 0 & 4 & 0 & 1 & 0 & 0 \\
\chi_{11} &4 & 0 & 1 & 0 & 0 & -4 i & 0 & -i & 0 & 0 & 4 i & 0 & i & 0 & 0 & -4 & 0 & -1 & 0 & 0 \\
\chi_{12} &4 & 0 & 1 & 0 & 0 & 4 i & 0 & i & 0 & 0 & -4 i & 0 & -i & 0 & 0 & -4 & 0 & -1 & 0 & 0 \\
\end{array}
$}\\[5pt]
\caption{Projective character table for the centralizer of class 4C.}
\end{table}

\newpage

{\bf Class 5A}

\medskip

$C_{M_{24}}(5A)\cong \ZZ_5\times A_4$
\hspace{40pt}
Order: $2^2\cdot 3\cdot 5 =60$
\hspace{40pt}
Linear rep.

\begin{table}[h]
\scalebox{0.8}{
$\begin{array}{lrrrrrrrrrrrrrrrrrrrr}
&60 & 15 & 15 & 20 & 15 & 15 & 15 & 15 & 20 & 20 & 15 & 15 & 60 & 20 & 20 & 15 & 15 & 60 & 60 & 60 \\ 
&\mathrm{1A} & \mathrm{3A_{1}} & \mathrm{3A_{2}} & \mathrm{10A_{1}} & \mathrm{15A_{1}} & \mathrm{15B_{1}} & \mathrm{15A_{2}} & \mathrm{15B_{2}} & \mathrm{10A_{2}} & \mathrm{10A_{3}} & \mathrm{15B_{3}} & \mathrm{15A_{3}} & \mathrm{5A_{1}} & \mathrm{2B_{1}} & \mathrm{10A_{4}} & \mathrm{15A_{4}} & \mathrm{15B_{4}} & \mathrm{5A_{2}} & \mathrm{5A_{3}} & \mathrm{5A_{4}}\\ \hline 
\chi_{1} & 1 & 1 & 1 & 1 & 1 & 1 & 1 & 1 & 1 & 1 & 1 & 1 & 1 & 1 & 1 & 1 & 1 & 1 & 1 & 1\\
\chi_{2} & 1 & 1 & 1 & \varepsilon_5^4 & \varepsilon_5^4 & \varepsilon_5^4 & \varepsilon_5^2 & \varepsilon_5^2 & \varepsilon_5^2 & \varepsilon_5 & \varepsilon_5 & \varepsilon_5 & \varepsilon_5^2 & 1 & \varepsilon_5^3 & \varepsilon_5^3 & \varepsilon_5^3 & \varepsilon_5^4 & \varepsilon_5^3 & \varepsilon_5\\
\chi_{3} & 1 & 1 & 1 & \varepsilon_5^3 & \varepsilon_5^3 & \varepsilon_5^3 & \varepsilon_5^4 & \varepsilon_5^4 & \varepsilon_5^4 & \varepsilon_5^2 & \varepsilon_5^2 & \varepsilon_5^2 & \varepsilon_5^4 & 1 & \varepsilon_5 & \varepsilon_5 & \varepsilon_5 & \varepsilon_5^3 & \varepsilon_5 & \varepsilon_5^2\\
\chi_{4} & 1 & 1 & 1 & \varepsilon_5^2 & \varepsilon_5^2 & \varepsilon_5^2 & \varepsilon_5 & \varepsilon_5 & \varepsilon_5 & \varepsilon_5^3 & \varepsilon_5^3 & \varepsilon_5^3 & \varepsilon_5 & 1 & \varepsilon_5^4 & \varepsilon_5^4 & \varepsilon_5^4 & \varepsilon_5^2 & \varepsilon_5^4 & \varepsilon_5^3\\
\chi_{5} & 1 & 1 & 1 & \varepsilon_5 & \varepsilon_5 & \varepsilon_5 & \varepsilon_5^3 & \varepsilon_5^3 & \varepsilon_5^3 & \varepsilon_5^4 & \varepsilon_5^4 & \varepsilon_5^4 & \varepsilon_5^3 & 1 & \varepsilon_5^2 & \varepsilon_5^2 & \varepsilon_5^2 & \varepsilon_5 & \varepsilon_5^2 & \varepsilon_5^4\\
\chi_{6} & 1 & -e_3^+ & -e_3^- & 1 & -e_3^+ & -e_3^- & -e_3^- & -e_3^+ & 1 & 1 & -e_3^- & -e_3^+ & 1 & 1 & 1 & -e_3^- & -e_3^+ & 1 & 1 & 1\\
\end{array}$
}\\[5pt]
\caption{Character table for the centralizer of class 5A.}
\end{table}

\newpage

{\bf Classes 6A and 6B}

\medskip

$C_{M_{24}}(6A)\cong C_{M_{24}}(6B)\cong \ZZ_6.\ZZ_2^2\cong \ZZ_3\times D_8$
\hspace{40pt}
Order: $24$
\hspace{40pt}
Linear rep.

\begin{table}[h]\scalebox{0.9}{
$\begin{array}{lrrrrrrrrrrrrrrr}
&24 & 12 & 12 & 24 & 24 & 12 & 12 & 12 & 12 & 12 & 12 & 12 & 24 & 24 & 24 \\ 
C_{M_{24}}(6A)
&\mathrm{1A} & \mathrm{2A_{1}} & \mathrm{6A_{1}} & \mathrm{3A_{1}} & \mathrm{3A_{2}} & \mathrm{6A_{2}} & \mathrm{2A_{2}} & \mathrm{4A_{1}} & \mathrm{12A_{1}} & \mathrm{6A_{3}} & \mathrm{12A_{2}} & \mathrm{6A_{4}} & \mathrm{6A_{5}} & \mathrm{2A_{3}} & \mathrm{6A_{6}}\\ 
C_{M_{24}}(6B)
 &\mathrm{1A } & \mathrm{  2B_{1} } & \mathrm{  6B_{2} } & \mathrm{  3B_{2} } & \mathrm{  3B_{1} } & \mathrm{  6B_{5} } & \mathrm{  2B_{2} } & \mathrm{  4C_{1} } & \mathrm{  12B_{1} } & \mathrm{  6B_{3} } & \mathrm{  12B_{2} } & \mathrm{  6B_{1} } & \mathrm{  6B_{4} } & \mathrm{  2B_{3} } & \mathrm{  6B_{6} } \\\hline 
\chi_{1} & 1 & 1 & 1 & 1 & 1 & 1 & 1 & 1 & 1 & 1 & 1 & 1 & 1 & 1 & 1\\
\chi_{2} & 1 & -1 & -1 & 1 & 1 & -1 & -1 & 1 & 1 & -1 & 1 & -1 & 1 & 1 & 1\\
\chi_{3} & 1 & -1 & -1 & 1 & 1 & -1 & 1 & -1 & -1 & 1 & -1 & 1 & 1 & 1 & 1\\
\chi_{4} & 1 & 1 & 1 & 1 & 1 & 1 & -1 & -1 & -1 & -1 & -1 & -1 & 1 & 1 & 1\\
\chi_{5} & 1 & -1 & e_3^- & -e_3^- & -e_3^+ & e_3^+ & -1 & 1 & -e_3^+ & e_3^+ & -e_3^- & e_3^- & -e_3^+ & 1 & -e_3^-\\
\end{array}$}\\[5pt]
\caption{Character table for the centralizer of classes 6AB. NOTE: There is an accidental isomorphism $C_{M_{24}}(6A)\cong C_{M_{24}}(6B)$, therefore the character table is given only once. However, following our conventions, we give different names to the conjugacy classes depending on whether the group is $C_{M_{24}}(6A)$ or $C_{M_{24}}(6B)$.
}
\end{table}

\newpage

{\bf Classes 7A and 7B}

\medskip

$C_{M_{24}}(7A)\cong C_{M_{24}}(7B)\cong S_3\times \ZZ_7$
\hspace{40pt}
Order: $2\cdot 3\cdot 7=42$
\hspace{40pt}
Linear rep.

\bigskip

\begin{table}[h]
\scalebox{0.7}{
$\begin{array}{lrrrrrrrrrrrrrrrrrrrrr}
&42 & 14 & 14 & 21 & 42 & 14 & 21 & 14 & 21 & 21 & 14 & 14 & 42 & 42 & 42 & 42 & 14 & 21 & 21 & 42 & 21 \\ 
C_{M_{24}}(\mathrm{7A}) &\mathrm{1A} & \mathrm{2A_{1}} & \mathrm{14B_{1}} & \mathrm{21B_{1}} & \mathrm{7B_{1}} & \mathrm{14B_{2}} & \mathrm{21B_{2}} & \mathrm{14A_{1}} & \mathrm{21A_{1}} & \mathrm{21A_{2}} & \mathrm{14A_{2}} & \mathrm{14A_{3}} & \mathrm{7A_{1}} & \mathrm{7A_{2}} & \mathrm{7A_{3}} & \mathrm{7B_{2}} & \mathrm{14B_{3}} & \mathrm{3B_{1}} & \mathrm{21B_{3}} & \mathrm{7B_{3}} & \mathrm{21A_{3}}\\ 
C_{M_{24}}(\mathrm{7B}) &\mathrm{1A} & \mathrm{2A_{1}} & \mathrm{14A_{1}} & \mathrm{21A_{1}} & \mathrm{7A_{1}} & \mathrm{14A_{2}} & \mathrm{21A_{2}} & \mathrm{14B_{1}} & \mathrm{21B_{1}} & \mathrm{21B_{2}} & \mathrm{14B_{2}} & \mathrm{14B_{3}} & \mathrm{7B_{1}} & \mathrm{7B_{2}} & \mathrm{7B_{3}} & \mathrm{7A_{2}} & \mathrm{14A_{3}} & \mathrm{3B_{1}} & \mathrm{21A_{3}} & \mathrm{7A_{3}} & \mathrm{21B_{3}}\\
\hline 
\chi_{1} & 1 & 1 & 1 & 1 & 1 & 1 & 1 & 1 & 1 & 1 & 1 & 1 & 1 & 1 & 1 & 1 & 1 & 1 & 1 & 1 & 1\\
\chi_{2} & 1 & -1 & -1 & 1 & 1 & -1 & 1 & -1 & 1 & 1 & -1 & -1 & 1 & 1 & 1 & 1 & -1 & 1 & 1 & 1 & 1\\
\chi_{3} & 1 & -1 & -\varepsilon_7 & \varepsilon_7 & \varepsilon_7 & -\varepsilon_7^4 & \varepsilon_7^4 & -\varepsilon_7^3 & \varepsilon_7^3 & \varepsilon_7^6 & -\varepsilon_7^6 & -\varepsilon_7^5 & \varepsilon_7^5 & \varepsilon_7^3 & \varepsilon_7^6 & \varepsilon_7^2 & -\varepsilon_7^2 & 1 & \varepsilon_7^2 & \varepsilon_7^4 & \varepsilon_7^5\\
\chi_{4} & 1 & -1 & -\varepsilon_7^2 & \varepsilon_7^2 & \varepsilon_7^2 & -\varepsilon_7 & \varepsilon_7 & -\varepsilon_7^6 & \varepsilon_7^6 & \varepsilon_7^5 & -\varepsilon_7^5 & -\varepsilon_7^3 & \varepsilon_7^3 & \varepsilon_7^6 & \varepsilon_7^5 & \varepsilon_7^4 & -\varepsilon_7^4 & 1 & \varepsilon_7^4 & \varepsilon_7 & \varepsilon_7^3\\
\chi_{5} & 1 & -1 & -\varepsilon_7^3 & \varepsilon_7^3 & \varepsilon_7^3 & -\varepsilon_7^5 & \varepsilon_7^5 & -\varepsilon_7^2 & \varepsilon_7^2 & \varepsilon_7^4 & -\varepsilon_7^4 & -\varepsilon_7 & \varepsilon_7 & \varepsilon_7^2 & \varepsilon_7^4 & \varepsilon_7^6 & -\varepsilon_7^6 & 1 & \varepsilon_7^6 & \varepsilon_7^5 & \varepsilon_7\\
\chi_{6} & 1 & -1 & -\varepsilon_7^4 & \varepsilon_7^4 & \varepsilon_7^4 & -\varepsilon_7^2 & \varepsilon_7^2 & -\varepsilon_7^5 & \varepsilon_7^5 & \varepsilon_7^3 & -\varepsilon_7^3 & -\varepsilon_7^6 & \varepsilon_7^6 & \varepsilon_7^5 & \varepsilon_7^3 & \varepsilon_7 & -\varepsilon_7 & 1 & \varepsilon_7 & \varepsilon_7^2 & \varepsilon_7^6\\
\chi_{7} & 1 & -1 & -\varepsilon_7^5 & \varepsilon_7^5 & \varepsilon_7^5 & -\varepsilon_7^6 & \varepsilon_7^6 & -\varepsilon_7 & \varepsilon_7 & \varepsilon_7^2 & -\varepsilon_7^2 & -\varepsilon_7^4 & \varepsilon_7^4 & \varepsilon_7 & \varepsilon_7^2 & \varepsilon_7^3 & -\varepsilon_7^3 & 1 & \varepsilon_7^3 & \varepsilon_7^6 & \varepsilon_7^4\\
\chi_{8} & 1 & -1 & -\varepsilon_7^6 & \varepsilon_7^6 & \varepsilon_7^6 & -\varepsilon_7^3 & \varepsilon_7^3 & -\varepsilon_7^4 & \varepsilon_7^4 & \varepsilon_7 & -\varepsilon_7 & -\varepsilon_7^2 & \varepsilon_7^2 & \varepsilon_7^4 & \varepsilon_7 & \varepsilon_7^5 & -\varepsilon_7^5 & 1 & \varepsilon_7^5 & \varepsilon_7^3 & \varepsilon_7^2\\
\chi_{9} & 1 & 1 & \varepsilon_7^6 & \varepsilon_7^6 & \varepsilon_7^6 & \varepsilon_7^3 & \varepsilon_7^3 & \varepsilon_7^4 & \varepsilon_7^4 & \varepsilon_7 & \varepsilon_7 & \varepsilon_7^2 & \varepsilon_7^2 & \varepsilon_7^4 & \varepsilon_7 & \varepsilon_7^5 & \varepsilon_7^5 & 1 & \varepsilon_7^5 & \varepsilon_7^3 & \varepsilon_7^2\\
\chi_{10} & 1 & 1 & \varepsilon_7^5 & \varepsilon_7^5 & \varepsilon_7^5 & \varepsilon_7^6 & \varepsilon_7^6 & \varepsilon_7 & \varepsilon_7 & \varepsilon_7^2 & \varepsilon_7^2 & \varepsilon_7^4 & \varepsilon_7^4 & \varepsilon_7 & \varepsilon_7^2 & \varepsilon_7^3 & \varepsilon_7^3 & 1 & \varepsilon_7^3 & \varepsilon_7^6 & \varepsilon_7^4\\
\chi_{11} & 1 & 1 & \varepsilon_7^4 & \varepsilon_7^4 & \varepsilon_7^4 & \varepsilon_7^2 & \varepsilon_7^2 & \varepsilon_7^5 & \varepsilon_7^5 & \varepsilon_7^3 & \varepsilon_7^3 & \varepsilon_7^6 & \varepsilon_7^6 & \varepsilon_7^5 & \varepsilon_7^3 & \varepsilon_7 & \varepsilon_7 & 1 & \varepsilon_7 & \varepsilon_7^2 & \varepsilon_7^6\\
\chi_{12} & 1 & 1 & \varepsilon_7^3 & \varepsilon_7^3 & \varepsilon_7^3 & \varepsilon_7^5 & \varepsilon_7^5 & \varepsilon_7^2 & \varepsilon_7^2 & \varepsilon_7^4 & \varepsilon_7^4 & \varepsilon_7 & \varepsilon_7 & \varepsilon_7^2 & \varepsilon_7^4 & \varepsilon_7^6 & \varepsilon_7^6 & 1 & \varepsilon_7^6 & \varepsilon_7^5 & \varepsilon_7\\
\chi_{13} & 1 & 1 & \varepsilon_7^2 & \varepsilon_7^2 & \varepsilon_7^2 & \varepsilon_7 & \varepsilon_7 & \varepsilon_7^6 & \varepsilon_7^6 & \varepsilon_7^5 & \varepsilon_7^5 & \varepsilon_7^3 & \varepsilon_7^3 & \varepsilon_7^6 & \varepsilon_7^5 & \varepsilon_7^4 & \varepsilon_7^4 & 1 & \varepsilon_7^4 & \varepsilon_7 & \varepsilon_7^3\\
\chi_{14} & 1 & 1 & \varepsilon_7 & \varepsilon_7 & \varepsilon_7 & \varepsilon_7^4 & \varepsilon_7^4 & \varepsilon_7^3 & \varepsilon_7^3 & \varepsilon_7^6 & \varepsilon_7^6 & \varepsilon_7^5 & \varepsilon_7^5 & \varepsilon_7^3 & \varepsilon_7^6 & \varepsilon_7^2 & \varepsilon_7^2 & 1 & \varepsilon_7^2 & \varepsilon_7^4 & \varepsilon_7^5\\
\chi_{15} & 2 & 0 & 0 & -1 & 2 & 0 & -1 & 0 & -1 & -1 & 0 & 0 & 2 & 2 & 2 & 2 & 0 & -1 & -1 & 2 & -1\\
\chi_{16} & 2 & 0 & 0 & -\varepsilon_7^3 & 2\varepsilon_7^3 & 0 & -\varepsilon_7^5 & 0 & -\varepsilon_7^2 & -\varepsilon_7^4 & 0 & 0 & 2\varepsilon_7 & 2\varepsilon_7^2 & 2\varepsilon_7^4 & 2\varepsilon_7^6 & 0 & -1 & -\varepsilon_7^6 & 2\varepsilon_7^5 & -\varepsilon_7\\
\chi_{17} & 2 & 0 & 0 & -\varepsilon_7^2 & 2\varepsilon_7^2 & 0 & -\varepsilon_7 & 0 & -\varepsilon_7^6 & -\varepsilon_7^5 & 0 & 0 & 2\varepsilon_7^3 & 2\varepsilon_7^6 & 2\varepsilon_7^5 & 2\varepsilon_7^4 & 0 & -1 & -\varepsilon_7^4 & 2\varepsilon_7 & -\varepsilon_7^3\\
\chi_{18} & 2 & 0 & 0 & -\varepsilon_7 & 2\varepsilon_7 & 0 & -\varepsilon_7^4 & 0 & -\varepsilon_7^3 & -\varepsilon_7^6 & 0 & 0 & 2\varepsilon_7^5 & 2\varepsilon_7^3 & 2\varepsilon_7^6 & 2\varepsilon_7^2 & 0 & -1 & -\varepsilon_7^2 & 2\varepsilon_7^4 & -\varepsilon_7^5\\
\chi_{19} & 2 & 0 & 0 & -\varepsilon_7^6 & 2\varepsilon_7^6 & 0 & -\varepsilon_7^3 & 0 & -\varepsilon_7^4 & -\varepsilon_7 & 0 & 0 & 2\varepsilon_7^2 & 2\varepsilon_7^4 & 2\varepsilon_7 & 2\varepsilon_7^5 & 0 & -1 & -\varepsilon_7^5 & 2\varepsilon_7^3 & -\varepsilon_7^2\\
\chi_{20} & 2 & 0 & 0 & -\varepsilon_7^5 & 2\varepsilon_7^5 & 0 & -\varepsilon_7^6 & 0 & -\varepsilon_7 & -\varepsilon_7^2 & 0 & 0 & 2\varepsilon_7^4 & 2\varepsilon_7 & 2\varepsilon_7^2 & 2\varepsilon_7^3 & 0 & -1 & -\varepsilon_7^3 & 2\varepsilon_7^6 & -\varepsilon_7^4\\
\chi_{21} & 2 & 0 & 0 & -\varepsilon_7^4 & 2\varepsilon_7^4 & 0 & -\varepsilon_7^2 & 0 & -\varepsilon_7^5 & -\varepsilon_7^3 & 0 & 0 & 2\varepsilon_7^6 & 2\varepsilon_7^5 & 2\varepsilon_7^3 & 2\varepsilon_7 & 0 & -1 & -\varepsilon_7 & 2\varepsilon_7^2 & -\varepsilon_7^6\\
\end{array}$
}\\[5pt]
\caption{Character table for the centralizer of classes 7AB.}
\end{table}

\newpage

{\bf Class 8A}

\medskip

$C_{M_{24}}(8A)\cong \ZZ_2\times \ZZ_8$
\hspace{40pt}   Order: $2^4=16$
\hspace{40pt}
%Central extension: $\ZZ_2. C_{M_{24}}(8A)\cong \ZZ_4\times \ZZ_8$
Linear rep.

\bigskip

\begin{table}[h]
$\begin{array}{lrrrrrrrrrrrrrrrr}
&16 & 16 & 16 & 16 & 16 & 16 & 16 & 16 & 16 & 16 & 16 & 16 & 16 & 16 & 16 & 16 \\ 
 &\mathrm{1A} & \mathrm{8A_{1}} & \mathrm{8A_{2}} & \mathrm{8A_{3}} & \mathrm{4B_{1}} & \mathrm{8A_{4}} & \mathrm{4B_{2}} & \mathrm{8A_{5}} & \mathrm{8A_{6}} & \mathrm{4B_{3}} & \mathrm{8A_{7}} & \mathrm{4B_{4}} & \mathrm{8A_{8}} & \mathrm{2A_{1}} & \mathrm{2B_{1}} & \mathrm{2B_{2}}\\ 
\hline 
\chi_{1} & 1 & 1 & 1 & 1 & 1 & 1 & 1 & 1 & 1 & 1 & 1 & 1 & 1 & 1 & 1 & 1\\
\chi_{2} & 1 & -1 & -1 & -1 & 1 & -1 & 1 & -1 & -1 & 1 & -1 & 1 & -1 & 1 & 1 & 1\\
\chi_{3} & 1 & -1 & 1 & -1 & -1 & -1 & 1 & -1 & 1 & -1 & 1 & 1 & 1 & 1 & -1 & -1\\
\chi_{4} & 1 & 1 & -1 & 1 & -1 & 1 & 1 & 1 & -1 & -1 & -1 & 1 & -1 & 1 & -1 & -1\\
\chi_{5} & 1 & -i & -i & i & -1 & i & -1 & -i & i & -1 & i & -1 & -i & 1 & 1 & 1\\
\chi_{6} & 1 & i & i & -i & -1 & -i & -1 & i & -i & -1 & -i & -1 & i & 1 & 1 & 1\\
\chi_{7} & 1 & -i & i & i & 1 & i & -1 & -i & -i & 1 & -i & -1 & i & 1 & -1 & -1\\
\chi_{8} & 1 & i & -i & -i & 1 & -i & -1 & i & i & 1 & i & -1 & -i & 1 & -1 & -1\\
\chi_{9} & 1 & -\varepsilon_8 & -\varepsilon_8 & \varepsilon_8^7 & -i & -\varepsilon_8^7 & i & \varepsilon_8 & -\varepsilon_8^7 & i & \varepsilon_8^7 & -i & \varepsilon_8 & -1 & -1 & 1\\
\chi_{10} & 1 & \varepsilon_8^7 & \varepsilon_8^7 & -\varepsilon_8 & i & \varepsilon_8 & -i & -\varepsilon_8^7 & \varepsilon_8 & -i & -\varepsilon_8 & i & -\varepsilon_8^7 & -1 & -1 & 1\\
\chi_{11} & 1 & -\varepsilon_8^7 & -\varepsilon_8^7 & \varepsilon_8 & i & -\varepsilon_8 & -i & \varepsilon_8^7 & -\varepsilon_8 & -i & \varepsilon_8 & i & \varepsilon_8^7 & -1 & -1 & 1\\
\chi_{12} & 1 & \varepsilon_8 & \varepsilon_8 & -\varepsilon_8^7 & -i & \varepsilon_8^7 & i & -\varepsilon_8 & \varepsilon_8^7 & i & -\varepsilon_8^7 & -i & -\varepsilon_8 & -1 & -1 & 1\\
\chi_{13} & 1 & -\varepsilon_8 & \varepsilon_8 & \varepsilon_8^7 & i & -\varepsilon_8^7 & i & \varepsilon_8 & \varepsilon_8^7 & -i & -\varepsilon_8^7 & -i & -\varepsilon_8 & -1 & 1 & -1\\
\chi_{14} & 1 & \varepsilon_8^7 & -\varepsilon_8^7 & -\varepsilon_8 & -i & \varepsilon_8 & -i & -\varepsilon_8^7 & -\varepsilon_8 & i & \varepsilon_8 & i & \varepsilon_8^7 & -1 & 1 & -1\\
\chi_{15} & 1 & -\varepsilon_8^7 & \varepsilon_8^7 & \varepsilon_8 & -i & -\varepsilon_8 & -i & \varepsilon_8^7 & \varepsilon_8 & i & -\varepsilon_8 & i & -\varepsilon_8^7 & -1 & 1 & -1\\
\chi_{16} & 1 & \varepsilon_8 & -\varepsilon_8 & -\varepsilon_8^7 & i & \varepsilon_8^7 & i & -\varepsilon_8 & -\varepsilon_8^7 & -i & \varepsilon_8^7 & -i & \varepsilon_8 & -1 & 1 & -1\\
\end{array}$\\[5pt]
\caption{Character table for the centralizer of class 8A.}
\end{table}

\newpage

{\bf Class 10A}

\medskip

$C_{M_{24}}(10A)\cong \ZZ_2\times \ZZ_{10}$%
\hspace{40pt}Order: $20$
\hspace{40pt}
%Central extension: $\ZZ_2. C_{M_{24}}(10A)\cong ???$
Linear rep.

\bigskip

\begin{table}[h]
\scalebox{0.8}{
$\begin{array}{ccccccccccccccccccccc}
&20 & 20 & 20 & 20 & 20 & 20 & 20 & 20 & 20 & 20 & 20 & 20 & 20 & 20 & 20 & 20 & 20 & 20 & 20 & 20 \\ 
&\mathrm{1A} & \mathrm{10A_{1}} & \mathrm{10A_{2}} & \mathrm{10A_{3}} & \mathrm{5A_{1}} & \mathrm{10A_{4}} & \mathrm{2B_{1}} & \mathrm{2B_{2}} & \mathrm{10A_{5}} & \mathrm{10A_{6}} & \mathrm{10A_{7}} & \mathrm{10A_{8}} & \mathrm{5A_{2}} & \mathrm{5A_{3}} & \mathrm{10A_{9}} & \mathrm{10A_{10}} & \mathrm{2B_{3}} & \mathrm{10A_{11}} & \mathrm{5A_{4}} & \mathrm{10A_{12}}\\ \hline 
 \chi _1 & 1 & 1 & 1 & 1 & 1 & 1 & 1 & 1 & 1 & 1 & 1 & 1 & 1 & 1 & 1 & 1 & 1 & 1 & 1 &
   1 \\
 \chi _2 & 1 & -1 & -1 & -1 & 1 & 1 & 1 & -1 & -1 & 1 & -1 & 1 & 1 & 1 & -1 & -1 & -1 &
   -1 & 1 & 1 \\
 \chi _3 & 1 & 1 & 1 & 1 & 1 & -1 & -1 & -1 & 1 & -1 & -1 & -1 & 1 & 1 & -1 & -1 & 1 &
   -1 & 1 & -1 \\
 \chi _4 & 1 & -1 & -1 & -1 & 1 & -1 & -1 & 1 & -1 & -1 & 1 & -1 & 1 & 1 & 1 & 1 & -1 &
   1 & 1 & -1 \\
 \chi _5 & 1 & \varepsilon _5^3 & \varepsilon _5^4 & \varepsilon _5^2 & \varepsilon
   _5^4 & \varepsilon _5^2 & 1 & 1 & \varepsilon _5 & \varepsilon _5^4 & \varepsilon
   _5^2 & \varepsilon _5^3 & \varepsilon _5^3 & \varepsilon _5 & \varepsilon _5 &
   \varepsilon _5^3 & 1 & \varepsilon _5^4 & \varepsilon _5^2 & \varepsilon _5 \\
 \chi _6 & 1 & -\varepsilon _5^3 & -\varepsilon _5^4 & -\varepsilon _5^2 & \varepsilon
   _5^4 & \varepsilon _5^2 & 1 & -1 & -\varepsilon _5 & \varepsilon _5^4 & -\varepsilon
   _5^2 & \varepsilon _5^3 & \varepsilon _5^3 & \varepsilon _5 & -\varepsilon _5 &
   -\varepsilon _5^3 & -1 & -\varepsilon _5^4 & \varepsilon _5^2 & \varepsilon _5 \\
 \chi _7 & 1 & \varepsilon _5^3 & \varepsilon _5^4 & \varepsilon _5^2 & \varepsilon
   _5^4 & -\varepsilon _5^2 & -1 & -1 & \varepsilon _5 & -\varepsilon _5^4 &
   -\varepsilon _5^2 & -\varepsilon _5^3 & \varepsilon _5^3 & \varepsilon _5 &
   -\varepsilon _5 & -\varepsilon _5^3 & 1 & -\varepsilon _5^4 & \varepsilon _5^2 &
   -\varepsilon _5 \\
 \chi _8 & 1 & -\varepsilon _5^3 & -\varepsilon _5^4 & -\varepsilon _5^2 & \varepsilon
   _5^4 & -\varepsilon _5^2 & -1 & 1 & -\varepsilon _5 & -\varepsilon _5^4 &
   \varepsilon _5^2 & -\varepsilon _5^3 & \varepsilon _5^3 & \varepsilon _5 &
   \varepsilon _5 & \varepsilon _5^3 & -1 & \varepsilon _5^4 & \varepsilon _5^2 &
   -\varepsilon _5 \\
 \chi _9 & 1 & \varepsilon _5 & \varepsilon _5^3 & \varepsilon _5^4 & \varepsilon _5^3
   & \varepsilon _5^4 & 1 & 1 & \varepsilon _5^2 & \varepsilon _5^3 & \varepsilon _5^4
   & \varepsilon _5 & \varepsilon _5 & \varepsilon _5^2 & \varepsilon _5^2 &
   \varepsilon _5 & 1 & \varepsilon _5^3 & \varepsilon _5^4 & \varepsilon _5^2 \\
 \chi _{10} & 1 & -\varepsilon _5 & -\varepsilon _5^3 & -\varepsilon _5^4 & \varepsilon
   _5^3 & \varepsilon _5^4 & 1 & -1 & -\varepsilon _5^2 & \varepsilon _5^3 &
   -\varepsilon _5^4 & \varepsilon _5 & \varepsilon _5 & \varepsilon _5^2 &
   -\varepsilon _5^2 & -\varepsilon _5 & -1 & -\varepsilon _5^3 & \varepsilon _5^4 &
   \varepsilon _5^2 \\
 \chi _{11} & 1 & \varepsilon _5 & \varepsilon _5^3 & \varepsilon _5^4 & \varepsilon
   _5^3 & -\varepsilon _5^4 & -1 & -1 & \varepsilon _5^2 & -\varepsilon _5^3 &
   -\varepsilon _5^4 & -\varepsilon _5 & \varepsilon _5 & \varepsilon _5^2 &
   -\varepsilon _5^2 & -\varepsilon _5 & 1 & -\varepsilon _5^3 & \varepsilon _5^4 &
   -\varepsilon _5^2 \\
 \chi _{12} & 1 & -\varepsilon _5 & -\varepsilon _5^3 & -\varepsilon _5^4 & \varepsilon
   _5^3 & -\varepsilon _5^4 & -1 & 1 & -\varepsilon _5^2 & -\varepsilon _5^3 &
   \varepsilon _5^4 & -\varepsilon _5 & \varepsilon _5 & \varepsilon _5^2 & \varepsilon
   _5^2 & \varepsilon _5 & -1 & \varepsilon _5^3 & \varepsilon _5^4 & -\varepsilon _5^2
   \\
 \chi _{13} & 1 & \varepsilon _5^4 & \varepsilon _5^2 & \varepsilon _5 & \varepsilon
   _5^2 & \varepsilon _5 & 1 & 1 & \varepsilon _5^3 & \varepsilon _5^2 & \varepsilon _5
   & \varepsilon _5^4 & \varepsilon _5^4 & \varepsilon _5^3 & \varepsilon _5^3 &
   \varepsilon _5^4 & 1 & \varepsilon _5^2 & \varepsilon _5 & \varepsilon _5^3 \\
 \chi _{14} & 1 & -\varepsilon _5^4 & -\varepsilon _5^2 & -\varepsilon _5 & \varepsilon
   _5^2 & \varepsilon _5 & 1 & -1 & -\varepsilon _5^3 & \varepsilon _5^2 & -\varepsilon
   _5 & \varepsilon _5^4 & \varepsilon _5^4 & \varepsilon _5^3 & -\varepsilon _5^3 &
   -\varepsilon _5^4 & -1 & -\varepsilon _5^2 & \varepsilon _5 & \varepsilon _5^3 \\
 \chi _{15} & 1 & \varepsilon _5^4 & \varepsilon _5^2 & \varepsilon _5 & \varepsilon
   _5^2 & -\varepsilon _5 & -1 & -1 & \varepsilon _5^3 & -\varepsilon _5^2 &
   -\varepsilon _5 & -\varepsilon _5^4 & \varepsilon _5^4 & \varepsilon _5^3 &
   -\varepsilon _5^3 & -\varepsilon _5^4 & 1 & -\varepsilon _5^2 & \varepsilon _5 &
   -\varepsilon _5^3 \\
 \chi _{16} & 1 & -\varepsilon _5^4 & -\varepsilon _5^2 & -\varepsilon _5 & \varepsilon
   _5^2 & -\varepsilon _5 & -1 & 1 & -\varepsilon _5^3 & -\varepsilon _5^2 &
   \varepsilon _5 & -\varepsilon _5^4 & \varepsilon _5^4 & \varepsilon _5^3 &
   \varepsilon _5^3 & \varepsilon _5^4 & -1 & \varepsilon _5^2 & \varepsilon _5 &
   -\varepsilon _5^3 \\
 \chi _{17} & 1 & \varepsilon _5^2 & \varepsilon _5 & \varepsilon _5^3 & \varepsilon _5
   & \varepsilon _5^3 & 1 & 1 & \varepsilon _5^4 & \varepsilon _5 & \varepsilon _5^3 &
   \varepsilon _5^2 & \varepsilon _5^2 & \varepsilon _5^4 & \varepsilon _5^4 &
   \varepsilon _5^2 & 1 & \varepsilon _5 & \varepsilon _5^3 & \varepsilon _5^4 \\
 \chi _{18} & 1 & -\varepsilon _5^2 & -\varepsilon _5 & -\varepsilon _5^3 & \varepsilon
   _5 & \varepsilon _5^3 & 1 & -1 & -\varepsilon _5^4 & \varepsilon _5 & -\varepsilon
   _5^3 & \varepsilon _5^2 & \varepsilon _5^2 & \varepsilon _5^4 & -\varepsilon _5^4 &
   -\varepsilon _5^2 & -1 & -\varepsilon _5 & \varepsilon _5^3 & \varepsilon _5^4 \\
 \chi _{19} & 1 & \varepsilon _5^2 & \varepsilon _5 & \varepsilon _5^3 & \varepsilon _5
   & -\varepsilon _5^3 & -1 & -1 & \varepsilon _5^4 & -\varepsilon _5 & -\varepsilon
   _5^3 & -\varepsilon _5^2 & \varepsilon _5^2 & \varepsilon _5^4 & -\varepsilon _5^4 &
   -\varepsilon _5^2 & 1 & -\varepsilon _5 & \varepsilon _5^3 & -\varepsilon _5^4 \\
 \chi _{20} & 1 & -\varepsilon _5^2 & -\varepsilon _5 & -\varepsilon _5^3 & \varepsilon
   _5 & -\varepsilon _5^3 & -1 & 1 & -\varepsilon _5^4 & -\varepsilon _5 & \varepsilon
   _5^3 & -\varepsilon _5^2 & \varepsilon _5^2 & \varepsilon _5^4 & \varepsilon _5^4 &
   \varepsilon _5^2 & -1 & \varepsilon _5 & \varepsilon _5^3 & -\varepsilon _5^4
\end{array}
$
}\\[5pt]
\caption{Character table for the centralizer of class 10A.}
\end{table}

\end{landscape}

\section{Decompositions of twisted sectors}
\label{app:decompositions}
In this section we give the decomposition of the first 20 representation spaces $\H^n_g$ for
all non-isomorphic twisted sectors.
In each table, the integer in the $i$'th row and the $r$'th column is the multiplicity $h_{i,r}$ of the irreducible
representation $\chi_i$ (numbered as in the corresponding character table) in the representation $\rho_{g,r}$ of 
eq.~\eqref{eqn:decomp}. 
When the centraliser $C_{M_{24}}(g)$ is a cyclic group $\ZZ_N$, $N=o(g)$ (classes 11A, 12A, 12B, 14AB, 15AB, 21AB, 23AB), 
the multiplicity $h_{i,r}$ of the (projective) irreducible characters $\chi_i$ (see eq.~\eqref{cyclicchar}) is only non-zero 
if $ N r+1/\ell(g)\equiv i\mod N$,  where $\ell(g)$ is the length of the shortest cycle of $g$ as a permutation of $24$ objects;
then, only these values of $h_{i,r}$ are given.

%\newpage
\begin{table}[h]
\scalebox{0.8}{
\begin{tabular}{|l|rrrrrrrrrrrrrrrrrrrrr|}
\hline
\diagbox[height=25pt]{$i$}{$2r$} & 0 & 1 & 2 & 3 & 4 & 5 & 6 & 7 & 8 & 9 & 10 & 11 & 12 & 13 & 14 & 15 & 16 & 17 & 18 & 19 & 20 \\
\hline
% i \\\hline
 1 & 1 & 0 & 0 & 0 & 0 & 0 & 0 & 0 & 4 & 0 & 2 & 0 & 14 & 0 & 18 & 0 & 52 & 0 & 96 & 0 & 204 \\
 2 & 0 & 0 & 0 & 0 & 0 & 0 & 0 & 0 & 0 & 4 & 0 & 6 & 0 & 14 & 0 & 34 & 0 & 72 & 0 & 132 & 0 \\
 3,4 & 0 & 1 & 0 & 0 & 0 & 0 & 0 & 1 & 0 & 2 & 0 & 6 & 0 & 18 & 0 & 32 & 0 & 72 & 0 & 137 & 0 \\
 5,6 & 0 & 0 & 1 & 0 & 0 & 0 & 2 & 0 & 4 & 0 & 16 & 0 & 28 & 0 & 72 & 0 & 141 & 0 & 306 & 0 & 570 \\
 7,8 & 0 & 0 & 0 & 0 & 0 & 2 & 0 & 2 & 0 & 10 & 0 & 18 & 0 & 50 & 0 & 96 & 0 & 216 & 0 & 408 & 0 \\
 9 & 0 & 0 & 0 & 0 & 2 & 0 & 2 & 0 & 12 & 0 & 22 & 0 & 68 & 0 & 130 & 0 & 304 & 0 & 578 & 0 & 1184 \\
 10 & 0 & 0 & 0 & 0 & 0 & 0 & 6 & 0 & 10 & 0 & 30 & 0 & 72 & 0 & 166 & 0 & 332 & 0 & 702 & 0 & 1346 \\
 11 & 0 & 0 & 0 & 2 & 0 & 2 & 0 & 6 & 0 & 20 & 0 & 48 & 0 & 106 & 0 & 240 & 0 & 486 & 0 & 972 & 0 \\
 12,13 & 0 & 0 & 0 & 0 & 0 & 2 & 0 & 8 & 0 & 18 & 0 & 48 & 0 & 108 & 0 & 236 & 0 & 486 & 0 & 978 & 0 \\
 14 & 0 & 0 & 0 & 0 & 2 & 0 & 4 & 0 & 14 & 0 & 36 & 0 & 80 & 0 & 184 & 0 & 392 & 0 & 788 & 0 & 1550 \\\hline
\end{tabular}}\\[5pt]
\caption{Multiplicities of irreducible representations for the 2A-twisted sector.}
\end{table}
\begin{table}[h]
\scalebox{0.8}{
\begin{tabular}{|l|llllllllllllllllllll|}
\hline
 \diagbox[height=25pt]{$i$}{$4r$} & 1 & 3 & 5 & 7 & 9 & 11 & 13 & 15 & 17 & 19 & 21 & 23 & 25 & 27 & 29 & 31 & 33 & 35 & 37 & 39 \\\hline
% i \\\hline
 1 & 0 & 0 & 0 & 0 & 0 & 0 & 2 & 0 & 2 & 0 & 8 & 0 & 16 & 0 & 42 & 0 & 76 & 0 & 172 & 0 \\
 2,3 & 1 & 0 & 0 & 0 & 0 & 0 & 1 & 0 & 4 & 0 & 6 & 0 & 21 & 0 & 36 & 0 & 84 & 0 & 160 & 0 \\
 4 & 0 & 0 & 0 & 0 & 0 & 0 & 2 & 0 & 2 & 0 & 8 & 0 & 16 & 0 & 42 & 0 & 76 & 0 & 172 & 0 \\
 5,8 & 0 & 0 & 1 & 0 & 0 & 0 & 6 & 0 & 10 & 0 & 32 & 0 & 68 & 0 & 162 & 0 & 318 & 0 & 670 & 0 \\
 6,7 & 0 & 0 & 0 & 0 & 2 & 0 & 4 & 0 & 14 & 0 & 26 & 0 & 76 & 0 & 150 & 0 & 334 & 0 & 648 & 0 \\
 9,10 & 0 & 1 & 0 & 0 & 0 & 2 & 0 & 8 & 0 & 18 & 0 & 47 & 0 & 108 & 0 & 224 & 0 & 466 & 0 & 924 \\
 11,14 & 0 & 0 & 0 & 0 & 2 & 0 & 4 & 0 & 16 & 0 & 38 & 0 & 90 & 0 & 192 & 0 & 414 & 0 & 814 & 0 \\
 12,13 & 0 & 0 & 1 & 0 & 2 & 0 & 6 & 0 & 14 & 0 & 40 & 0 & 86 & 0 & 198 & 0 & 406 & 0 & 826 & 0 \\
 15,16 & 0 & 0 & 0 & 0 & 2 & 0 & 6 & 0 & 18 & 0 & 46 & 0 & 106 & 0 & 234 & 0 & 490 & 0 & 986 & 0 \\
 17 & 0 & 0 & 0 & 2 & 0 & 4 & 0 & 14 & 0 & 40 & 0 & 94 & 0 & 210 & 0 & 456 & 0 & 932 & 0 & 1836 \\
 18 & 0 & 0 & 0 & 2 & 0 & 8 & 0 & 22 & 0 & 58 & 0 & 140 & 0 & 318 & 0 & 682 & 0 & 1396 & 0 & 2760 \\
\hline
\end{tabular}}\\[5pt]
\caption{Multiplicities of irreducible representations for the 2B-twisted sector.}
\end{table}

\begin{table}[h]\scalebox{0.9}{
\begin{tabular}{|l|llllllllllllllllllll|}
\hline
 \diagbox[height=25pt]{$i$}{$3r$} & 0 & 1 & 2 & 3 & 4 & 5 & 6 & 7 & 8 & 9 & 10 & 11 & 12 & 13 & 14 & 15 & 16 & 17 & 18 & 19 \\\hline
% i \\\hline
 1 & 0 & 0 & 0 & 0 & 0 & 0 & 0 & 0 & 2 & 0 & 0 & 2 & 0 & 0 & 10 & 0 & 0 & 16 & 0 & 0 \\
 2,3 & 0 & 1 & 0 & 0 & 0 & 0 & 0 & 2 & 0 & 0 & 6 & 0 & 0 & 20 & 0 & 0 & 41 & 0 & 0 & 104 \\
 4,5 & 0 & 0 & 0 & 0 & 0 & 0 & 1 & 0 & 0 & 6 & 0 & 0 & 10 & 0 & 0 & 34 & 0 & 0 & 74 & 0 \\
 6,7 & 0 & 0 & 0 & 0 & 0 & 0 & 0 & 0 & 6 & 0 & 0 & 14 & 0 & 0 & 42 & 0 & 0 & 92 & 0 & 0 \\
 8 & 0 & 0 & 0 & 0 & 2 & 0 & 0 & 2 & 0 & 0 & 16 & 0 & 0 & 30 & 0 & 0 & 94 & 0 & 0 & 194 \\
 9 & 1 & 0 & 0 & 0 & 0 & 0 & 4 & 0 & 0 & 6 & 0 & 0 & 28 & 0 & 0 & 60 & 0 & 0 & 162 & 0 \\
 10,11 & 0 & 0 & 0 & 0 & 0 & 2 & 0 & 0 & 8 & 0 & 0 & 24 & 0 & 0 & 64 & 0 & 0 & 154 & 0 & 0 \\
 12 & 0 & 0 & 2 & 0 & 0 & 2 & 0 & 0 & 10 & 0 & 0 & 24 & 0 & 0 & 74 & 0 & 0 & 172 & 0 & 0 \\
 13 & 0 & 0 & 0 & 0 & 2 & 0 & 0 & 6 & 0 & 0 & 20 & 0 & 0 & 50 & 0 & 0 & 134 & 0 & 0 & 298 \\
 14 & 0 & 0 & 0 & 0 & 0 & 0 & 4 & 0 & 0 & 12 & 0 & 0 & 40 & 0 & 0 & 94 & 0 & 0 & 236 & 0 \\
 15 & 0 & 0 & 0 & 0 & 0 & 4 & 0 & 0 & 8 & 0 & 0 & 32 & 0 & 0 & 76 & 0 & 0 & 200 & 0 & 0 \\
 16 & 0 & 0 & 0 & 0 & 2 & 0 & 0 & 10 & 0 & 0 & 30 & 0 & 0 & 88 & 0 & 0 & 216 & 0 & 0 & 510 \\
 17 & 0 & 0 & 0 & 2 & 0 & 0 & 6 & 0 & 0 & 22 & 0 & 0 & 62 & 0 & 0 & 164 & 0 & 0 & 382 & 0 \\\hline
\end{tabular}}\\[5pt]
\caption{Multiplicities of irreducible representations for the 3A-twisted sector.}
\end{table}

\begin{table}[h]\scalebox{0.9}{
\begin{tabular}{|l|lllllllllllllllllllll|}
\hline
\diagbox[height=25pt]{$i$}{$ 9 r$} & 2 & 5 & 8 & 11 & 14 & 17 & 20 & 23 & 26 & 29 & 32 & 35 & 38 & 41 & 44 & 47 & 50 & 53 & 56 & 59 & 62 \\\hline
% i \\\hline
 1 & 0 & 0 & 0 & 0 & 0 & 2 & 0 & 0 & 2 & 0 & 0 & 10 & 0 & 0 & 18 & 0 & 0 & 54 & 0 & 0 & 108 \\
 2 & 0 & 0 & 0 & 0 & 0 & 0 & 0 & 2 & 0 & 0 & 4 & 0 & 0 & 18 & 0 & 0 & 36 & 0 & 0 & 90 & 0 \\
 3 & 0 & 0 & 0 & 0 & 0 & 0 & 0 & 0 & 0 & 6 & 0 & 0 & 10 & 0 & 0 & 30 & 0 & 0 & 62 & 0 & 0 \\
 4,5 & 0 & 0 & 0 & 0 & 0 & 2 & 0 & 0 & 10 & 0 & 0 & 22 & 0 & 0 & 65 & 0 & 0 & 146 & 0 & 0 & 344 \\
 6,7 & 0 & 0 & 0 & 0 & 2 & 0 & 0 & 5 & 0 & 0 & 18 & 0 & 0 & 44 & 0 & 0 & 118 & 0 & 0 & 254 & 0 \\
 8,9 & 1 & 0 & 0 & 0 & 0 & 0 & 4 & 0 & 0 & 10 & 0 & 0 & 36 & 0 & 0 & 82 & 0 & 0 & 202 & 0 & 0 \\
 10 & 0 & 0 & 0 & 0 & 0 & 6 & 0 & 0 & 14 & 0 & 0 & 50 & 0 & 0 & 122 & 0 & 0 & 308 & 0 & 0 & 666 \\
 11 & 0 & 2 & 0 & 0 & 2 & 0 & 0 & 12 & 0 & 0 & 32 & 0 & 0 & 96 & 0 & 0 & 220 & 0 & 0 & 526 & 0 \\
 12 & 0 & 0 & 0 & 2 & 0 & 0 & 6 & 0 & 0 & 26 & 0 & 0 & 64 & 0 & 0 & 174 & 0 & 0 & 390 & 0 & 0 \\
 13 & 0 & 0 & 2 & 0 & 0 & 4 & 0 & 0 & 20 & 0 & 0 & 54 & 0 & 0 & 150 & 0 & 0 & 346 & 0 & 0 & 794 \\
 14 & 0 & 0 & 0 & 0 & 4 & 0 & 0 & 12 & 0 & 0 & 42 & 0 & 0 & 104 & 0 & 0 & 268 & 0 & 0 & 600 & 0 \\
 15 & 0 & 0 & 0 & 2 & 0 & 0 & 10 & 0 & 0 & 26 & 0 & 0 & 80 & 0 & 0 & 194 & 0 & 0 & 468 & 0 & 0 \\
 16 & 0 & 0 & 2 & 0 & 0 & 6 & 0 & 0 & 22 & 0 & 0 & 64 & 0 & 0 & 168 & 0 & 0 & 400 & 0 & 0 & 902 \\
 17 & 0 & 0 & 0 & 0 & 4 & 0 & 0 & 14 & 0 & 0 & 46 & 0 & 0 & 122 & 0 & 0 & 304 & 0 & 0 & 690 & 0 \\
 18 & 0 & 0 & 0 & 2 & 0 & 0 & 10 & 0 & 0 & 32 & 0 & 0 & 90 & 0 & 0 & 224 & 0 & 0 & 530 & 0 & 0 \\\hline
\end{tabular}}\\[5pt]
\caption{Multiplicities of irreducible representations for the 3B-twisted sector.}
\end{table}

\begin{table}[h]\scalebox{0.9}{
\begin{tabular}{|l|llllllllllllllllllll|}
\hline
\diagbox[height=25pt]{$i$}{$ 8 r$} & 1 & 3 & 5 & 7 & 9 & 11 & 13 & 15 & 17 & 19 & 21 & 23 & 25 & 27 & 29 & 31 & 33 & 35 & 37 & 39 \\\hline
%i \\\hline
 1,2 & 1 & 0 & 0 & 0 & 0 & 0 & 0 & 0 & 4 & 0 & 0 & 0 & 10 & 0 & 0 & 0 & 26 & 0 & 0 & 0 \\
 3,4 & 0 & 0 & 0 & 0 & 0 & 0 & 2 & 0 & 0 & 0 & 4 & 0 & 0 & 0 & 16 & 0 & 0 & 0 & 44 & 0 \\
 5 & 0 & 0 & 0 & 0 & 2 & 0 & 0 & 0 & 6 & 0 & 0 & 0 & 18 & 0 & 0 & 0 & 54 & 0 & 0 & 0 \\
 6 & 0 & 0 & 0 & 0 & 0 & 0 & 2 & 0 & 0 & 0 & 12 & 0 & 0 & 0 & 32 & 0 & 0 & 0 & 84 & 0 \\
 7,11 & 0 & 1 & 0 & 0 & 0 & 2 & 0 & 0 & 0 & 7 & 0 & 0 & 0 & 26 & 0 & 0 & 0 & 68 & 0 & 0 \\
 8,12 & 0 & 0 & 0 & 0 & 0 & 0 & 0 & 4 & 0 & 0 & 0 & 14 & 0 & 0 & 0 & 40 & 0 & 0 & 0 & 110 \\
 9,10 & 0 & 0 & 0 & 0 & 2 & 0 & 0 & 0 & 8 & 0 & 0 & 0 & 28 & 0 & 0 & 0 & 80 & 0 & 0 & 0 \\
 13 & 0 & 0 & 2 & 0 & 0 & 0 & 4 & 0 & 0 & 0 & 16 & 0 & 0 & 0 & 48 & 0 & 0 & 0 & 130 & 0 \\
 14 & 0 & 0 & 0 & 0 & 0 & 0 & 4 & 0 & 0 & 0 & 16 & 0 & 0 & 0 & 48 & 0 & 0 & 0 & 128 & 0 \\
 15 & 0 & 0 & 0 & 2 & 0 & 0 & 0 & 8 & 0 & 0 & 0 & 28 & 0 & 0 & 0 & 84 & 0 & 0 & 0 & 216 \\
 16 & 0 & 0 & 0 & 0 & 0 & 4 & 0 & 0 & 0 & 16 & 0 & 0 & 0 & 48 & 0 & 0 & 0 & 136 & 0 & 0 \\\hline
\end{tabular}}\\[5pt]
\caption{Multiplicities of irreducible representations for the 4A-twisted sector.}
\end{table}
\begin{table}[h]\scalebox{0.9}{
\begin{tabular}{|l|llllllllllllllllllll|}
 \hline
\diagbox[width=60pt,height=25pt]{$i$}{$ 4 r$} & 0 & 1 & 2 & 3 & 4 & 5 & 6 & 7 & 8 & 9 & 10 & 11 & 12 & 13 & 14 & 15 & 16 & 17 & 18 & 19 \\\hline
% i \\\hline
 1,6 & 0 & 1 & 0 & 0 & 0 & 3 & 0 & 0 & 0 & 10 & 0 & 0 & 0 & 33 & 0 & 0 & 0 & 90 & 0 & 0 \\
 2,3,4,5,7,8 & 0 & 0 & 0 & 0 & 0 & 2 & 0 & 0 & 0 & 10 & 0 & 0 & 0 & 32 & 0 & 0 & 0 & 90 & 0 & 0 \\
 9,10 & 0 & 0 & 0 & 2 & 0 & 0 & 0 & 10 & 0 & 0 & 0 & 36 & 0 & 0 & 0 & 110 & 0 & 0 & 0 & 290 \\
 11 & 1 & 0 & 0 & 0 & 2 & 0 & 0 & 0 & 16 & 0 & 0 & 0 & 46 & 0 & 0 & 0 & 146 & 0 & 0 & 0 \\
 12 & 0 & 0 & 0 & 0 & 4 & 0 & 0 & 0 & 12 & 0 & 0 & 0 & 52 & 0 & 0 & 0 & 136 & 0 & 0 & 0 \\
 13 & 0 & 0 & 0 & 0 & 0 & 0 & 8 & 0 & 0 & 0 & 24 & 0 & 0 & 0 & 88 & 0 & 0 & 0 & 224 & 0 \\
 14 & 0 & 0 & 2 & 0 & 0 & 0 & 6 & 0 & 0 & 0 & 30 & 0 & 0 & 0 & 80 & 0 & 0 & 0 & 236 & 0 \\\hline
\end{tabular}}\\[5pt]
\caption{Multiplicities of irreducible representations for the 4B-twisted sector.}
\end{table}
\begin{table}[h]\scalebox{0.9}{
\begin{tabular}{|l|llllllllllllllllllll|}
\hline
\diagbox[height=25pt]{$i$}{$ 16 r$} & 3 & 7 & 11 & 15 & 19 & 23 & 27 & 31 & 35 & 39 & 43 & 47 & 51 & 55 & 59 & 63 & 67 & 71 & 75 & 79 \\\hline
%i \\\hline
 1,2 & 0 & 0 & 0 & 2 & 0 & 0 & 0 & 8 & 0 & 0 & 0 & 30 & 0 & 0 & 0 & 90 & 0 & 0 & 0 & 228 \\
 3,4 & 0 & 0 & 0 & 0 & 0 & 4 & 0 & 0 & 0 & 18 & 0 & 0 & 0 & 52 & 0 & 0 & 0 & 144 & 0 & 0 \\
 5,6 & 0 & 0 & 1 & 0 & 0 & 0 & 7 & 0 & 0 & 0 & 22 & 0 & 0 & 0 & 68 & 0 & 0 & 0 & 184 & 0 \\
 7,8 & 1 & 0 & 0 & 0 & 2 & 0 & 0 & 0 & 12 & 0 & 0 & 0 & 41 & 0 & 0 & 0 & 112 & 0 & 0 & 0 \\
 9 & 0 & 0 & 0 & 4 & 0 & 0 & 0 & 18 & 0 & 0 & 0 & 60 & 0 & 0 & 0 & 176 & 0 & 0 & 0 & 460 \\
 10 & 0 & 2 & 0 & 0 & 0 & 8 & 0 & 0 & 0 & 32 & 0 & 0 & 0 & 106 & 0 & 0 & 0 & 288 & 0 & 0 \\
 11 & 0 & 0 & 2 & 0 & 0 & 0 & 12 & 0 & 0 & 0 & 46 & 0 & 0 & 0 & 136 & 0 & 0 & 0 & 364 & 0 \\
 12 & 0 & 0 & 0 & 0 & 6 & 0 & 0 & 0 & 24 & 0 & 0 & 0 & 80 & 0 & 0 & 0 & 228 & 0 & 0 & 0 \\\hline
\end{tabular}}\\[5pt]
\caption{Multiplicities of irreducible representations for the 4C-twisted sector.}
\end{table}

\begin{table}[h]\scalebox{0.9}{
\begin{tabular}{|l|lllllllllllllllllllll|}
\hline
\diagbox[height=25pt]{$i$}{$ 5 r$} & 0 & 1 & 2 & 3 & 4 & 5 & 6 & 7 & 8 & 9 & 10 & 11 & 12 & 13 & 14 & 15 & 16 & 17 & 18 & 19 & 20 \\\hline
%i \\\hline
 1 & 0 & 0 & 0 & 0 & 0 & 0 & 0 & 0 & 0 & 8 & 0 & 0 & 0 & 0 & 20 & 0 & 0 & 0 & 0 & 60 & 0 \\
 2 & 0 & 0 & 0 & 0 & 0 & 0 & 0 & 2 & 0 & 0 & 0 & 0 & 14 & 0 & 0 & 0 & 0 & 42 & 0 & 0 & 0 \\
 3 & 1 & 0 & 0 & 0 & 0 & 2 & 0 & 0 & 0 & 0 & 6 & 0 & 0 & 0 & 0 & 28 & 0 & 0 & 0 & 0 & 74 \\
 4 & 0 & 0 & 0 & 2 & 0 & 0 & 0 & 0 & 4 & 0 & 0 & 0 & 0 & 16 & 0 & 0 & 0 & 0 & 50 & 0 & 0 \\
 5 & 0 & 0 & 0 & 0 & 0 & 0 & 2 & 0 & 0 & 0 & 0 & 10 & 0 & 0 & 0 & 0 & 30 & 0 & 0 & 0 & 0 \\
 6,7 & 0 & 0 & 0 & 0 & 1 & 0 & 0 & 0 & 0 & 6 & 0 & 0 & 0 & 0 & 20 & 0 & 0 & 0 & 0 & 63 & 0 \\
 8,12 & 0 & 0 & 0 & 0 & 0 & 0 & 0 & 4 & 0 & 0 & 0 & 0 & 12 & 0 & 0 & 0 & 0 & 42 & 0 & 0 & 0 \\
 9,13 & 0 & 0 & 0 & 0 & 0 & 2 & 0 & 0 & 0 & 0 & 7 & 0 & 0 & 0 & 0 & 26 & 0 & 0 & 0 & 0 & 74 \\
 10,14 & 0 & 0 & 0 & 0 & 0 & 0 & 0 & 0 & 4 & 0 & 0 & 0 & 0 & 18 & 0 & 0 & 0 & 0 & 48 & 0 & 0 \\
 11,15 & 0 & 1 & 0 & 0 & 0 & 0 & 2 & 0 & 0 & 0 & 0 & 10 & 0 & 0 & 0 & 0 & 32 & 0 & 0 & 0 & 0 \\
 16 & 0 & 0 & 0 & 0 & 4 & 0 & 0 & 0 & 0 & 16 & 0 & 0 & 0 & 0 & 64 & 0 & 0 & 0 & 0 & 180 & 0 \\
 17 & 0 & 0 & 0 & 2 & 0 & 0 & 0 & 0 & 14 & 0 & 0 & 0 & 0 & 46 & 0 & 0 & 0 & 0 & 154 & 0 & 0 \\
 18 & 0 & 0 & 0 & 0 & 0 & 0 & 8 & 0 & 0 & 0 & 0 & 28 & 0 & 0 & 0 & 0 & 100 & 0 & 0 & 0 & 0 \\
 19 & 0 & 0 & 2 & 0 & 0 & 0 & 0 & 8 & 0 & 0 & 0 & 0 & 40 & 0 & 0 & 0 & 0 & 118 & 0 & 0 & 0 \\
 20 & 1 & 0 & 0 & 0 & 0 & 4 & 0 & 0 & 0 & 0 & 24 & 0 & 0 & 0 & 0 & 76 & 0 & 0 & 0 & 0 & 230 \\\hline
\end{tabular}}\\[5pt]
\caption{Multiplicities of irreducible representations for the 5A-twisted sector.}
\end{table}

\begin{table}[h]\scalebox{0.9}{
\begin{tabular}{|l|llllllllllllllllllll|}
\hline
\diagbox[height=25pt]{$i$}{$ 6 r$} & 0 & 1 & 2 & 3 & 4 & 5 & 6 & 7 & 8 & 9 & 10 & 11 & 12 & 13 & 14 & 15 & 16 & 17 & 18 & 19 \\\hline
% i \\\hline
 1,2 & 0 & 0 & 0 & 0 & 0 & 4 & 0 & 0 & 0 & 0 & 0 & 14 & 0 & 0 & 0 & 0 & 0 & 56 & 0 & 0 \\
 3,4 & 0 & 0 & 0 & 0 & 0 & 2 & 0 & 0 & 0 & 0 & 0 & 16 & 0 & 0 & 0 & 0 & 0 & 52 & 0 & 0 \\
 5,12 & 0 & 1 & 0 & 0 & 0 & 0 & 0 & 5 & 0 & 0 & 0 & 0 & 0 & 25 & 0 & 0 & 0 & 0 & 0 & 76 \\
 6,11 & 0 & 0 & 0 & 1 & 0 & 0 & 0 & 0 & 0 & 10 & 0 & 0 & 0 & 0 & 0 & 34 & 0 & 0 & 0 & 0 \\
 7,10 & 0 & 0 & 0 & 0 & 0 & 0 & 0 & 6 & 0 & 0 & 0 & 0 & 0 & 22 & 0 & 0 & 0 & 0 & 0 & 80 \\
 8,9 & 0 & 0 & 0 & 2 & 0 & 0 & 0 & 0 & 0 & 8 & 0 & 0 & 0 & 0 & 0 & 38 & 0 & 0 & 0 & 0 \\
 13 & 0 & 0 & 2 & 0 & 0 & 0 & 0 & 0 & 14 & 0 & 0 & 0 & 0 & 0 & 58 & 0 & 0 & 0 & 0 & 0 \\
 14 & 1 & 0 & 0 & 0 & 0 & 0 & 8 & 0 & 0 & 0 & 0 & 0 & 38 & 0 & 0 & 0 & 0 & 0 & 130 & 0 \\
 15 & 0 & 0 & 0 & 0 & 4 & 0 & 0 & 0 & 0 & 0 & 24 & 0 & 0 & 0 & 0 & 0 & 88 & 0 & 0 & 0 \\\hline
\end{tabular}}\\[5pt]
\caption{Multiplicities of irreducible representations for the 6A-twisted sector.}
\end{table}
\clearpage

\begin{table}[h]
\scalebox{0.8}{
\begin{tabular}{|l|llllllllllllllllllllllll|}
\hline
\diagbox[height=25pt]{$i$}{$ 36 r$} & 5 & 11 & 17 & 23 & 29 & 35 & 41 & 47 & 53 & 59 & 65 & 71 & 77 & 83 & 89 & 95 & 101 & 107 & 113 & 119 & 125 & 131 & 137 & 143 \\\hline
%i \\\hline
 1,2 & 0 & 0 & 0 & 0 & 2 & 0 & 0 & 0 & 0 & 0 & 16 & 0 & 0 & 0 & 0 & 0 & 50 & 0 & 0 & 0 & 0 & 0 & 158 & 0 \\
 3,4 & 0 & 0 & 0 & 0 & 3 & 0 & 0 & 0 & 0 & 0 & 13 & 0 & 0 & 0 & 0 & 0 & 54 & 0 & 0 & 0 & 0 & 0 & 152 & 0 \\
 5,12 & 0 & 0 & 0 & 0 & 0 & 0 & 6 & 0 & 0 & 0 & 0 & 0 & 22 & 0 & 0 & 0 & 0 & 0 & 78 & 0 & 0 & 0 & 0 & 0 \\
 6,11 & 0 & 0 & 2 & 0 & 0 & 0 & 0 & 0 & 8 & 0 & 0 & 0 & 0 & 0 & 36 & 0 & 0 & 0 & 0 & 0 & 106 & 0 & 0 & 0 \\
 7,10 & 1 & 0 & 0 & 0 & 0 & 0 & 4 & 0 & 0 & 0 & 0 & 0 & 24 & 0 & 0 & 0 & 0 & 0 & 74 & 0 & 0 & 0 & 0 & 0 \\
 8,9 & 0 & 0 & 1 & 0 & 0 & 0 & 0 & 0 & 10 & 0 & 0 & 0 & 0 & 0 & 33 & 0 & 0 & 0 & 0 & 0 & 112 & 0 & 0 & 0 \\
 13 & 0 & 2 & 0 & 0 & 0 & 0 & 0 & 14 & 0 & 0 & 0 & 0 & 0 & 56 & 0 & 0 & 0 & 0 & 0 & 182 & 0 & 0 & 0 & 0 \\
 14 & 0 & 0 & 0 & 0 & 0 & 8 & 0 & 0 & 0 & 0 & 0 & 36 & 0 & 0 & 0 & 0 & 0 & 126 & 0 & 0 & 0 & 0 & 0 & 366 \\
 15 & 0 & 0 & 0 & 4 & 0 & 0 & 0 & 0 & 0 & 22 & 0 & 0 & 0 & 0 & 0 & 86 & 0 & 0 & 0 & 0 & 0 & 260 & 0 & 0 \\\hline
\end{tabular}
}\\[5pt]
\caption{Multiplicities of irreducible representations for the 6B-twisted sector.}
\end{table}

\begin{table}[h]\scalebox{0.8}{
\begin{tabular}{|l|llllllllllllllllllll|}
\hline
\diagbox[height=25pt]{$i$}{$ 7 r$} & 0 & 1 & 2 & 3 & 4 & 5 & 6 & 7 & 8 & 9 & 10 & 11 & 12 & 13 & 14 & 15 & 16 & 17 & 18 & 19 \\\hline
%i \\\hline
 1 & 0 & 0 & 0 & 0 & 0 & 0 & 2 & 0 & 0 & 0 & 0 & 0 & 0 & 8 & 0 & 0 & 0 & 0 & 0 & 0 \\
 2 & 0 & 0 & 0 & 0 & 0 & 0 & 2 & 0 & 0 & 0 & 0 & 0 & 0 & 10 & 0 & 0 & 0 & 0 & 0 & 0 \\
 3 & 0 & 0 & 0 & 0 & 0 & 0 & 0 & 0 & 0 & 4 & 0 & 0 & 0 & 0 & 0 & 0 & 15 & 0 & 0 & 0 \\
 4 & 0 & 0 & 0 & 0 & 0 & 2 & 0 & 0 & 0 & 0 & 0 & 0 & 6 & 0 & 0 & 0 & 0 & 0 & 0 & 28 \\
 5 & 0 & 1 & 0 & 0 & 0 & 0 & 0 & 0 & 2 & 0 & 0 & 0 & 0 & 0 & 0 & 14 & 0 & 0 & 0 & 0 \\
 6 & 0 & 0 & 0 & 0 & 1 & 0 & 0 & 0 & 0 & 0 & 0 & 6 & 0 & 0 & 0 & 0 & 0 & 0 & 22 & 0 \\
 7 & 0 & 0 & 0 & 0 & 0 & 0 & 0 & 3 & 0 & 0 & 0 & 0 & 0 & 0 & 10 & 0 & 0 & 0 & 0 & 0 \\
\hline
\end{tabular}}\\[5pt]
\caption{Multiplicities of irreducible representations for the 7AB-twisted sectors.}
\end{table}

\begin{table}[h]\scalebox{0.8}{
\begin{tabular}{|l|llllllllllllllllllll|}
\hline
\diagbox[height=25pt]{$i$}{$8 r$} & 0 & 1 & 2 & 3 & 4 & 5 & 6 & 7 & 8 & 9 & 10 & 11 & 12 & 13 & 14 & 15 & 16 & 17 & 18 & 19 \\\hline
%i \\\hline
 1,4 & 0 & 0 & 0 & 0 & 0 & 0 & 0 & 4 & 0 & 0 & 0 & 0 & 0 & 0 & 0 & 24 & 0 & 0 & 0 & 0 \\
 2,3 & 0 & 0 & 0 & 2 & 0 & 0 & 0 & 0 & 0 & 0 & 0 & 12 & 0 & 0 & 0 & 0 & 0 & 0 & 0 & 46 \\
 5 & 0 & 0 & 0 & 0 & 0 & 2 & 0 & 0 & 0 & 0 & 0 & 0 & 0 & 16 & 0 & 0 & 0 & 0 & 0 & 0 \\
 6 & 0 & 1 & 0 & 0 & 0 & 0 & 0 & 0 & 0 & 8 & 0 & 0 & 0 & 0 & 0 & 0 & 0 & 34 & 0 & 0 \\
 7 & 0 & 0 & 0 & 0 & 0 & 4 & 0 & 0 & 0 & 0 & 0 & 0 & 0 & 16 & 0 & 0 & 0 & 0 & 0 & 0 \\
 8 & 0 & 0 & 0 & 0 & 0 & 0 & 0 & 0 & 0 & 8 & 0 & 0 & 0 & 0 & 0 & 0 & 0 & 32 & 0 & 0 \\
 9,13 & 1 & 0 & 0 & 0 & 0 & 0 & 0 & 0 & 6 & 0 & 0 & 0 & 0 & 0 & 0 & 0 & 28 & 0 & 0 & 0 \\
 10,14 & 0 & 0 & 1 & 0 & 0 & 0 & 0 & 0 & 0 & 0 & 9 & 0 & 0 & 0 & 0 & 0 & 0 & 0 & 40 & 0 \\
 11,15 & 0 & 0 & 0 & 0 & 0 & 0 & 4 & 0 & 0 & 0 & 0 & 0 & 0 & 0 & 20 & 0 & 0 & 0 & 0 & 0 \\
 12,16 & 0 & 0 & 0 & 0 & 2 & 0 & 0 & 0 & 0 & 0 & 0 & 0 & 14 & 0 & 0 & 0 & 0 & 0 & 0 & 0 \\\hline
\end{tabular}}\\[5pt]
\caption{Multiplicities of irreducible representations for the 8A-twisted sector.}
\end{table}

\begin{table}[h]\scalebox{0.8}{
\begin{tabular}{|l|llllllllllllllllllll|}
\hline
\diagbox[height=25pt]{$i$}{$ 20 r$} & 3 & 5 & 7 & 9 & 11 & 13 & 15 & 17 & 19 & 21 & 23 & 25 & 27 & 29 & 31 & 33 & 35 & 37 & 39 & 41 \\\hline
%i \\\hline
 1,2 & 0 & 0 & 0 & 0 & 0 & 0 & 0 & 0 & 4 & 0 & 0 & 0 & 0 & 0 & 0 & 0 & 0 & 0 & 22 & 0 \\
 3,4 & 0 & 0 & 0 & 2 & 0 & 0 & 0 & 0 & 0 & 0 & 0 & 0 & 0 & 10 & 0 & 0 & 0 & 0 & 0 & 0 \\
 5,6 & 0 & 0 & 0 & 0 & 2 & 0 & 0 & 0 & 0 & 0 & 0 & 0 & 0 & 0 & 12 & 0 & 0 & 0 & 0 & 0 \\
 7,8 & 0 & 0 & 0 & 0 & 0 & 0 & 0 & 0 & 0 & 6 & 0 & 0 & 0 & 0 & 0 & 0 & 0 & 0 & 0 & 24 \\
 9,10 & 1 & 0 & 0 & 0 & 0 & 0 & 0 & 0 & 0 & 0 & 6 & 0 & 0 & 0 & 0 & 0 & 0 & 0 & 0 & 0 \\
 11,12 & 0 & 0 & 0 & 0 & 0 & 2 & 0 & 0 & 0 & 0 & 0 & 0 & 0 & 0 & 0 & 14 & 0 & 0 & 0 & 0 \\
 13,14 & 0 & 0 & 0 & 0 & 0 & 0 & 3 & 0 & 0 & 0 & 0 & 0 & 0 & 0 & 0 & 0 & 16 & 0 & 0 & 0 \\
 15,16 & 0 & 1 & 0 & 0 & 0 & 0 & 0 & 0 & 0 & 0 & 0 & 7 & 0 & 0 & 0 & 0 & 0 & 0 & 0 & 0 \\
 17,18 & 0 & 0 & 1 & 0 & 0 & 0 & 0 & 0 & 0 & 0 & 0 & 0 & 9 & 0 & 0 & 0 & 0 & 0 & 0 & 0 \\
 19,20 & 0 & 0 & 0 & 0 & 0 & 0 & 0 & 4 & 0 & 0 & 0 & 0 & 0 & 0 & 0 & 0 & 0 & 18 & 0 & 0 \\\hline
\end{tabular}}\\[5pt]
\caption{Multiplicities of irreducible representations for the 10A-twisted sector.}
\end{table}

%\begin{table}\scalebox{0.8}{
%\begin{tabular}{|l|llllllllllllllllllll|}
%\hline
%\diagbox[height=25pt]{$i$}{$ 11 r$} & 0 & 1 & 2 & 3 & 4 & 5 & 6 & 7 & 8 & 9 & 10 & 11 & 12 & 13 & 14 & 15 & 16 & 17 & 18 & 19 \\\hline
%%i \\\hline
% 1 & 0 & 0 & 0 & 0 & 0 & 0 & 0 & 0 & 0 & 0 & 8 & 0 & 0 & 0 & 0 & 0 & 0 & 0 & 0 & 0 \\
% 2 & 2 & 0 & 0 & 0 & 0 & 0 & 0 & 0 & 0 & 0 & 0 & 8 & 0 & 0 & 0 & 0 & 0 & 0 & 0 & 0 \\
% 3 & 0 & 0 & 0 & 0 & 0 & 0 & 0 & 0 & 0 & 0 & 0 & 0 & 12 & 0 & 0 & 0 & 0 & 0 & 0 & 0 \\
% 4 & 0 & 0 & 2 & 0 & 0 & 0 & 0 & 0 & 0 & 0 & 0 & 0 & 0 & 10 & 0 & 0 & 0 & 0 & 0 & 0 \\
% 5 & 0 & 0 & 0 & 2 & 0 & 0 & 0 & 0 & 0 & 0 & 0 & 0 & 0 & 0 & 14 & 0 & 0 & 0 & 0 & 0 \\
% 6 & 0 & 0 & 0 & 0 & 2 & 0 & 0 & 0 & 0 & 0 & 0 & 0 & 0 & 0 & 0 & 16 & 0 & 0 & 0 & 0 \\
% 7 & 0 & 0 & 0 & 0 & 0 & 2 & 0 & 0 & 0 & 0 & 0 & 0 & 0 & 0 & 0 & 0 & 20 & 0 & 0 & 0 \\
% 8 & 0 & 0 & 0 & 0 & 0 & 0 & 4 & 0 & 0 & 0 & 0 & 0 & 0 & 0 & 0 & 0 & 0 & 20 & 0 & 0 \\
% 9 & 0 & 0 & 0 & 0 & 0 & 0 & 0 & 4 & 0 & 0 & 0 & 0 & 0 & 0 & 0 & 0 & 0 & 0 & 26 & 0 \\
% 10 & 0 & 0 & 0 & 0 & 0 & 0 & 0 & 0 & 6 & 0 & 0 & 0 & 0 & 0 & 0 & 0 & 0 & 0 & 0 & 26 \\
% 11 & 0 & 0 & 0 & 0 & 0 & 0 & 0 & 0 & 0 & 6 & 0 & 0 & 0 & 0 & 0 & 0 & 0 & 0 & 0 & 0 \\\hline
%\end{tabular}}\\[5pt]
\begin{table}\scalebox{0.9}{
\begin{tabular}{|l|llllllllllllllllllll|}
\hline
$ 11 r\equiv i-1$ & 0 & 1 & 2 & 3 & 4 & 5 & 6 & 7 & 8 & 9 & 10 & 11 & 12 & 13 & 14 & 15 & 16 & 17 & 18 & 19 \\\hline
%i \\\hline
$h_{i,r}$  
 & 2 & 0 & 2 & 2 & 2 & 2 & 4 & 4 & 6 & 6 & 8 & 8 & 12 & 10 & 14 & 16 & 20 & 20 & 26 & 26\\
 \hline
\end{tabular}}\\[5pt]
\caption{Multiplicities of irreducible representations for the 11A-twisted sector.}
\end{table}

%\begin{table}[h]\scalebox{0.9}{
%\begin{tabular}{|l|llllllllllllllllllll|}
%\hline
%\diagbox[height=25pt]{$i$}{$ 24 r$} & 3 & 5 & 7 & 9 & 11 & 13 & 15 & 17 & 19 & 21 & 23 & 25 & 27 & 29 & 31 & 33 & 35 & 37 & 39 & 41 \\\hline
%%i \\\hline
% 1 & 0 & 0 & 0 & 0 & 0 & 0 & 0 & 0 & 0 & 0 & 8 & 0 & 0 & 0 & 0 & 0 & 0 & 0 & 0 & 0 \\
% 2 & 0 & 0 & 0 & 0 & 2 & 0 & 0 & 0 & 0 & 0 & 0 & 0 & 0 & 0 & 0 & 0 & 18 & 0 & 0 & 0 \\
% 3 & 0 & 0 & 2 & 0 & 0 & 0 & 0 & 0 & 0 & 0 & 0 & 0 & 0 & 0 & 12 & 0 & 0 & 0 & 0 & 0 \\
% 4 & 0 & 0 & 0 & 0 & 0 & 0 & 0 & 0 & 6 & 0 & 0 & 0 & 0 & 0 & 0 & 0 & 0 & 0 & 0 & 0 \\
% 5 & 0 & 0 & 0 & 0 & 0 & 0 & 4 & 0 & 0 & 0 & 0 & 0 & 0 & 0 & 0 & 0 & 0 & 0 & 22 & 0 \\
% 6 & 1 & 0 & 0 & 0 & 0 & 0 & 0 & 0 & 0 & 0 & 0 & 0 & 10 & 0 & 0 & 0 & 0 & 0 & 0 & 0 \\
% 7 & 0 & 2 & 0 & 0 & 0 & 0 & 0 & 0 & 0 & 0 & 0 & 0 & 0 & 12 & 0 & 0 & 0 & 0 & 0 & 0 \\
% 8 & 0 & 0 & 0 & 0 & 0 & 0 & 0 & 4 & 0 & 0 & 0 & 0 & 0 & 0 & 0 & 0 & 0 & 0 & 0 & 24 \\
% 9 & 0 & 0 & 0 & 0 & 0 & 4 & 0 & 0 & 0 & 0 & 0 & 0 & 0 & 0 & 0 & 0 & 0 & 20 & 0 & 0 \\
% 10 & 0 & 0 & 0 & 0 & 0 & 0 & 0 & 0 & 0 & 0 & 0 & 8 & 0 & 0 & 0 & 0 & 0 & 0 & 0 & 0 \\
% 11 & 0 & 0 & 0 & 0 & 0 & 0 & 0 & 0 & 0 & 6 & 0 & 0 & 0 & 0 & 0 & 0 & 0 & 0 & 0 & 0 \\
% 12 & 0 & 0 & 0 & 2 & 0 & 0 & 0 & 0 & 0 & 0 & 0 & 0 & 0 & 0 & 0 & 16 & 0 & 0 & 0 & 0 \\\hline
%\end{tabular}}\\[5pt]
\begin{table}[h]\scalebox{0.9}{
\begin{tabular}{|l|llllllllllllllllllll|}
\hline
$ 24 r\equiv 2i-1$ & 3 & 5 & 7 & 9 & 11 & 13 & 15 & 17 & 19 & 21 & 23 & 25 & 27 & 29 & 31 & 33 & 35 & 37 & 39 & 41 \\\hline
%i \\\hline
 $h_{i,r}$ 
     & 1 & 2 & 2 & 2 & 2 & 4 & 4 & 4 & 6 & 6 & 8 & 8 & 10 & 12 & 12 & 16 & 18 & 20 & 22 & 24 \\
\hline
\end{tabular}}\\[5pt]
\caption{Multiplicities of irreducible representations for the 12A-twisted sector.}
\end{table}

\begin{table}[h]
\scalebox{0.8}{
%\begin{tabular}{|l|lllllllllllllllllllllll|}
%\hline
%\diagbox[height=25pt]{$i$}{$ 144 r$} & 23 & 35 & 47 & 59 & 71 & 83 & 95 & 107 & 119 & 131 & 143 & 155 & 167 & 179 & 191 & 203 & 215 & 227 & 239 & 251 & 263 & 275 & 287 \\\hline
%%i \\\hline
% 1 & 0 & 0 & 0 & 0 & 0 & 0 & 0 & 0 & 0 & 0 & 8 & 0 & 0 & 0 & 0 & 0 & 0 & 0 & 0 & 0 & 0 & 0 & 38 \\
% 2 & 0 & 0 & 0 & 0 & 2 & 0 & 0 & 0 & 0 & 0 & 0 & 0 & 0 & 0 & 0 & 0 & 18 & 0 & 0 & 0 & 0 & 0 & 0 \\
% 3 & 0 & 0 & 2 & 0 & 0 & 0 & 0 & 0 & 0 & 0 & 0 & 0 & 0 & 0 & 14 & 0 & 0 & 0 & 0 & 0 & 0 & 0 & 0 \\
% 4 & 0 & 0 & 0 & 0 & 0 & 0 & 0 & 0 & 6 & 0 & 0 & 0 & 0 & 0 & 0 & 0 & 0 & 0 & 0 & 0 & 30 & 0 & 0 \\
% 5 & 0 & 0 & 0 & 0 & 0 & 0 & 4 & 0 & 0 & 0 & 0 & 0 & 0 & 0 & 0 & 0 & 0 & 0 & 24 & 0 & 0 & 0 & 0 \\
% 6 & 2 & 0 & 0 & 0 & 0 & 0 & 0 & 0 & 0 & 0 & 0 & 0 & 10 & 0 & 0 & 0 & 0 & 0 & 0 & 0 & 0 & 0 & 0 \\
% 7 & 0 & 2 & 0 & 0 & 0 & 0 & 0 & 0 & 0 & 0 & 0 & 0 & 0 & 12 & 0 & 0 & 0 & 0 & 0 & 0 & 0 & 0 & 0 \\
% 8 & 0 & 0 & 0 & 0 & 0 & 0 & 0 & 6 & 0 & 0 & 0 & 0 & 0 & 0 & 0 & 0 & 0 & 0 & 0 & 26 & 0 & 0 & 0 \\
% 9 & 0 & 0 & 0 & 0 & 0 & 4 & 0 & 0 & 0 & 0 & 0 & 0 & 0 & 0 & 0 & 0 & 0 & 20 & 0 & 0 & 0 & 0 & 0 \\
% 10 & 0 & 0 & 0 & 0 & 0 & 0 & 0 & 0 & 0 & 0 & 0 & 10 & 0 & 0 & 0 & 0 & 0 & 0 & 0 & 0 & 0 & 0 & 0 \\
% 11 & 0 & 0 & 0 & 0 & 0 & 0 & 0 & 0 & 0 & 6 & 0 & 0 & 0 & 0 & 0 & 0 & 0 & 0 & 0 & 0 & 0 & 34 & 0 \\
% 12 & 0 & 0 & 0 & 2 & 0 & 0 & 0 & 0 & 0 & 0 & 0 & 0 & 0 & 0 & 0 & 16 & 0 & 0 & 0 & 0 & 0 & 0 & 0 \\\hline
%\end{tabular}}\\[5pt]
\begin{tabular}{|l|llllllllllllllllllll|}
\hline
$ 144 r\equiv 12i-1$ & 23 & 35 & 47 & 59 & 71 & 83 & 95 & 107 & 119 & 131 & 143 & 155 & 167 & 179 & 191 & 203 & 215 & 227 & 239 & 251   \\\hline
%i \\\hline
 $h_{i,r}$  & 2 & 2 & 2 & 2 & 2 & 4 & 4 & 6 & 6 & 6 & 8 & 10 & 10 & 12 & 14 & 16 & 18 & 20 & 24 & 26  \\
\hline
\end{tabular}}\\[5pt]
\caption{Multiplicities of irreducible representations for the 12B-twisted sector.}
\end{table}

\begin{table}[h]\scalebox{0.9}{
%\begin{tabular}{|l|llllllllllllllllllll|}
%\hline
%\diagbox[height=25pt]{$i$}{$ 14 r$} & 0 & 1 & 2 & 3 & 4 & 5 & 6 & 7 & 8 & 9 & 10 & 11 & 12 & 13 & 14 & 15 & 16 & 17 & 18 & 19 \\\hline
%%i \\\hline
% 1 & 0 & 0 & 0 & 0 & 0 & 0 & 0 & 0 & 0 & 0 & 0 & 0 & 0 & 6 & 0 & 0 & 0 & 0 & 0 & 0 \\
% 2 & 0 & 0 & 0 & 0 & 0 & 0 & 2 & 0 & 0 & 0 & 0 & 0 & 0 & 0 & 0 & 0 & 0 & 0 & 0 & 0 \\
% 3 & 0 & 0 & 0 & 0 & 0 & 0 & 0 & 2 & 0 & 0 & 0 & 0 & 0 & 0 & 0 & 0 & 0 & 0 & 0 & 0 \\
% 4 & 1 & 0 & 0 & 0 & 0 & 0 & 0 & 0 & 0 & 0 & 0 & 0 & 0 & 0 & 7 & 0 & 0 & 0 & 0 & 0 \\
% 5 & 0 & 0 & 0 & 0 & 0 & 0 & 0 & 0 & 0 & 0 & 0 & 0 & 0 & 0 & 0 & 8 & 0 & 0 & 0 & 0 \\
% 6 & 0 & 0 & 0 & 0 & 0 & 0 & 0 & 0 & 3 & 0 & 0 & 0 & 0 & 0 & 0 & 0 & 0 & 0 & 0 & 0 \\
% 7 & 0 & 0 & 0 & 0 & 0 & 0 & 0 & 0 & 0 & 4 & 0 & 0 & 0 & 0 & 0 & 0 & 0 & 0 & 0 & 0 \\
% 8 & 0 & 0 & 1 & 0 & 0 & 0 & 0 & 0 & 0 & 0 & 0 & 0 & 0 & 0 & 0 & 0 & 8 & 0 & 0 & 0 \\
% 9 & 0 & 0 & 0 & 2 & 0 & 0 & 0 & 0 & 0 & 0 & 0 & 0 & 0 & 0 & 0 & 0 & 0 & 10 & 0 & 0 \\
% 10 & 0 & 0 & 0 & 0 & 0 & 0 & 0 & 0 & 0 & 0 & 4 & 0 & 0 & 0 & 0 & 0 & 0 & 0 & 0 & 0 \\
% 11 & 0 & 0 & 0 & 0 & 0 & 0 & 0 & 0 & 0 & 0 & 0 & 4 & 0 & 0 & 0 & 0 & 0 & 0 & 0 & 0 \\
% 12 & 0 & 0 & 0 & 0 & 1 & 0 & 0 & 0 & 0 & 0 & 0 & 0 & 0 & 0 & 0 & 0 & 0 & 0 & 12 & 0 \\
% 13 & 0 & 0 & 0 & 0 & 0 & 2 & 0 & 0 & 0 & 0 & 0 & 0 & 0 & 0 & 0 & 0 & 0 & 0 & 0 & 12 \\
% 14 & 0 & 0 & 0 & 0 & 0 & 0 & 0 & 0 & 0 & 0 & 0 & 0 & 6 & 0 & 0 & 0 & 0 & 0 & 0 & 0 \\\hline
%\end{tabular}}\\[5pt]
\begin{tabular}{|l|llllllllllllllllllll|}
\hline
$ 14 r\equiv i-1$ & 0 & 1 & 2 & 3 & 4 & 5 & 6 & 7 & 8 & 9 & 10 & 11 & 12 & 13 & 14 & 15 & 16 & 17 & 18 & 19 \\\hline
%i \\\hline
 $h_{i,r}$ & 1 & 0 & 1 & 2 & 1 & 2 & 2 & 2 & 3 & 4 & 4 & 4 &  6 & 6 & 7 & 8 & 8 & 10 & 12 & 12 \\
\hline
\end{tabular}}\\[5pt]
\caption{Multiplicities of irreducible representations for the 14AB-twisted sectors.}
\end{table}

%\begin{table}[h]\scalebox{0.9}{
%\begin{tabular}{|l|llllllllllllllllllll|}
%\hline
%\diagbox[height=25pt]{$i$}{$ 15 r$} & 0 & 1 & 2 & 3 & 4 & 5 & 6 & 7 & 8 & 9 & 10 & 11 & 12 & 13 & 14 & 15 & 16 & 17 & 18 & 19 \\\hline
%%i \\\hline
% 1 & 0 & 0 & 0 & 0 & 0 & 0 & 0 & 0 & 0 & 0 & 0 & 0 & 0 & 0 & 6 & 0 & 0 & 0 & 0 & 0 \\
% 2 & 0 & 0 & 0 & 0 & 2 & 0 & 0 & 0 & 0 & 0 & 0 & 0 & 0 & 0 & 0 & 0 & 0 & 0 & 0 & 10 \\
% 3 & 0 & 0 & 0 & 0 & 0 & 0 & 0 & 0 & 0 & 2 & 0 & 0 & 0 & 0 & 0 & 0 & 0 & 0 & 0 & 0 \\
% 4 & 0 & 0 & 0 & 0 & 0 & 1 & 0 & 0 & 0 & 0 & 0 & 0 & 0 & 0 & 0 & 0 & 0 & 0 & 0 & 0 \\
% 5 & 0 & 0 & 0 & 0 & 0 & 0 & 0 & 0 & 0 & 0 & 4 & 0 & 0 & 0 & 0 & 0 & 0 & 0 & 0 & 0 \\
% 6 & 1 & 0 & 0 & 0 & 0 & 0 & 0 & 0 & 0 & 0 & 0 & 0 & 0 & 0 & 0 & 6 & 0 & 0 & 0 & 0 \\
% 7 & 0 & 0 & 0 & 0 & 0 & 0 & 0 & 0 & 0 & 0 & 0 & 4 & 0 & 0 & 0 & 0 & 0 & 0 & 0 & 0 \\
% 8 & 0 & 0 & 0 & 0 & 0 & 0 & 0 & 0 & 0 & 0 & 0 & 0 & 0 & 0 & 0 & 0 & 8 & 0 & 0 & 0 \\
% 9 & 0 & 0 & 0 & 0 & 0 & 0 & 2 & 0 & 0 & 0 & 0 & 0 & 0 & 0 & 0 & 0 & 0 & 0 & 0 & 0 \\
% 10 & 0 & 0 & 1 & 0 & 0 & 0 & 0 & 0 & 0 & 0 & 0 & 0 & 0 & 0 & 0 & 0 & 0 & 7 & 0 & 0 \\
% 11 & 0 & 0 & 0 & 0 & 0 & 0 & 0 & 2 & 0 & 0 & 0 & 0 & 0 & 0 & 0 & 0 & 0 & 0 & 0 & 0 \\
% 12 & 0 & 0 & 0 & 0 & 0 & 0 & 0 & 0 & 0 & 0 & 0 & 0 & 5 & 0 & 0 & 0 & 0 & 0 & 0 & 0 \\
% 13 & 0 & 0 & 0 & 0 & 0 & 0 & 0 & 0 & 3 & 0 & 0 & 0 & 0 & 0 & 0 & 0 & 0 & 0 & 0 & 0 \\
% 14 & 0 & 0 & 0 & 0 & 0 & 0 & 0 & 0 & 0 & 0 & 0 & 0 & 0 & 4 & 0 & 0 & 0 & 0 & 0 & 0 \\
% 15 & 0 & 0 & 0 & 1 & 0 & 0 & 0 & 0 & 0 & 0 & 0 & 0 & 0 & 0 & 0 & 0 & 0 & 0 & 10 & 0 \\\hline
%\end{tabular}}\\[5pt]
\begin{table}[h]\scalebox{0.9}{
\begin{tabular}{|l|llllllllllllllllllll|}
\hline
$ 15 r\equiv i-1$ & 0 & 1 & 2 & 3 & 4 & 5 & 6 & 7 & 8 & 9 & 10 & 11 & 12 & 13 & 14 & 15 & 16 & 17 & 18 & 19 \\\hline
%i \\\hline
 $h_{i,r}$ & 1 & 0 & 1 & 1 & 2 & 1 & 2 & 2 & 3 & 2 & 4 & 4 & 5 & 4 & 6 & 6 & 8 & 7 & 10 & 10 \\
\hline
\end{tabular}}\\[5pt]
\caption{Multiplicities of irreducible representations for the 15AB-twisted sectors.}
\end{table}

\begin{table}[h]\scalebox{0.8}{
\begin{tabular}{|l|llllllllllllllllllll|}
\hline
$ 63 r\equiv 3i-1$ & 8 & 11 & 14 & 17 & 20 & 23 & 26 & 29 & 32 & 35 & 38 & 41 & 44 & 47 & 50 & 53 & 56 & 59 & 62 & 65 \\\hline
%i \\\hline
  $h_{i,r}$ & 7 & 7 & 13 & 0 & 14 & 7 & 14 & 7 & 14 & 26 & 14 & 14 & 21 & 14 & 28 & 21 & 58 & 28 & 588 & 28 \\
 \hline
\end{tabular}}\\[5pt]
\caption{Multiplicities of irreducible representations for the 21AB-twisted sectors.}
\end{table}

\begin{table}[h]\scalebox{0.9}{
\begin{tabular}{|l|llllllllllllllllllll|}
\hline
$ 23 r\equiv i-1$ & 0 & 1 & 2 & 3 & 4 & 5 & 6 & 7 & 8 & 9 & 10 & 11 & 12 & 13 & 14 & 15 & 16 & 17 & 18 & 19 \\\hline
%i \\\hline
  $h_{i,r}$ & 1 & 0 & 0 & 1 & 1 & 0 & 1 & 0 & 2 & 1 & 2 & 2 & 2 & 1 & 2 & 2 & 2 & 2 & 3 & 2 \\
 \hline
\end{tabular}}\\[5pt]
\caption{Multiplicities of irreducible representations for the 23AB-twisted sectors.}
\end{table}
\clearpage

%\newpage

\end{document}